\documentclass[12pt]{article}

\pdfoutput=1
\usepackage{jheppub}
\usepackage{graphicx,epsfig,amsmath,amssymb}
\usepackage{subcaption}
\usepackage{booktabs}
\graphicspath{{feyndiags/}}

\newcommand\TwoFigBottom{-2}

\newcommand{\als}{\ensuremath{\alpha_s}}

\def\be{\begin{equation}}
\def\ee{\end{equation}}
\def\bea{\begin{eqnarray}}
\def\eea{\end{eqnarray}}
\def\nn{\nonumber}

\newcommand{\gosam}{\textsc{GoSam}{}}

\newcommand{\cg}{c_{ggh}}
\newcommand{\cgg}{c_{gghh}}
\newcommand{\ctt}{c_{tt}}
\newcommand{\ct}{c_{t}}

\newcommand{\chhh}{c_{hhh}}
\newcommand{\mhh}{m_{hh}}
\newcommand{\pth}{p_{T,h}}
\newcommand{\ftapprox}{FT$_{\mathrm{approx}}$}

\title{Higgs boson pair production in non-linear Effective Field Theory with full $m_t$-dependence at NLO QCD}

\author[a]{G.~Buchalla,}
\author[b]{M.~Capozi,}
\author[a]{A.~Celis,}
\author[b]{G.~Heinrich,}
\author[b]{L.~Scyboz}

\affiliation[a]{Ludwig-Maximilians-Universit{\"a}t M{\"u}nchen, Fakult\"at f\"ur Physik,
Arnold Sommerfeld Center for Theoretical Physics, 80333 M\"unchen, Germany}
\affiliation[b]{Max Planck Institute for Physics, F\"ohringer Ring 6,
  80805 M\"unchen, Germany}

\emailAdd{gerhard.buchalla@physik.uni-muenchen.de}
\emailAdd{mcapozi@mpp.mpg.de}
\emailAdd{alejandro.celis@physik.uni-muenchen.de}
\emailAdd{gudrun@mpp.mpg.de}
\emailAdd{scyboz@mpp.mpg.de}

\preprint{{\small  LMU-ASC 34/18, MPP-2018-127}}

\abstract{
 We present a calculation of the NLO QCD corrections to Higgs boson
 pair production within the framework of a non-linearly realised
 Effective Field Theory in the Higgs sector, described by the electroweak chiral Lagrangian.
 We analyse how the NLO corrections affect distributions in the Higgs
 boson pair invariant mass and the transverse momentum of one of the Higgs bosons.      
We find that these corrections lead to significant and non-homogeneous K-factors in certain regions of the parameter space.     
We also provide an analytical parametrisation for the total
cross-section and the $m_{hh}$ distribution as a function of the anomalous Higgs couplings 
that includes NLO corrections.  Such a parametrisation can be useful for phenomenological studies.    
}


\begin{document}

\maketitle

\section{Introduction}

Exploring the Higgs sector and the mechanism of electroweak symmetry breaking is one of the primary goals for the current and future LHC program as well as other planned experiments. 
While some of the properties of the Higgs boson, like its mass, spin and couplings to electroweak bosons, have been measured meanwhile impressively well~\cite{Khachatryan:2016vau}, other parameters, like the couplings to (light) fermions, and in particular the self-coupling, are still largely unconstrained and leave room for physics beyond the Standard Model, see e.g. Ref.~\cite{Brooijmans:2018xbu} for a recent review.

In the Standard Model (SM) the strength of all Higgs boson couplings is 
predicted; however, effects of physics beyond the Standard Model (BSM) may 
lead to deviations which, once firmly established, are a clear sign of 
New Physics. Since higher-order QCD corrections are known to be important in 
Higgs boson production processes, they need to be taken into account 
to improve the sensitivity to New Physics effects.

Given the energy gap between the electroweak scale at $v\simeq 250$\,GeV and a New Physics scale $\Lambda$ which is supposed to be in the TeV range, it is natural to parametrise the BSM effects in a model-independent way in an Effective Field Theory (EFT) framework.  Such a framework can be formulated in various ways, where we can distinguish two main categories, often called ``linear EFT'' and ``non-linear EFT''. The linear EFTs~\cite{Buchmuller:1985jz,Grzadkowski:2010es}, 
also known as ``SMEFT''~\cite{Berthier:2015oma,Ghezzi:2015vva,deBlas:2017wmn,Brivio:2017bnu,Ellis:2018gqa}, 
are organised by canonical dimensions, formulated as power series in the dimensionful parameter $1/\Lambda$. The non-linear EFTs are organised by chiral dimensions.
The corresponding formalism, including a light Higgs boson, has been developed in Refs.~\cite{Feruglio:1992wf,Bagger:1993zf,Koulovassilopoulos:1993pw,Burgess:1999ha,Wang:2006im,Grinstein:2007iv,Contino:2010mh,Contino:2010rs,Alonso:2012px,
Buchalla:2013rka,Delgado:2013hxa,Buchalla:2013eza,Buchalla:2015qju,deBlas:2018tjm} and usually goes by the name ``Electroweak Chiral Lagrangian'' (EWChL).
We will work in the ``non-linear EFT'' framework, where the Higgs field is an electroweak singlet. 
The main benefit of this approach is that the anomalous Higgs couplings
are singled out, in a systematic way, as the dominant New Physics effects
in the electroweak sector. 

Higgs boson pair production in gluon fusion is the most promising process to find out whether the Higgs boson self-coupling is Standard-Model-like.
Early studies of Higgs boson pair production within an EFT framework can be found in Refs.~\cite{Pierce:2006dh,Contino:2012xk,Baglio:2012np,Dawson:2012mk}.
Many phenomenological investigations about the potential of this process to reveal New Physics have been performed since, see e.g. Refs.~\cite{Dolan:2012rv,Goertz:2014qta,Barr:2014sga,Azatov:2015oxa,Dolan:2015zja,Behr:2015oqq,Maltoni:2016yxb,Kling:2016lay,Cao:2016zob,DiVita:2017eyz,DiLuzio:2017tfn,Corbett:2017ieo,Dawson:2017vgm,Alves:2017ued,Adhikary:2017jtu,Kim:2018uty,Goncalves:2018qas}.

In the SM, Higgs boson pair production has been calculated at leading order in Refs.~\cite{Eboli:1987dy,Glover:1987nx,Plehn:1996wb}.
As it is a loop-induced process, higher order calculations with full top quark mass dependence involve multi-scale two-loop integrals. Therefore, the NLO calculations until recently have been performed in the $m_t\to\infty$ limit~\cite{Dawson:1998py} 
also called HEFT (``Higgs Effective Field Theory''),\footnote{Sometimes the electroweak chiral Lagrangian with a light Higgs
boson is also referred to as {\it Higgs Effective Field Theory (HEFT)\/} in the
literature. The two EFTs are unrelated and should be carefully
distinguished. Here we employ the term 
{\it electroweak chiral Lagrangian\/} for the non-linear EFT of physics beyond
the SM, and reserve the expression {\it HEFT\/} for the heavy-top limit
in Higgs interactions.}
and then rescaled by a factor $B_{FT}/B_{HEFT}$, $B_{FT}$
denoting the leading order matrix element squared in the full theory.
This procedure is called ``Born-improved HEFT'' in the following.
In Refs.~\cite{Frederix:2014hta,Maltoni:2014eza}, an approximation called
``\ftapprox'' was introduced, which contains the full top quark
mass dependence in the real radiation, while the virtual part is
calculated in the HEFT approximation and rescaled at the event level by
the re-weighting factor $B_{FT}/B_{HEFT}$.

In addition, the HEFT results at NLO and NNLO have been improved by an
expansion in $1/m_t^{2}$ in
Refs.~\cite{Grigo:2013rya,Grigo:2014jma,Grigo:2015dia,Degrassi:2016vss}.
The NNLO QCD corrections in the heavy-top limit have been computed in
Refs.~\cite{deFlorian:2013uza,deFlorian:2013jea,Grigo:2014jma,deFlorian:2016uhr},
and they have been supplemented by an expansion in $1/m_t^2$ in
Ref.~\cite{Grigo:2015dia} and by threshold resummation, at NLO+NNLL in
Ref.~\cite{Shao:2013bz} and at NNLO+NNLL in
Ref.~\cite{deFlorian:2015moa}, leading to K-factors of about 1.2
relative to the Born-improved HEFT result.

The full NLO corrections, including the top quark mass
dependence also in the virtual two-loop amplitudes, have been
calculated in Ref.~\cite{Borowka:2016ehy}.
Phenomenological studies at 14\,TeV and 100\,TeV, including variations of the Higgs boson self-coupling, have been presented in Ref.~\cite{Borowka:2016ypz}.  
The full NLO calculation was supplemented by NLL transverse momentum resummation in Ref.~\cite{Ferrera:2016prr}. 
It also has been matched to parton shower Monte Carlo programs~\cite{Heinrich:2017kxx,Jones:2017giv}, 
where the matched result of Ref.~\cite{Heinrich:2017kxx} is publicly available within the {\tt POWHEG-BOX-V2} framework.

Recent work also includes a combination of an analytic threshold expansion and a large-$m_t$ expansion together with a Pad{\'e} approximation framework~\cite{Grober:2017uho}, 
and analytic results based on a high energy expansion for the planar part of the two-loop amplitude~\cite{Davies:2018ood}.

Very recently, top quark mass effects have been incorporated in the NNLO HEFT calculation, including the full NLO result and combining
one-loop double-real corrections with full top mass dependence with suitably reweighted
real-virtual and double-virtual contributions evaluated in the large-$m_t$ approximation~\cite{Grazzini:2018bsd}.

Within a non-linear EFT framework, higher order QCD corrections have been 
performed in the $m_t\to\infty$ limit.
The NLO QCD corrections have been calculated in Ref.~\cite{Grober:2015cwa}, recently also supplemented with the case of CP-violating Higgs sectors~\cite{Grober:2017gut}. 
The NNLO QCD corrections in the $m_t\to\infty$ limit including dimension~6 operators have been presented in Ref.~\cite{deFlorian:2017qfk}.
These calculations found rather flat K-factors, which however could be an artefact of the $m_t\to\infty$ limit. 
One of the main goals of the present paper is to investigate whether this feature is preserved once the full top quark mass dependence is taken into account.
We calculate the NLO QCD corrections to Higgs boson pair production in gluon fusion within the non-linear EFT framework, 
retaining the full top quark mass dependence, 
based on the numerical approach developed in Ref.~\cite{Borowka:2016ehy}.
In order to quantify the different effects of the five operators and corresponding couplings that can lead to deviations from the SM in the Higgs sector, we give results for the total NLO cross section parametrised in terms of 23 coefficients of all possible combinations of these couplings, as introduced at LO in Refs.~\cite{Azatov:2015oxa,Carvalho:2015ttv}. 
We also show differential distributions for 
12 benchmarks points which should be characteristic for clusters of BSM scenarios. Such clusters were identified in  Refs.~\cite{Carvalho:2015ttv,Carvalho:2016rys,Carvalho:2017vnu} at leading order and represent partitions of the BSM parameter space according to the shape of the differential distributions.
We demonstrate that there are regions where the NLO corrections lead to substantial and non-homogenous K-factors and provide numbers for the 
parametrisation of the NLO cross section, which can be used in subsequent phenomenological studies.

This paper is organised as follows. 
In Section \ref{sec:calculation}, we explain the framework of the calculation. 
In particular, we introduce the Higgs-electroweak chiral Lagrangian and describe how it is applied to Higgs boson pair production, including the NLO QCD corrections. Section \ref{sec:results} is dedicated to the phenomenological results. We provide a parametrisation of the NLO cross section in terms of coefficients of all combinations of couplings occurring in the NLO cross section. Based on this parametrisation we show heat maps both at LO and at NLO, where we vary two couplings while keeping the others fixed to the SM values. Then we give results for total cross sections and differential distributions at twelve benchmark points and discuss their implications before we conclude. 
An appendix explains the conventions used for the tables containing the {\em differential} coefficients of the couplings in the Higgs boson pair invariant mass distribution. The values are available in {\tt csv} format as ancillary files to the arXiv submission and the JHEP publication. 
A further appendix compares the treatment of Higgs-pair production 
   in the Higgs-electroweak chiral Lagrangian and in SMEFT.

\section{Details of the calculation}
\label{sec:calculation}

\subsection{The Higgs-electroweak chiral Lagrangian}
\label{sec:ewchl}

In the present analysis, we will describe the potential impact of physics
beyond the Standard Model through the electroweak chiral Lagrangian
including a light Higgs 
boson~\cite{Buchalla:2013rka,Buchalla:2013eza,Buchalla:2017jlu}.
This framework provides us with a consistent effective field theory (EFT)
for New Physics in the Higgs sector, as we will summarise in the following.

To leading order the Lagrangian is given by
\begin{eqnarray}\label{l2}
{\cal L}_2 &=& -\frac{1}{2} \langle G_{\mu\nu}G^{\mu\nu}\rangle
-\frac{1}{2}\langle W_{\mu\nu}W^{\mu\nu}\rangle
-\frac{1}{4} B_{\mu\nu}B^{\mu\nu}
+\sum_{\psi=q_L,l_L,u_R,d_R,e_R}\bar \psi i\!\not\!\! D\psi
\nonumber\\
&& +\frac{v^2}{4}\ \langle D_\mu U^\dagger D^\mu U\rangle\, \left( 1+F_U(h)\right)
+\frac{1}{2} \partial_\mu h \partial^\mu h - V(h) \nonumber\\
&& - v \left[ \bar q_L \left( Y_u +
       \sum^\infty_{n=1} Y^{(n)}_u \left(\frac{h}{v}\right)^n \right) U P_+ q_R
+ \bar q_L \left( Y_d +
     \sum^\infty_{n=1} Y^{(n)}_d \left(\frac{h}{v}\right)^n \right) U P_- q_R
  \right. \nonumber\\
&& \quad\quad\left. + \bar l_L \left( Y_e +
   \sum^\infty_{n=1} Y^{(n)}_e \left(\frac{h}{v}\right)^n \right) U P_- l_R
+ {\rm h.c.}\right]\;.
\end{eqnarray}
The first line is the unbroken SM, the remainder represents the Higgs sector. 
Here $h$ is the Higgs field and $U = \exp(2i\varphi^a T^a/v)$ encodes the 
electroweak Goldstone fields $\varphi^a$, with $T^a$ the generators 
of $SU(2)$. $v$ is the electroweak vacuum expectation value, $P_\pm = 1/2\pm T_3$, and
\begin{equation}\label{dcovu}
D_\mu U=\partial_\mu U+i g W_\mu U -i g' B_\mu U T_3\;.
\end{equation}
The trace of a matrix $A$ is denoted by $\langle A\rangle$. The left-handed
doublets of quarks and leptons are written as $q_L$ and $l_L$, the
right-handed singlets as $u_R$, $d_R$, $e_R$. Generation indices are omitted.
In the Yukawa terms the right-handed quark and lepton fields are collected
into $q_R=(u_R,d_R)^T$ and $l_R=(0,e_R)^T$, respectively. In general,
different flavour couplings $Y^{(n)}_{u,d,e}$ can arise at every order in
the Higgs field $h^n$, in addition to the usual Yukawa matrices $Y_{u,d,e}$.
The $h$-dependent functions are
\begin{equation}\label{fuv}
F_U(h)=\sum^\infty_{n=1} f_{U,n} \left(\frac{h}{v}\right)^n,\qquad
V(h)=v^4\sum^\infty_{n=2} f_{V,n} \left(\frac{h}{v}\right)^n\;.
\end{equation}

In the limit where
\begin{equation}\label{smlimit}
f_{U,1}=2,\quad f_{U,2}=1,
\quad f_{V,2}=f_{V,3}=\frac{m^2_h}{2v^2}, \quad f_{V,4}=\frac{m^2_h}{8v^2},
\quad Y^{(1)}_f=Y_f,
\end{equation}
and all other couplings $f_{U,n}$, $f_{V,n}$, $Y^{(n)}_f$ equal to zero,
the Lagrangian in (\ref{l2}) reduces to the usual SM. For generic values of
those parameters, the Lagrangian describes the SM with arbitrary modifications 
in the Higgs couplings. While the deviations of these couplings from their 
SM values could, in principle, be of order unity, the parametrisation in
(\ref{l2}) remains relevant as long as the anomalous Higgs couplings are 
the dominant New Physics effects at electroweak energies. 
Employing (\ref{l2}), we assume that this is the case. Such a hypothesis
remains to be tested experimentally. We emphasise, however, that the assumption
is well motivated by the current status of Higgs coupling measurements.
These still allow deviations from the SM of  $10$\,--\,$20\%$ or more,
considerably larger than the typical precision of $1\%$ reached in the
electroweak gauge sector \cite{ALEPH:2005ab}.
A useful property of the Lagrangian (\ref{l2}) is therefore that it 
allows us to concentrate on anomalous Higgs couplings in a 
systematic way~\cite{Buchalla:2015wfa,Buchalla:2015qju}.

In fact, the intuitive picture of introducing (\ref{l2}) as the SM with
modified Higgs couplings can be formulated as a consistent EFT. Because of 
the need to write the modified Higgs couplings in a gauge-invariant way,
the Higgs field has to be represented as an electroweak singlet $h$,
independent of the Goldstone matrix $U = \exp(2i\varphi^a T^a/v)$.
The latter transforms as $U\to g_L U g^\dagger_Y$ under the SM gauge group.
The symmetry is non-linearly realised on the Goldstone fields $\varphi^a$.
The Lagrangian (\ref{l2}) is then nonrenormalisable (in the traditional sense)
as it contains interaction terms of arbitrary canonical dimension.
The EFT is therefore not organised by the canonical dimension of operators,
but rather by {\it chiral counting} in analogy to the chiral perturbation 
of pions in QCD.
Chiral counting is equivalent to an expansion in loop orders $L$, which can
be conveniently counted by assigning {\it chiral dimensions} $d_\chi\equiv 2L+2$
to fields and weak couplings. This assignment is simply $0$ for bosons, 
and $1$ for each derivative, fermion bilinear and weak coupling:
\begin{equation}\label{chidim}
d_\chi(A_\mu,\varphi,h)=0\, ,\qquad d_\chi(\partial,\bar\psi\psi,g,y)=1\;.
\end{equation}
Here $A_\mu$ represents a generic gauge field, $\varphi$ the Goldstone
bosons, and $h$ the Higgs scalar. $g$ denotes any of the SM gauge couplings
$g$, $g'$, $g_s$, and $y$ any other weak coupling, such as the Yukawa
couplings or the square-roots of the parameters $f_{V,n}$ in the Higgs potential.

Based on this counting, the leading-order expression (\ref{l2}) can be 
constructed from the SM field content and symmetries as the most general
Lagrangian of chiral dimension 2. Leading processes are described by
tree-level amplitudes from (\ref{l2}). Next-to-leading order effects come
from one-loop contributions of (\ref{l2}) and from tree-level terms of
the NLO Lagrangian ${\cal L}_4$. Both are considered to be of 
`one-loop order', or chiral dimension $d_\chi=4$.

\vspace*{0.4cm}

We next apply this framework to Higgs-pair production $gg\to hh$. 
Since this process is loop-induced, at leading order
both one-loop diagrams built from the LO interactions, as
well as tree contributions from the NLO Lagrangian have to be taken into 
account. The relevant terms from the effective Lagrangian 
${\cal L}_2 + {\cal L}_4$ are given by  \cite{deFlorian:2016spz}
\begin{align}
{\cal L}\supset 
-m_t\left(c_t\frac{h}{v}+c_{tt}\frac{h^2}{v^2}\right)\,\bar{t}\,t -
c_{hhh} \frac{m_h^2}{2v} h^3+\frac{\alpha_s}{8\pi} \left( c_{ggh} \frac{h}{v}+
c_{gghh}\frac{h^2}{v^2}  \right)\, G^a_{\mu \nu} G^{a,\mu \nu}\;.
\label{eq:ewchl}
\end{align}
The first three couplings, $c_t$, $c_{tt}$, $c_{hhh}$, are from ${\cal L}_2$,
the Higgs-gluon couplings $c_{ggh}$ and $c_{gghh}$ from ${\cal L}_4$
\cite{Buchalla:2013rka,Buchalla:2015wfa}.
To lowest order in the SM $c_t=c_{hhh}=1$ and $c_{tt}=c_{ggh}=c_{gghh}=0$.
In general, all couplings may have arbitrary values of ${\cal O}(1)$.
Note that we have extracted a loop factor from the definition of the
Higgs-gluon couplings.  

\begin{figure*}[h]
\begin{center}
\includegraphics[width=13cm]{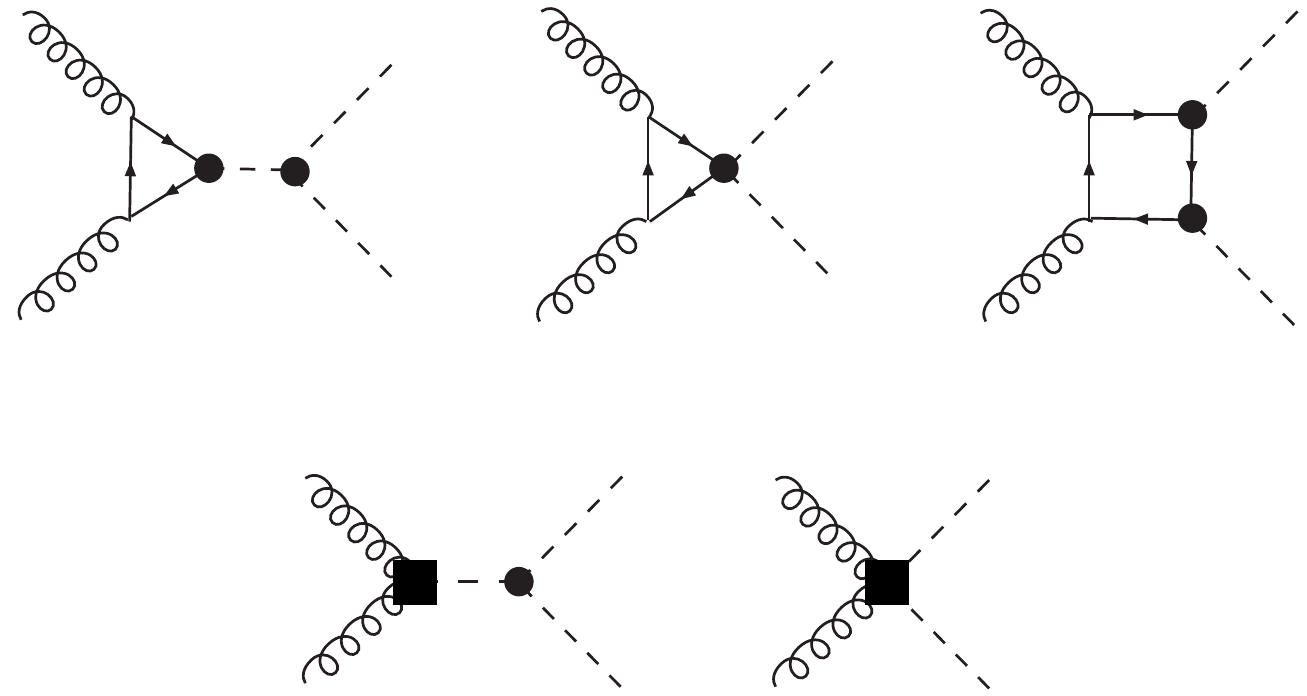}
\end{center}
\caption{Higgs-pair production in gluon fusion at leading order
in the chiral Lagrangian. The black dots indicate vertices from ${\cal L}_2$,
the black squares denote local terms from ${\cal L}_4$.}
\label{fig:hprocess}
\end{figure*}

The leading-order diagrams are shown in Fig.~\ref{fig:hprocess}.
All diagrams are at the same order in the chiral counting
(chiral dimension~4, equivalent to one-loop order).
They illustrate the interplay between leading order anomalous couplings
(black dots) within loops, and next-to-leading order terms
(black squares) at tree level. All the five couplings defined 
in (\ref{eq:ewchl}) appear in Fig.~\ref{fig:hprocess}.
In the following section we discuss the extension of this analysis to
the next order in QCD.



\subsection{Calculation of the NLO QCD corrections}
\label{sec:2loop}

Within the framework of the electroweak chiral Lagrangian, 
the calculation of the $gg\to hh$ amplitude can be extended to
the next order in the loop expansion, that is to two-loop order,
or chiral dimension 6. In full generality, this would require to
also include two-loop electroweak corrections and local terms
from the Lagrangian at chiral dimension 6. The latter introduce
additional couplings, parametrising subleading new-physics effects.
Such effects are beyond the experimental sensitivity in the foreseeable
future, given that even the determination of the LO couplings in (\ref{eq:ewchl})
remains a substantial challenge. On the other hand, radiative corrections from
QCD are known to be very important for $gg\to hh$ and similar processes.

\begin{figure*}[t]
\begin{center}
\includegraphics[width=14cm]{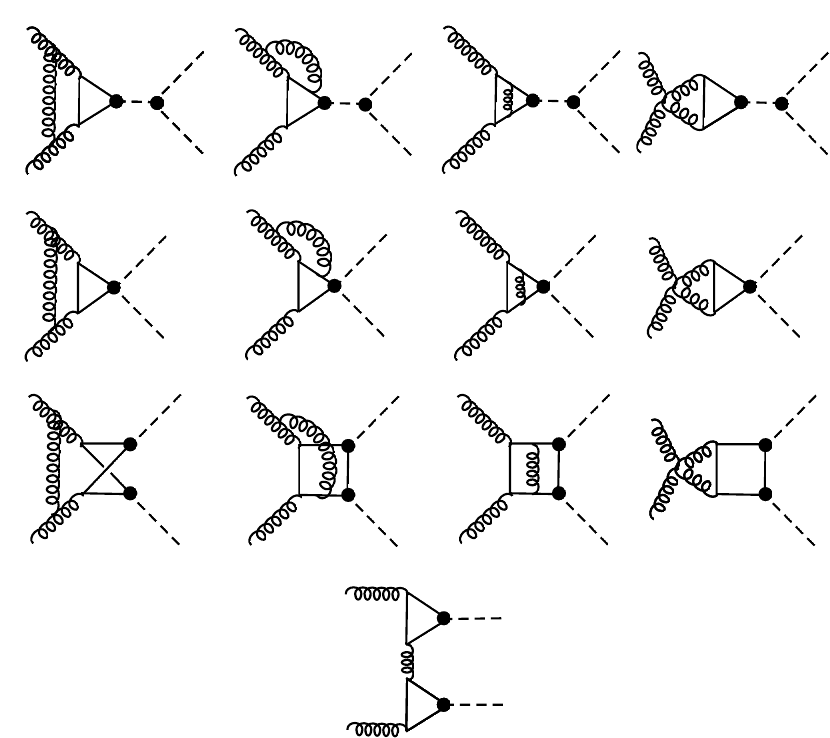}
\end{center}
\caption{Higgs-pair production in gluon fusion at NLO:
Examples for virtual two-loop diagrams at order $g^4_s$.}
\label{fig:hpnlov2}
\end{figure*}

\begin{figure*}[t]
\begin{center}
\includegraphics[width=10cm]{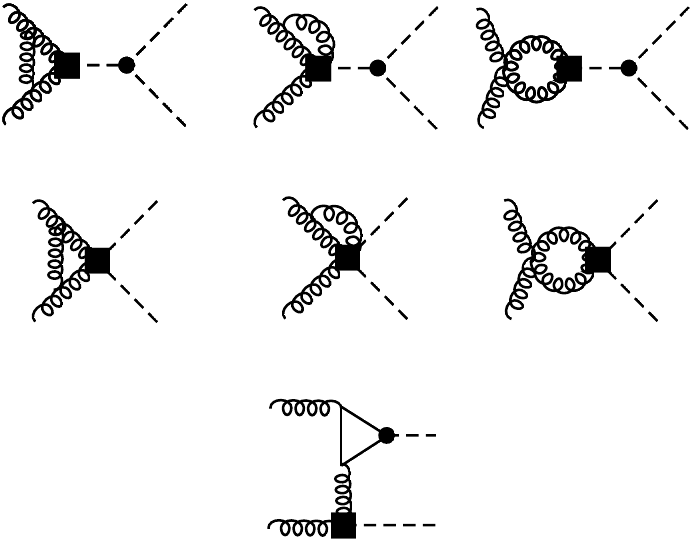}
\end{center}
\caption{Higgs-pair production in gluon fusion at NLO:
Examples for virtual one-loop diagrams at order $g^4_s$.}
\label{fig:hpnlov1}
\end{figure*}

\begin{figure*}[t]
\begin{center}
\includegraphics[width=3cm]{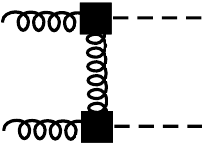}
\end{center}
\caption{Higgs-pair production in gluon fusion at NLO:
Tree diagram at order $g^4_s$.}
\label{fig:hpnlov0}
\end{figure*}

For this reason, we extend the calculation of $gg\to hh$ to the next order in
the non-linear EFT, but restrict the NLO corrections to the effects from QCD.
Within the systematics of the EWChL this approximation corresponds to 
including those corrections at chiral dimension 6 that come with a relative 
factor of the QCD coupling $g^2_s$. This procedure is consistent without
introducing further anomalous couplings, beyond the ones in (\ref{eq:ewchl}),
because this effective Lagrangian is renormalisable with respect to 
QCD \cite{Buchalla:2015qju}. Since the LO amplitude for $gg\to hh$ scales
as $\sim g^2_s$, the NLO virtual corrections of interest to us comprise
all the diagrams at two-loop order carrying a factor of $g^4_s$. They exist
as two-loop, one-loop and tree topologies, as illustrated in
Figs.~\ref{fig:hpnlov2}, \ref{fig:hpnlov1} and  \ref{fig:hpnlov0},
respectively. 

\begin{figure*}[t]
\begin{center}
\includegraphics[width=14cm]{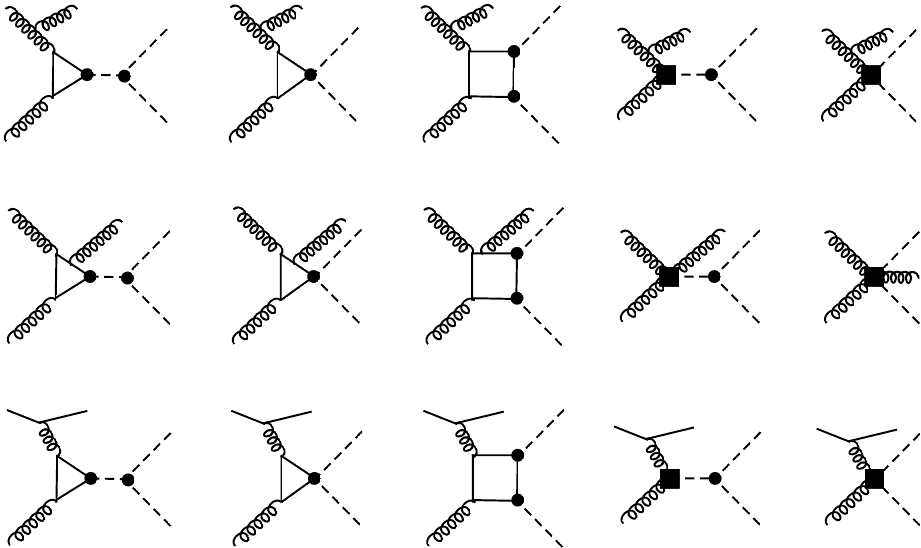}
\end{center}
\caption{Higgs-pair production in gluon fusion at NLO:
Examples for real-emission diagrams at order $g^3_s$.}
\label{fig:hpnlore}
\end{figure*}

In addition, real emission diagrams at ${\cal O}(g^3_s)$
have to be included as shown in Fig. \ref{fig:hpnlore}.

To further clarify our approximation with respect to the full chiral
expansion at NLO, we give in Fig. \ref{fig:hpdrop} a few examples of
higher-order effects that are consistently neglected in our scheme:
\begin{figure*}[t]
\begin{center}
\includegraphics[width=10cm]{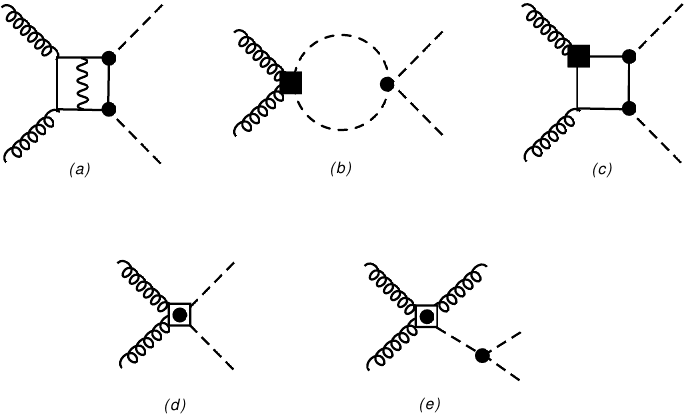}
\end{center}
\caption{Higgs-pair production in gluon fusion at NLO:
Examples for contributions that are consistently
neglected within our approximation. The dotted square indicates
a local term at chiral dimension 6 (two-loop order). See text for
further explanation.}
\label{fig:hpdrop}
\end{figure*} 

Example (a) shows a correction from electroweak-boson
exchange. It is of two-loop order, but scales as $g^2_s g^2$, rather
than $g^4_s$. It is not a NLO QCD effect and we neglect it here.

Similarly, the one-loop topology in (b) counts as two-loop order,
but scales only as $g^2_s c_{hhhh}$, with $c_{hhhh}$ the (anomalous)
quartic Higgs coupling.

In example (c) we consider an anomalous top-gluon coupling
of the form $Q_{ttG}=y_t g_s\bar t_L\sigma_{\mu\nu}G^{\mu\nu}t_R$, where
the top Yukawa coupling reflects the change in chirality.
This operator is therefore (at least) of chiral dimension 4 
(one-loop order) and the diagram in Fig. \ref{fig:hpdrop} (c) 
of two-loop order, but again not of order $g^4_s$.
Since (\ref{l2}) assumes that the top quark is weakly coupled to
the (possibly strongly interacting) new-physics sector, it is more
likely that the operator comes with further weak couplings from
$t_L$ and $t_R$ and thus carries chiral dimension 6. In this case,
diagram (c) is of three-loop order and clearly negligible.
The effect of the chromomagnetic operator on single Higgs boson
production has been calculated recently in the context of SMEFT in Ref.~\cite{Deutschmann:2017qum}.

Example (d) illustrates the effect of a local Higgs-gluon interaction
of chiral dimension~6, which enters at two-loop order as a tree-level
topology. A possible operator would be 
$g^2_s G^a_{\mu\nu} G^{a,\mu\nu}\partial_\lambda h \partial^\lambda h$.
However, this effect, although of two-loop order, does not scale as~$g^4_s$.

Finally, we may have an operator
$g^3_s f^{abc} G^a_{\mu\nu} G^{b,\nu}_{\ \ \lambda}  G^{c,\lambda\mu} h$, also 
of chiral dimension~6. Diagram (e) then amounts to a two-loop order
interaction with real emission, which is beyond our approximation.

\vspace*{0.5cm}

At the technical level, the NLO QCD corrections have been calculated building on the setup described in Refs.~\cite{Borowka:2016ehy,Borowka:2016ypz}, summarised briefly below.

\subsubsection{Virtual corrections}
The virtual part of order $\alpha_s^3$ consists of genuine two-loop diagrams as well as one-loop and tree-level diagrams, 
see Figs.~\ref{fig:hpnlov2},~\ref{fig:hpnlov1} and \ref{fig:hpnlov0}.

For the two-loop part, we made use of the numerical results for the two-loop virtual diagrams in the Standard Model (SM) by dividing them into two classes: 
diagrams containing the Higgs-boson self-coupling (``triangle-type''), and diagrams without (``box-type''). 
The $t\bar{t}hh$ coupling generates new two-loop topologies, see e.g. the second line of Fig.~\ref{fig:hpnlov2}.
The results for these diagrams however can be obtained from the SM
triangle-type diagrams by multiplying them with the inverse Higgs boson propagator and rescaling the couplings, i.e. multiplying with $\ctt/\chhh$. 
The other two-loop diagrams occurring in our calculation have the same topologies as in the SM and therefore can be derived from the SM results by rescaling of the couplings $\ct$ and $\chhh$.

The one-loop part containing the Higgs-gluon contact interactions has been calculated in two ways: first, using \gosam~\cite{Cullen:2011ac,Cullen:2014yla} in combination with a model file in {\sc ufo} format~\cite{Degrande:2011ua}, derived from an effective Lagrangian using {\sc FeynRules}~\cite{Alloul:2013bka}, and second analytically as a cross-check.

As we are only considering QCD corrections, the renormalisation procedure is the same as in the SM and is described in Ref.~\cite{Borowka:2016ypz}.

\subsubsection{Real radiation}

The real corrections consist of 5-point one-loop topologies with closed top quark loops as well as tree-level diagrams, see Fig.~\ref{fig:hpnlore}. Both classes of diagrams have been generated with \gosam{} and arranged such that interferences between the two classes are properly taken into account.

In order to isolate the singularities due to unresolved radiation, 
we use the same framework as in Ref.~\cite{Borowka:2016ypz}, i.e. we use the Catani-Seymour dipole formalism~\cite{Catani:1996vz}, 
combined with a phase space restriction parameter $\alpha$ as suggested in Ref.~\cite{Nagy:2003tz}.

The various building blocks are assembled in a {\tt C++} program  and integrated over the phase space using 
the {\sc Vegas} algorithm~\cite{Lepage:1980dq} as implemented in the {\sc Cuba} library~\cite{Hahn:2004fe}.

\subsubsection{Parametrisation of the total cross section}

To parametrise the deviations of the total cross section from the one 
in the SM, we write the LO cross section in terms of the 15 coefficients 
$A_1, \ldots, A_{15}$, following Refs.~\cite{Azatov:2015oxa,Carvalho:2015ttv}.
\begin{align}
\label{eq:Acoeffs}
\sigma/\sigma_{SM}  &=   A_1\, c_t^4 + A_2 \, c_{tt}^2  + A_3\,  c_t^2 \chhh^2  + 
A_4 \, \cg^2 \chhh^2  + A_5\,  \cgg^2  + 
A_6\, c_{tt} c_t^2 + A_7\,  c_t^3 \chhh \nn\\
& + A_8\,  c_{tt} c_t\, \chhh  + A_9\, c_{tt} \cg \chhh + A_{10}\, c_{tt} \cgg + 
A_{11}\,  c_t^2 \cg \chhh + A_{12}\, c_t^2 \cgg \nn\\
& + A_{13}\, c_t \chhh^2 \cg  + A_{14}\, c_t \chhh \cgg +
A_{15}\, \cg \chhh \cgg  \,.
\end{align}
At NLO the coefficients $A_1,\ldots, A_{15}$ are modified and new
terms appear. We find: 
\begin{align}
\label{eq:AcoeffsNLO}
\Delta\sigma/\sigma_{SM}  &= A_{16}\, c^3_t \cg + A_{17}\,  c_t c_{tt} \cg 
+ A_{18}\, c_t \cg^2 \chhh + A_{19}\, c_t \cg \cgg 
\nn\\
&+ A_{20}\,  c_t^2 \cg^2 + A_{21}\, c_{tt} \cg^2 
+ A_{22}\, \cg^3 \chhh + A_{23}\, \cg^2 \cgg  \,.
\end{align}

\subsubsection{Validation of the calculation}
\label{sec:checks}

To validate our results, we have compared the Born-improved NLO HEFT results calculated with our setup with the ones from Ref.~\cite{Grober:2015cwa}, 
where we find agreement if we use $\mu_r=\mu_f=m_{hh}$ and MSTW2008~\cite{Martin:2009iq} PDFs at LO/NLO for the LO/NLO calculation, 
along with the corresponding $\als$ value.\footnote{Our default settings are to use PDF4LHC15~\cite{Butterworth:2015oua} PDFs for both the LO and the NLO results.}

We also have cross-checked the results by using two independent codes, where the only common parts are the {\sc ufo} model files and the SM virtual two-loop corrections.

In addition, we have compared the leading order distributions, benchmark points and fits of the coupling coefficients in the total cross section (see Eq.~(\ref{eq:Acoeffs})) with the ones given in Refs.~\cite{Azatov:2015oxa,Carvalho:2015ttv,Carvalho:2016rys}.
We find agreement with Ref.~\cite{Azatov:2015oxa} for all $A_i$
coefficients at the 1\% level.  Comparing to Refs.~\cite{Carvalho:2015ttv,Carvalho:2016rys}, we systematically find values that differ by 15-20\%  for coefficients
linear in $\cg$ and by $\sim 40$\% for the coefficient quadratic in $\cg$.    
We also compared our results with the distributions shown in Refs.~\cite{Carvalho:2015ttv,Carvalho:2016rys}, finding agreement for all benchmark points except for benchmark point 8.  While in Refs.~\cite{Carvalho:2015ttv,Carvalho:2016rys} a dip in the leading
order distribution is found for benchmark point 8, we find no such dip. This is why we chose a different point of
cluster 8 which does show a dip, and which we call 8a.


\section{Phenomenological results}
\label{sec:results}

In this section we present numerical results for benchmark points
which were identified in Ref.~\cite{Carvalho:2015ttv} to represent
partitions of the BSM parameter space 
according to characteristic shapes of differential distributions, in
particular the Higgs boson pair invariant mass distributions.
All our results are for a centre-of-mass energy of $\sqrt{s}=14$\,TeV.

The results were computed using the
PDF4LHC15{\tt\_}nlo{\tt\_}100{\tt\_}pdfas~\cite{Butterworth:2015oua,CT14,MMHT14,NNPDF}
parton distribution functions interfaced  via
LHAPDF~\cite{Buckley:2014ana}, along with the corresponding value for
$\alpha_s(\mu)$, with $\alpha_s(M_Z)=0.118$.  The masses of the Higgs boson and the top quark have been
set  to $m_h=125$\,GeV and $m_t=173$\,GeV (pole mass), respectively. 
The widths of the top quark (and the Higgs boson) have been set to zero.
Bottom quarks are treated as massless and therefore are
not included in the fermion loops.
The scale uncertainties are estimated by varying the factorisation scale $\mu_{F}$ and the
renormalisation scale $\mu_{R}$ 
around the central scale $\mu_0 =\mhh/2$, using the envelope of a 7-point scale variation.
The latter means that we use $\mu_{R,F}=c_{R,F}\,\mu_0$, where
$c_R,c_F\in \{2,1,0.5\}$, and consider each combination except the two
extreme ones $c_R=0.5,c_F=2$ and $c_R=2,c_F=0.5$.
In the SM case, the
combinations $c_R=c_F=0.5$ and $c_R=c_F=2$ always coincided with the
envelope of the 7 combinations to vary $c_R,c_F$.

\subsection{NLO cross sections and heat maps}
\label{sec:heatmaps}

In this section we will provide results for the coefficients defined
in Eqs.~(\ref{eq:Acoeffs}) and (\ref{eq:AcoeffsNLO}), i.e. for the expression
\begin{align}
\label{eq:Acoeffs_all}
\sigma^{\rm{NLO}}/\sigma^{\rm{NLO}}_{SM}  &=   A_1\, c_t^4 + A_2 \, c_{tt}^2  + A_3\,  c_t^2 \chhh^2  + 
A_4 \, \cg^2 \chhh^2  + A_5\,  \cgg^2  + 
A_6\, c_{tt} c_t^2 + A_7\,  c_t^3 \chhh \nn\\
& + A_8\,  c_{tt} c_t\, \chhh  + A_9\, c_{tt} \cg \chhh + A_{10}\, c_{tt} \cgg + 
A_{11}\,  c_t^2 \cg \chhh + A_{12}\, c_t^2 \cgg \nn\\
& + A_{13}\, c_t \chhh^2 \cg  + A_{14}\, c_t \chhh \cgg +
A_{15}\, \cg \chhh \cgg \nn\\ 
& + A_{16}\, c^3_t \cg + A_{17}\,  c_t c_{tt} \cg 
+ A_{18}\, c_t \cg^2 \chhh + A_{19}\, c_t \cg \cgg 
\nn\\
&+ A_{20}\,  c_t^2 \cg^2 + A_{21}\, c_{tt} \cg^2 
+ A_{22}\, \cg^3 \chhh + A_{23}\, \cg^2 \cgg\,.
\end{align}

We evaluated the coefficients in two different ways: determination via
projections and performing a fit, finding agreement of the results
within their uncertainties.
The results of the projection method, including uncertainties, are summarised in Table~\ref{tab:Ai}.

\begin{table}[htb]
\begin{center}
\begin{tabular}{|c|c|c|c|c|}
\hline
A coeff & LO  value & LO  uncertainty & NLO  value & NLO  uncertainty\\ 
\hline
$   A_1$ & 2.08059 & 0.00163127 & 2.23389 & 0.0100989\\
\hline
$   A_2$ & 10.2011 & 0.00809032 & 12.4598 & 0.0424131\\
\hline 
$   A_3$ & 0.27814 & 0.00187658& 0.342248  & 0.0153637 \\ 
\hline
$  A_4$ & 0.314043& 0.000312416  & 0.346822 & 0.00327358\\
\hline 
$    A_5$ & 12.2731 & 0.0101351 & 13.0087  & 0.0962361 \\ 
\hline
$   A_6$ & $-8.49307$& 0.00885261  & $-9.6455$ & 0.0503776  \\
\hline 
$   A_7$ & $-1.35873$ & 0.00148022 & $-1.57553$ & 0.0136033 \\
\hline 
$   A_8$ & 2.80251 & 0.0130855 & 3.43849 &  0.0771694\\
\hline 
$   A_9$ & 2.48018 & 0.0127927 & 2.86694 & 0.0772341 \\
\hline 
$   A_{10}$ & 14.6908 & 0.0311171 & 16.6912  & 0.178501 \\
\hline 
$   A_{11}$ & $-1.15916$ & 0.00307598 & $-1.25293$  & 0.0291153 \\
\hline 
$   A_{12}$ & $-5.51183$ & 0.0131254 & $-5.81216$ & 0.134029 \\
\hline 
$   A_{13}$ & 0.560503 & 0.00339209 & 0.649714 & 0.0287388 \\
\hline 
$   A_{14}$ & 2.47982 & 0.0190299 & 2.85933 & 0.193023 \\
\hline 
$   A_{15}$ & 2.89431 & 0.0157818 & 3.14475 & 0.148658\\
\hline
$   A_{16}$ &   &  & $-0.00816241$ & 0.000224985\\
\hline 
$   A_{17}$ &   &  & 0.0208652  & 0.000398929 \\
\hline 
$   A_{18}$ &   &  &  0.0168157  & 0.00078306\\
\hline 
$   A_{19}$ &   &  & 0.0298576    & 0.000829474\\
\hline 
$   A_{20}$ &   &  &  $-0.0270253$  & 0.000701919\\
\hline 
$   A_{21}$ &   &  & 0.0726921  & 0.0012875\\
\hline 
$   A_{22}$ &   &   &  0.0145232 & 0.000703893\\
\hline 
$   A_{23}$ &   &   & 0.123291 & 0.00650551 \\
\hline 
\end{tabular}
\end{center}
\caption{Results for the coefficients defined in  Eq.~(\ref{eq:Acoeffs_all}).
The uncertainties are obtained from the uncertainties on the total
cross sections entering the projections, using error propagation which
neglects  correlations between these cross sections.\label{tab:Ai}}
\end{table}
In the following we show heat maps for the ratio $\sigma/\sigma_{SM}$, 
based on the results for $A_1,\ldots, A_{23}$.
For the fixed parameters the SM values are used. Further we use
$\sigma_{SM}^{\rm{LO}}=19.85$\,fb,
$\sigma_{SM}^{\rm{NLO}}=32.95$\,fb.

The couplings are varied in a range which seems reasonable when taking into account the current constraints on the 
Higgs coupling measurements~\cite{Aad:2015gba,Khachatryan:2016vau,CMS-PAS-HIG-17-031}, 
as well as recent limits on the di-Higgs production cross section~\cite{CMS:2018obr,Aaboud:2018knk,Sirunyan:2018iwt,Aaboud:2018ftw}.

\begin{figure}[htb]
\begin{center}
\begin{subfigure}{0.495\textwidth}
\includegraphics[width=8cm]{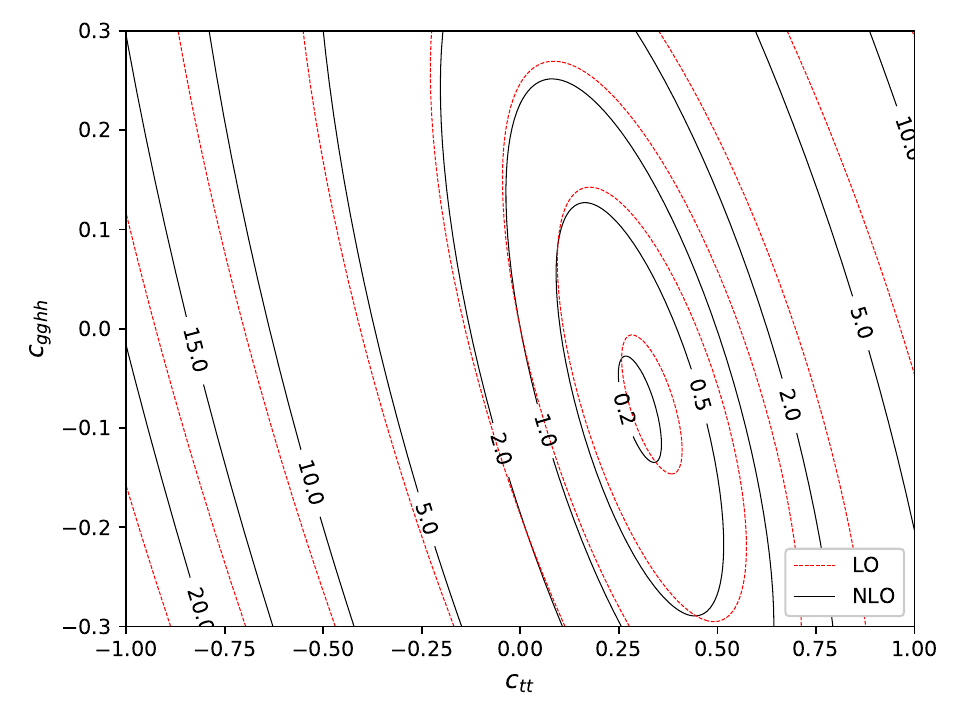}
    \caption{\label{fig:ctt_cgghh}}
  \end{subfigure}
  \hfill
  \begin{subfigure}{0.495\textwidth}
\includegraphics[width=8cm]{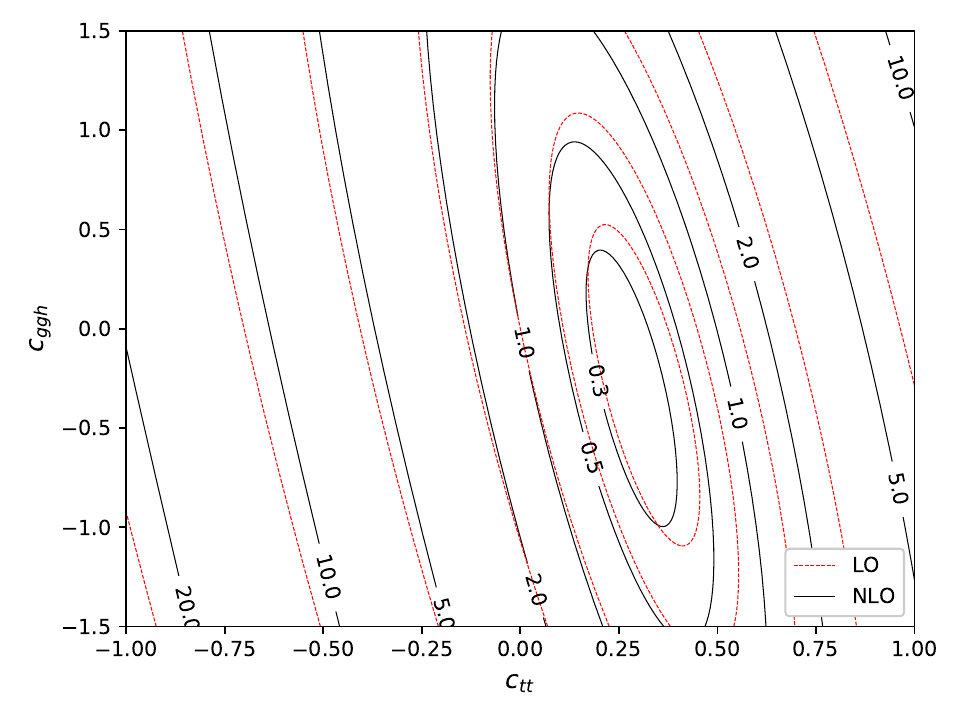}
    \caption{\label{fig:ctt_cggh}}
  \end{subfigure}
\end{center}
\caption{Iso-contours of $\sigma/\sigma_{SM}$: (a) $\cgg$ and (b) $\cg$ versus $\ctt$.}
\label{fig:CTT}
\end{figure}
In Fig.~\ref{fig:CTT} we display heat maps where the anomalous coupling $\ctt$ is varied in combination with the Higgs-gluon contact interactions $\cgg$ and $\cg$. 
We show the ratio to the SM total cross section both at LO and at NLO. 
We can see that the NLO corrections can lead to a significant shift in the iso-contours. 
It also becomes apparent that the cross sections are more sensitive to variations of $\ctt$ than to variations of the contact interaction $\cg$.

\begin{figure}[htb]
\begin{center}
\begin{subfigure}{0.495\textwidth}
   \includegraphics[width=8cm]{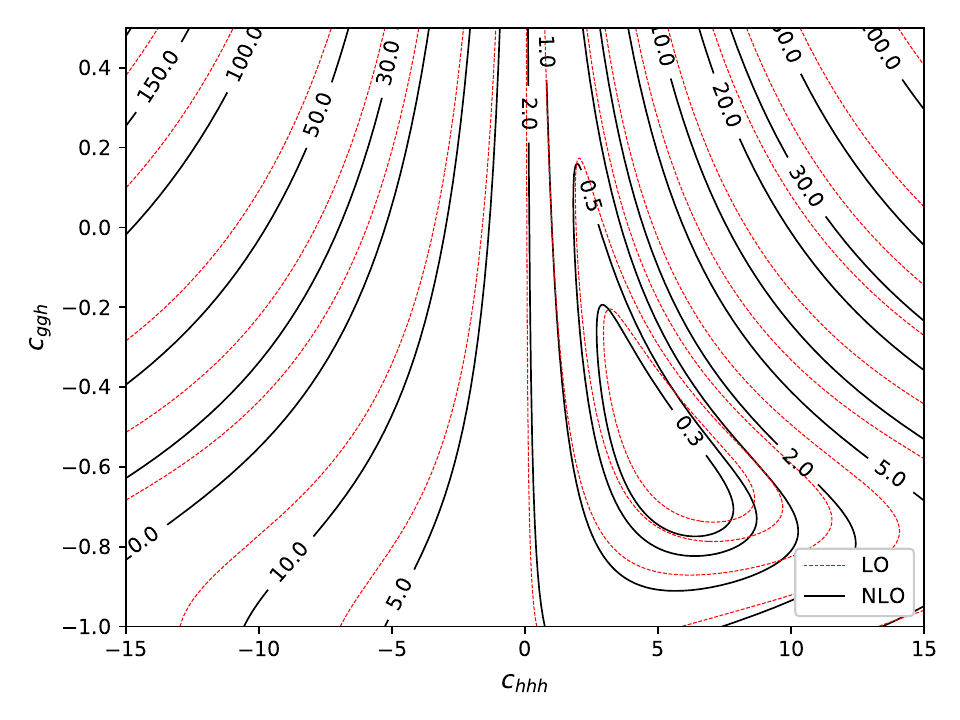}    
   \caption{\label{fig:chhh_cg}}
 \end{subfigure}
  \hfill
  \begin{subfigure}{0.495\textwidth}
\includegraphics[width=8cm]{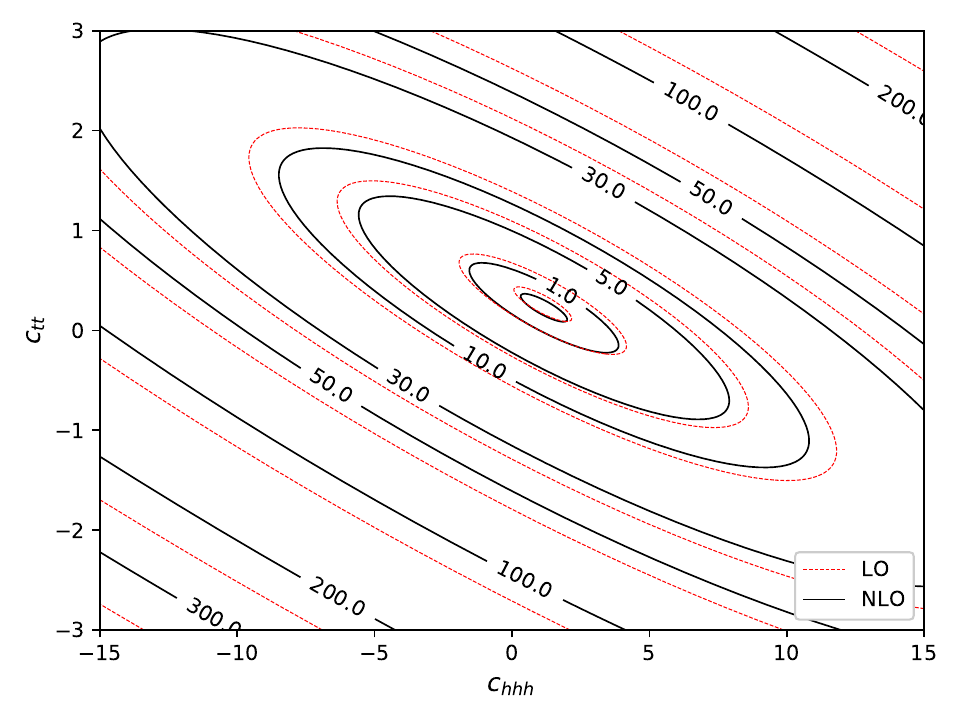}
  \caption{\label{fig:chhh_ctt}}
 \end{subfigure}
\end{center}
\caption{Iso-contours of $\sigma/\sigma_{SM}$: (a) $\cg$ versus $\chhh$ and (b) $\ctt$ versus $\chhh$.}
\label{fig:chhh_cg_ctt}
\end{figure}

Fig.~\ref{fig:chhh_cg_ctt} shows variations of the triple Higgs coupling $\chhh$ in combination with $\cg$ and $\ctt$.
We observe that the deviations from the SM cross section can be substantial, and
again we see a rapid variation of the cross section when changing $\ctt$.

\begin{figure}[htb]
\begin{center}
  \begin{subfigure}{0.495\textwidth}
   \includegraphics[width=8cm]{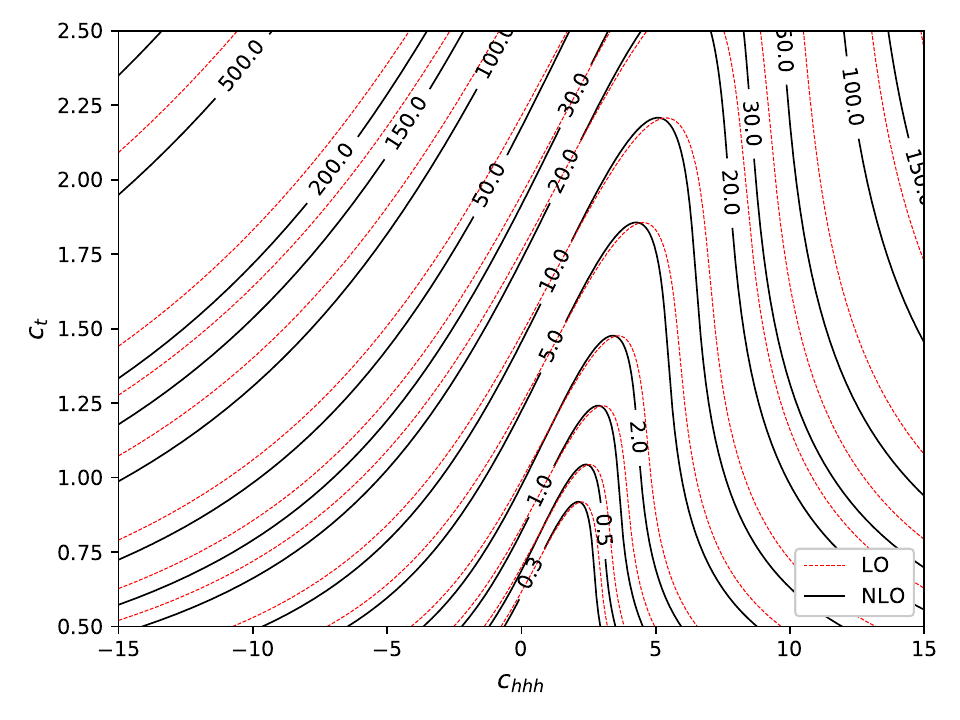}
   \caption{\label{fig:chhh_ct}}
 \end{subfigure}
  \hfill
  \begin{subfigure}{0.495\textwidth}
  \includegraphics[width=8cm]{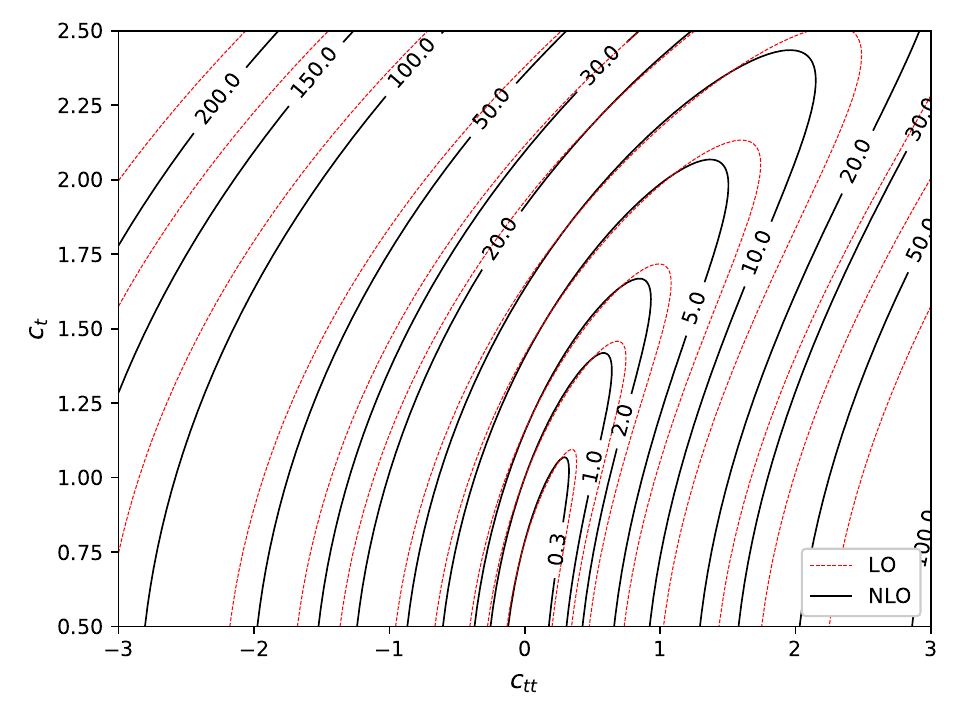}
   \caption{\label{fig:ctt_ct}}
 \end{subfigure}
\end{center}
\caption{Iso-contours of $\sigma/\sigma_{SM}$: (a) $\ct$ versus $\chhh$ and (b) $\ct$ versus $\ctt$.}
\label{fig:chhh_ct_ctt}
\end{figure}
In Fig.~\ref{fig:chhh_ct_ctt} we display  variations of $\ct$ versus $\chhh$, and  variations of $\ct$ versus $\ctt$.
We see that values of  $\ct$ around 2.0 in combination with large
negative values of $\chhh$ can enhance the cross section by two orders
of magnitude. Current experimental limits suggest that the total cross
section for Higgs boson pair production does not exceed about 13--24
times the SM value, assuming a SM-like shape in the
distributions~\cite{Sirunyan:2018iwt,Aaboud:2018knk}. Together with
the prospects that $\ct$ will be increasingly well constrained in the
future, e.g. from measurements of $t\bar{t}H$
production~\cite{Sirunyan:2018hoz,Aaboud:2018urx}, this should allow to constrain
some of the parameter space for $\chhh$.\footnote{Note that $\ct$ and
  $\cg$ already receive indirect constraints from single Higgs boson
  processes, as they enter in $gg\to h$ and $h\to\gamma\gamma$.}
\begin{figure}[htb]
\begin{center}
 \begin{subfigure}{0.495\textwidth}
\includegraphics[width=8cm]{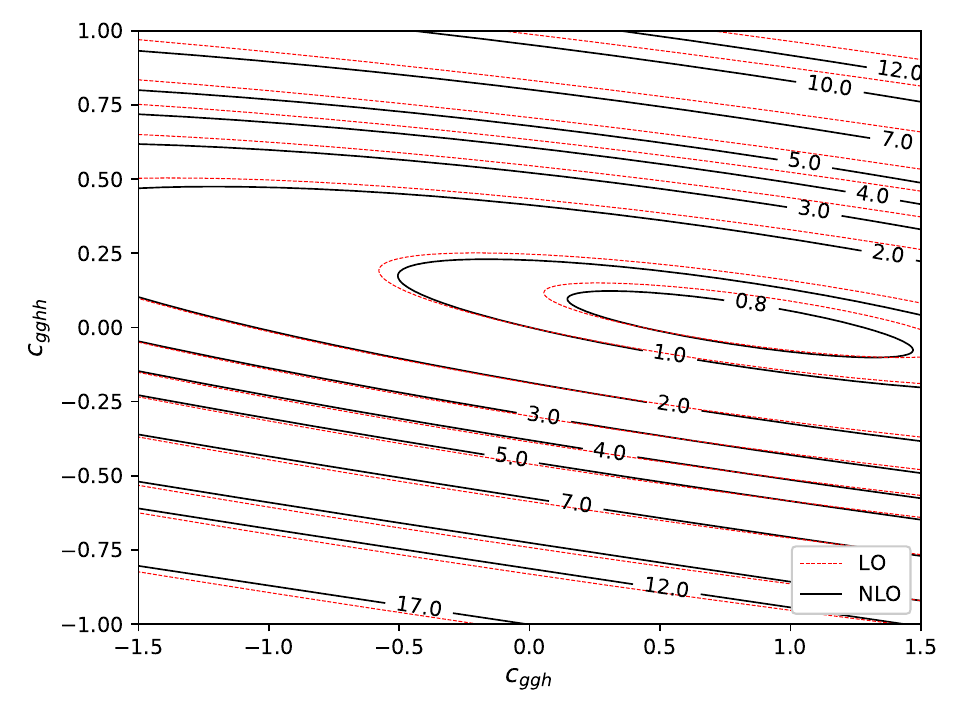}
   \caption{\label{fig:cggh_cgghh}}
 \end{subfigure}
\begin{subfigure}{0.495\textwidth}
\includegraphics[width=8cm]{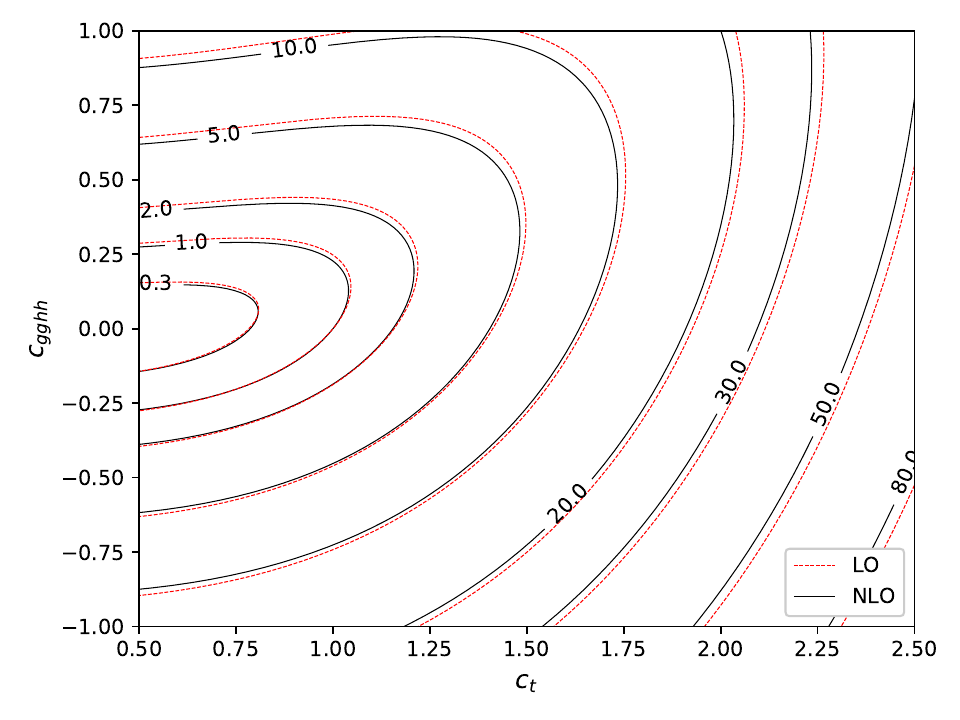}
   \caption{\label{fig:ctt_cgg}}
 \end{subfigure}
\end{center}
\caption{Iso-contours of $\sigma/\sigma_{SM}$: (a)  $\cgg$ versus $\cg$ and (b) $\cgg$ versus $\ct$.}
\label{fig:chhh_ctt_cgg}
\end{figure}
Fig.~\ref{fig:chhh_ctt_cgg} shows variations of $\cgg$ versus $\cg$ and $\ct$. 
We observe that the impact on the NLO corrections is milder in this case.

\begin{figure}[htb]
\begin{center}
\includegraphics[width=11cm]{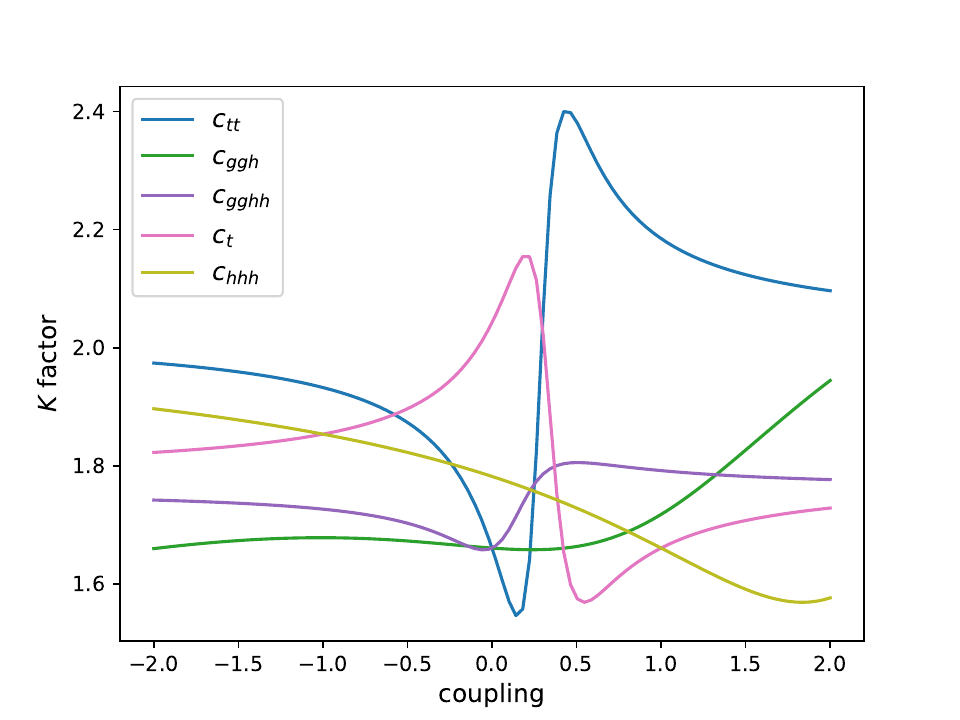}
\end{center}
\caption{K-factors for the total cross section as a function of the
  different couplings.}
\label{fig:project_ctt}
\end{figure}

In Fig.~\ref{fig:project_ctt} we show the K-factors as a function of
the coupling parameters, with the others fixed to their SM values. 
It shows that the rather flat K-factors which have been found~\cite{Grober:2015cwa,deFlorian:2017qfk}
in the $m_t\to\infty$ limit (flat with respect to variations of one of
the coupling parameters) show a much stronger dependence on the
coupling parameters once the full top quark mass dependence is taken into account.

\subsection{Cross sections and distributions at several benchmark points}
\label{subsec:benchmarks}

In the following we will show results for the benchmark points defined in Ref.~\cite{Carvalho:2015ttv}, 
except for benchmark point 8, where we choose a different one 
(denoted as ``outlier'' number 5 for cluster 8 in Ref.~\cite{Carvalho:2016rys}) which has a more characteristic shape,
and which we call 8a.

The conventions for the definition of the couplings between our
Lagrangian, given in Eq.~(\ref{eq:ewchl}), and the one of Ref.~\cite{Carvalho:2015ttv} 
are slightly different. In Table~\ref{tab:conventions} we list the 
conversion factors to translate between the conventions.

\begin{table}[htb]
\begin{center}
\begin{tabular}{ |c | c | }
\hline
EWChL Eq.~(\ref{eq:ewchl})& Ref.~\cite{Carvalho:2015ttv}\\
\hline
$c_{hhh}$ & $\kappa_{\lambda}$  \\
\hline
$c_t$ &$ \kappa_t$  \\
\hline
$ c_{tt} $ & $ c_{2}$  \\
\hline
$c_{ggh}$ &$ \frac{2}{3}c_g $  \\
\hline
$c_{gghh}$ & $-\frac{1}{3}c_{2g}$ \\
\hline
\end{tabular}
\end{center}
\caption{Translation between the conventions for the definition of the anomalous couplings.\label{tab:conventions}}
\end{table}

The benchmark points translated to our conventions are given in Table~\ref{tab:benchmarks}.
\begin{table}[htb]
\begin{center}
\begin{tabular}{| c | c  c  c  c  c |}
\hline
Benchmark & $c_{hhh}$ & $c_t$ & $c_{tt}$ & $c_{ggh}$ & $c_{gghh}$ \\
\hline
1 & 7.5 & 1.0 & $-1.0$ & 0.0 & 0.0 \\
\hline
2 & 1.0 & 1.0 & 0.5 & $-\frac{1.6}{3}$ & $-0.2$ \\
\hline
3 & 1.0 & 1.0 & $-1.5$ & 0.0 & $\frac{0.8}{3}$  \\
\hline
4 & $-3.5$ & 1.5 & $-3.0$ & 0.0 & 0.0 \\ 
\hline
5 & 1.0 & 1.0 & 0.0 &   $\frac{1.6}{3}$ & $\frac{1.0}{3}$\\
\hline
6 & 2.4 & 1.0 & 0.0 & $\frac{0.4}{3}$ & $\frac{0.2}{3}$  \\
\hline
7 & 5.0 & 1.0 & 0.0 & $\frac{0.4}{3}$ & $\frac{0.2}{3}$  \\
\hline
8a & 1.0 & 1.0 & 0.5 & $\frac{0.8}{3}$ & 0.0\\
\hline
9 & 1.0 & 1.0 & 1.0 &  $-0.4$ &  $-0.2$ \\
\hline
10 & 10.0 & 1.5 & $-1.0$ & 0.0 & 0.0 \\
\hline
11 & 2.4 & 1.0 & 0.0 & $\frac{2.0}{3}$ & $\frac{1.0}{3}$ \\
\hline
12 & 15.0 & 1.0 & 1.0 & 0.0 & 0.0 \\
\hline
SM & 1.0 & 1.0 & 0.0 & 0.0 & 0.0 \\
\hline
\end{tabular}
\end{center}
\caption{Benchmark points used for the distributions shown below.\label{tab:benchmarks}}
\end{table}

\subsubsection{Total cross sections}

We first show the values for the total cross sections, together with their statistical uncertainties and the uncertainties from scale variations.
We should point out that the cross sections for benchmark points $B_3, B_4$ and $B_{12}$ are larger than the limits measured in the $b\bar{b}\gamma\gamma$ decay channel~\cite{Sirunyan:2018iwt,Aaboud:2018ftw}. However, within the same cluster~\cite{Carvalho:2016rys}, i.e. the set of couplings which lead to a similar {\em shape} of the $m_{hh}$ distribution, one can easily find combinations of couplings where the value of the total cross section is below the experimental exclusion bound. 
For example, taking the point
$\chhh=1,\ct=1,\ctt=0,\cg=4/15,\cgg=-0.2$ in cluster 4 leads to a
cross section of about 1.8 times the SM cross section, still far from
being excluded, see Fig.~\ref{fig:benchmark4a}.


\begin{table}[htb]
\begin{center}
\begin{tabular}{|c|c|c|c|c|c|}
\hline
Benchmark & $\sigma_{NLO}$ [fb] & K-factor & scale uncert.  [\%] &
stat. uncert. [\%]  &$\frac{\sigma_{NLO}}{\sigma_{NLO,SM}}$ \\ 
\hline
$   B_1$ & 194.89 & 1.88 &$ {+19 \atop -15}$ & 1.6 & 5.915 \\
\hline
$   B_2$ & 14.55 & 1.88& ${+5\atop -13}$ & 0.56 & 0.4416 \\
\hline 
$   B_3$ & 1047.37 & 1.98& ${+21\atop -16}$ & 0.15 & 31.79  \\ 
\hline
$  B_4$ & 8922.75 & 1.98& ${+19\atop -16}$  & 0.39 & 270.8 \\
\hline 
$    B_5$ & 59.325 & 1.83& ${+4\atop -15}$ & 0.36 & 1.801   \\ 
\hline
$   B_6$ & 24.69 &  1.89& ${+2\atop -11}$  & 2.1 & 0.7495  \\
\hline 
$   B_7$ & 169.41 & 2.07& ${+9\atop -12}$ & 2.2 & 5.142 \\
\hline 
$   B_{8a}$ & 41.70 & 2.34& ${+6\atop -9}$ & 0.63 & 1.266 \\
\hline 
$   B_9$ & 146.00 & 2.30& ${+22\atop -16}$ & 0.31 & 4.431 \\
\hline 
$   B_{10}$ & 575.86 & 2.00& ${+17\atop -14}$ & 3.2 & 17.48   \\
\hline 
$   B_{11}$ & 174.70 & 1.92& ${+24\atop -8}$ & 1.2 & 5.303  \\
\hline 
$   B_{12}$ & 3618.53 & 2.07& ${+16\atop -15}$ & 1.2 & 109.83  \\
\hline 
$   SM $ & 32.95 & 1.66 & ${+14\atop -13}$ & 0.1 & 1\\
\hline
\end{tabular}
\end{center}
\caption{Total cross sections at NLO (second column) including the K-factor (third column), scale
  uncertainties (4th column) and statistical uncertainties (5th
  column) and the ratio to the SM total NLO cross section (6th column).\label{sigmatot}}
\end{table}
The large differences in the statistical uncertainties for the different benchmark points are due to the fact that 
 the results for the virtual two-loop part are based on rescaling of the SM numerical results, which are distributed differently in the phase space. Therefore the statistical uncertainties are largest for benchmark points where the distribution in phase space is very different from the SM case. For example, benchmark 10 has a large differential cross section at low $\mhh$ values, where the SM statistics is very low. This translates into the large statistical uncertainty for benchmark 10.

\subsubsection{$m_{hh}$ and $p_{T,h}$ distributions}

Now we consider differential cross sections for the 12 benchmark points. 
We show the Higgs boson pair invariant mass distribution  and the transverse
momentum distribution of one (any) of the Higgs bosons.
For each benchmark point we show the full NLO result in red, and compare it to the two approximations ``Born-improved NLO HEFT'' (purple) and \ftapprox{} (green).
The leading order (yellow) as well as the SM results are also shown (blue NLO, black LO).
The lower ratio plot shows the ratio of the two approximate results to the full NLO result.
The upper ratio plot shows the differential BSM K-factor, i.e. NLO$_{\rm{BSM}}$/LO$_{\rm{BSM}}$, both evaluated with the same PDFs.
\begin{figure}[tbp!]
  \centering
  \begin{subfigure}{0.495\textwidth}
    \includegraphics[width=\textwidth]{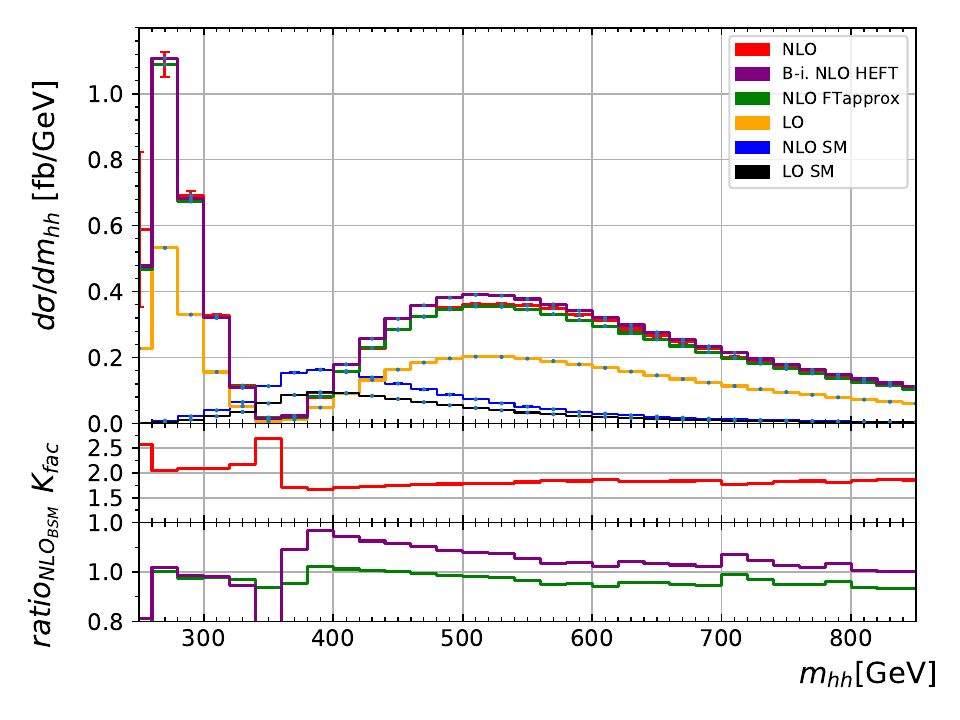}
    \vspace{\TwoFigBottom em}
    \caption{\label{fig:B1_mhh}}
  \end{subfigure}
  \hfill
  \begin{subfigure}{0.495\textwidth}
    \includegraphics[width=\textwidth]{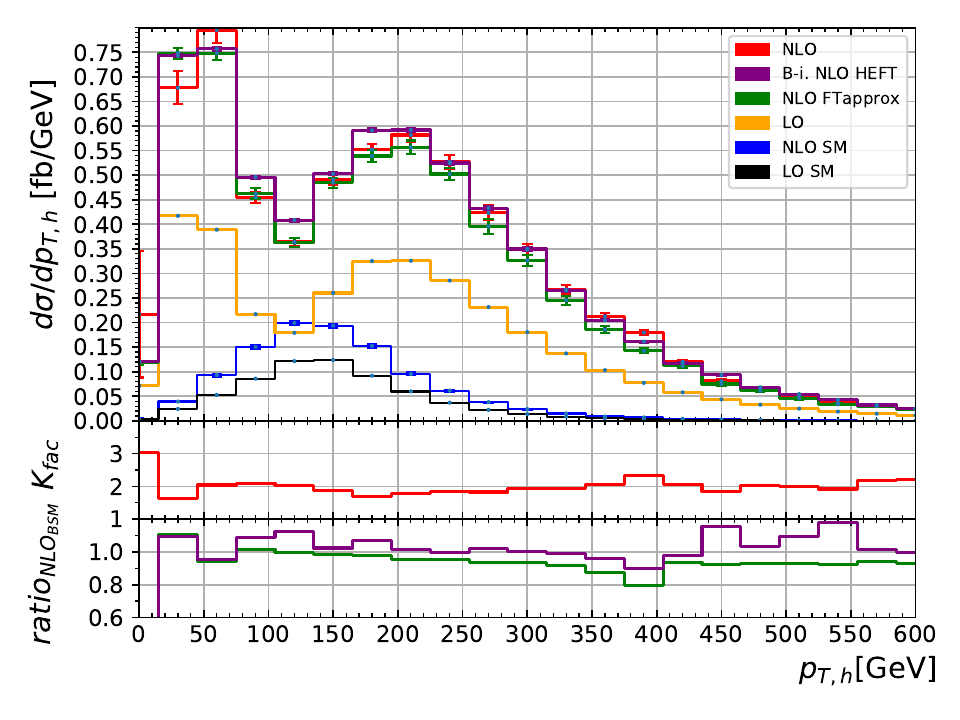}
    \vspace{\TwoFigBottom em}
    \caption{\label{fig:B1_pth}}
  \end{subfigure}
\caption{Higgs boson pair invariant mass distribution and transverse
  momentum distribution of one of the Higgs bosons for benchmark point
  1, $\chhh=7.5,\ct= 1, \ctt=-1, \cg=\cgg=0$. 
The ratio plot with the K-factor shows NLO$_{\rm{BSM}}$/LO$_{\rm{BSM}}$. The
  lower ratio plot shows the ratios (Born-improved NLO HEFT)/NLO$_{\rm{BSM}}$
  (purple) and \ftapprox/NLO$_{\rm{BSM}}$ (green).}
\label{fig:benchmark1}
\end{figure}

\begin{figure}[htb]
  \centering
  \begin{subfigure}{0.495\textwidth}
    \includegraphics[width=\textwidth]{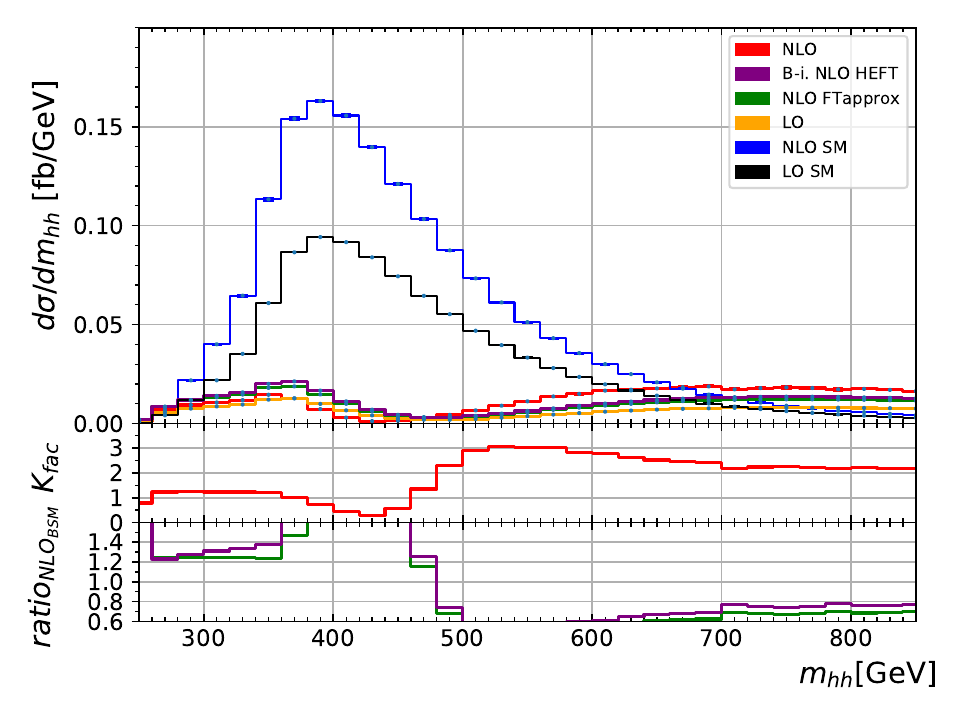}
    \vspace{\TwoFigBottom em}
    \caption{\label{fig:B2_mhh}}
  \end{subfigure}
  \hfill
  \begin{subfigure}{0.495\textwidth}
    \includegraphics[width=\textwidth]{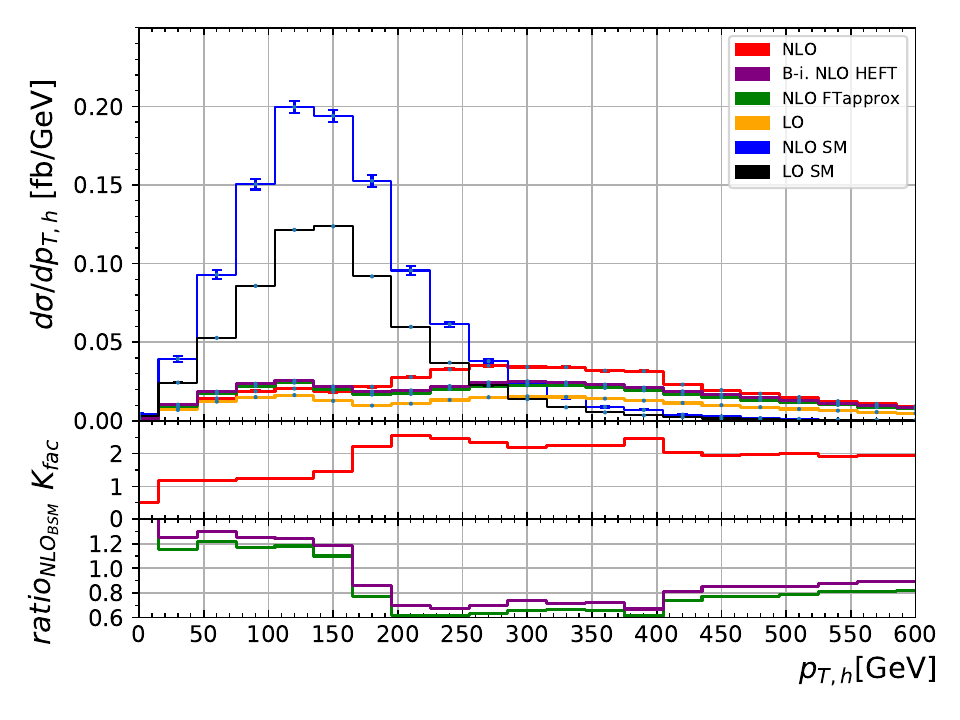}
    \vspace{\TwoFigBottom em}
    \caption{\label{fig:B2_pth}}
  \end{subfigure}
\caption{Same as Fig.~\ref{fig:benchmark1} but  for benchmark point 2, $\chhh=1,\ct= 1, \ctt=0.5, \cg=-8/15,\cgg=-0.2$.}
\label{fig:benchmark2}
\end{figure}
Fig.~\ref{fig:benchmark1} corresponds to a benchmark point with no Higgs-gluon contact interactions, but an enhanced triple Higgs coupling and a nonzero $t\bar{t}hh$ interaction with $\ctt<0$. 
The total cross section is about 6 times the SM cross section, and the shape of the $\mhh$ distribution is completely different from the SM. 
In fact, one can show analytically that the LO cross section in the $m_t\to \infty$ limit exactly vanishes near $\mhh=364$\,GeV, which relates to the dip in the distribution. 
The huge enhancement at low $\mhh$ values is due to the large value of $\chhh$.

Fig.~\ref{fig:benchmark2}, corresponding to benchmark 2, shows a very different behaviour. 
The result is very much suppressed in the region where the SM shows a peak, while there is a large enhancement in the tail of both the $\mhh$ and the $\pth$ distributions. The enhancement in the tail is mainly due to the nonzero $\cgg$ value, 
as the amplitude proportional to $\cgg$ grows like $\hat{s}$~\cite{Azatov:2015oxa}.
We also notice that the approximations ``Born-improved NLO HEFT''  and
\ftapprox{} cannot describe the pattern around the $2m_t$ threshold,
where the nonzero value of $\ctt$ seems to play a significant role. 
The K-factor for benchmark 2 is very non-homogeneous around the dip in
the $\mhh$ distribution, and can reach up to a factor of three. 
This is a clear example where rescaling the LO result with a K-factor
obtained from higher order calculations in the HEFT approximation
would lead to very different results.

\begin{figure}[htb]
  \centering
  \begin{subfigure}{0.495\textwidth}
    \includegraphics[width=\textwidth]{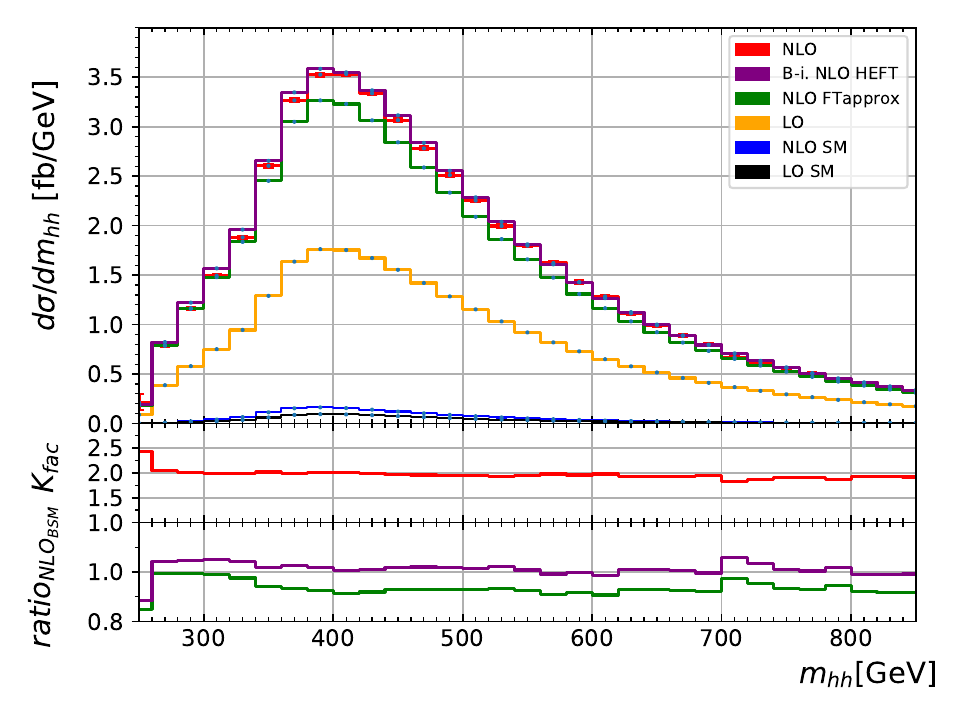}
    \vspace{\TwoFigBottom em}
    \caption{\label{fig:B3_mhh}}
  \end{subfigure}
  \hfill
  \begin{subfigure}{0.495\textwidth}
    \includegraphics[width=\textwidth]{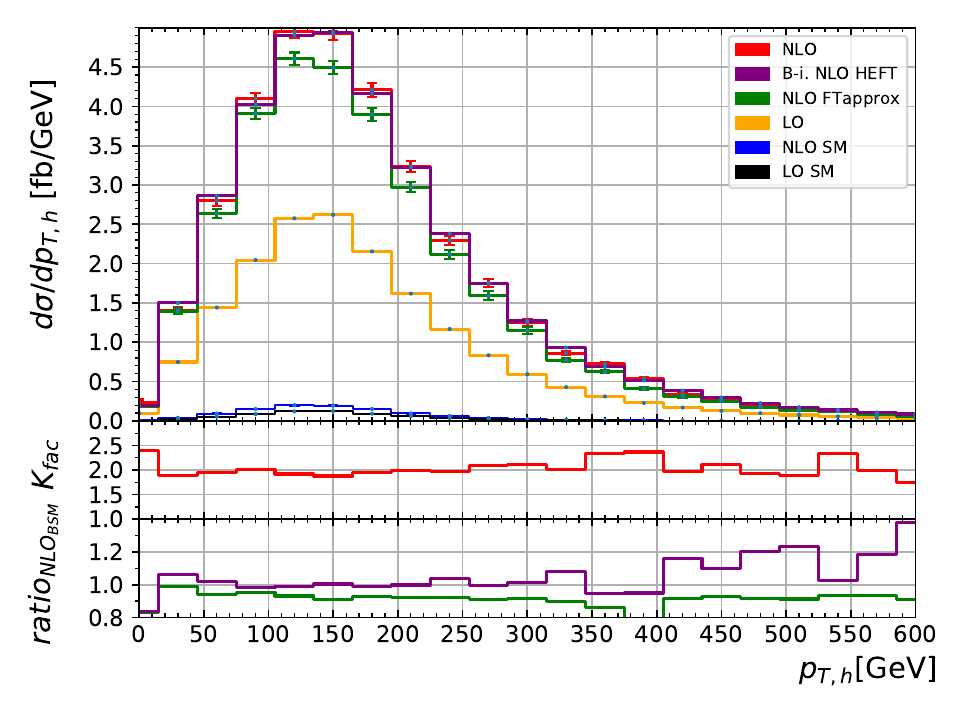}
    \vspace{\TwoFigBottom em}
    \caption{\label{fig:B3_pth}}
  \end{subfigure}
\caption{Same as Fig.~\ref{fig:benchmark1} but  for benchmark point 3, $\chhh=1,\ct= 1, \ctt=-1.5, \cg=0,\cgg=4/15$.}
\label{fig:benchmark3}
\end{figure}
Benchmark point 3, shown in Fig.~\ref{fig:benchmark3}, has the same values for $\chhh$ and $\ct$ as benchmark point 2 
(the SM values), but the distributions show a very different behaviour. 
As in the SM, there is a peak around the $2m_t$ threshold, but the
cross section is largely enhanced, not only in the peak region. As
mentioned above, with a total cross section of about 32 times the SM
NLO cross section, this parameter point is above the current limit  
deduced at 95\% CL from the measured $pp\to HH\to \gamma\gamma
b\bar{b}$ cross section~\cite{Sirunyan:2018iwt,Aaboud:2018ftw}.

\begin{figure}[htb]
  \centering
  \begin{subfigure}{0.495\textwidth}
    \includegraphics[width=\textwidth]{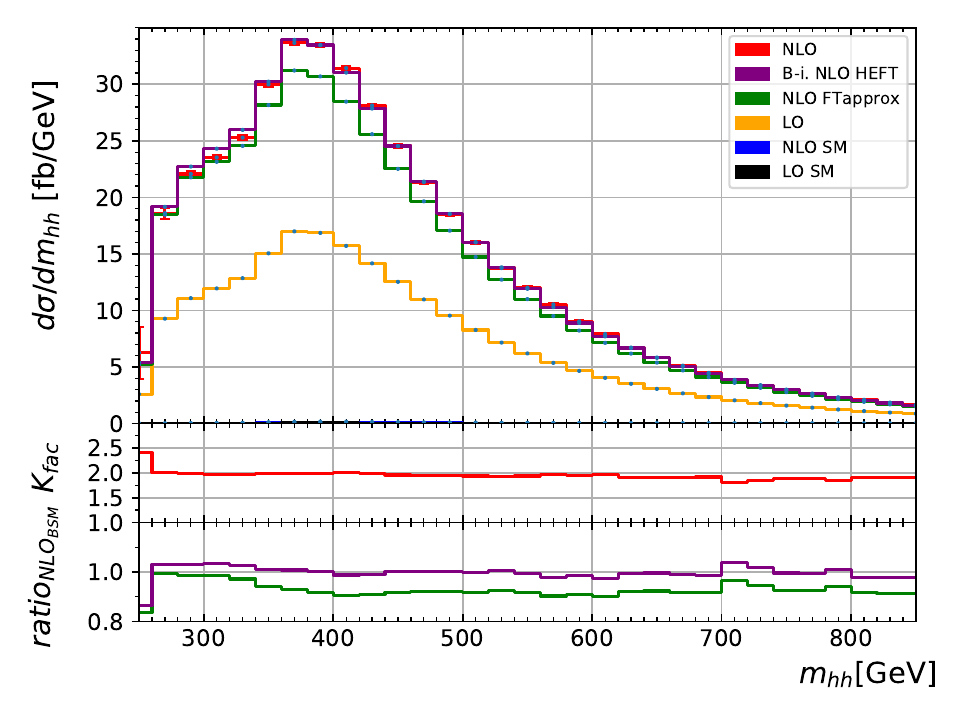}
    \vspace{\TwoFigBottom em}
    \caption{\label{fig:B4_mhh}}
  \end{subfigure}
  \hfill
  \begin{subfigure}{0.495\textwidth}
    \includegraphics[width=\textwidth]{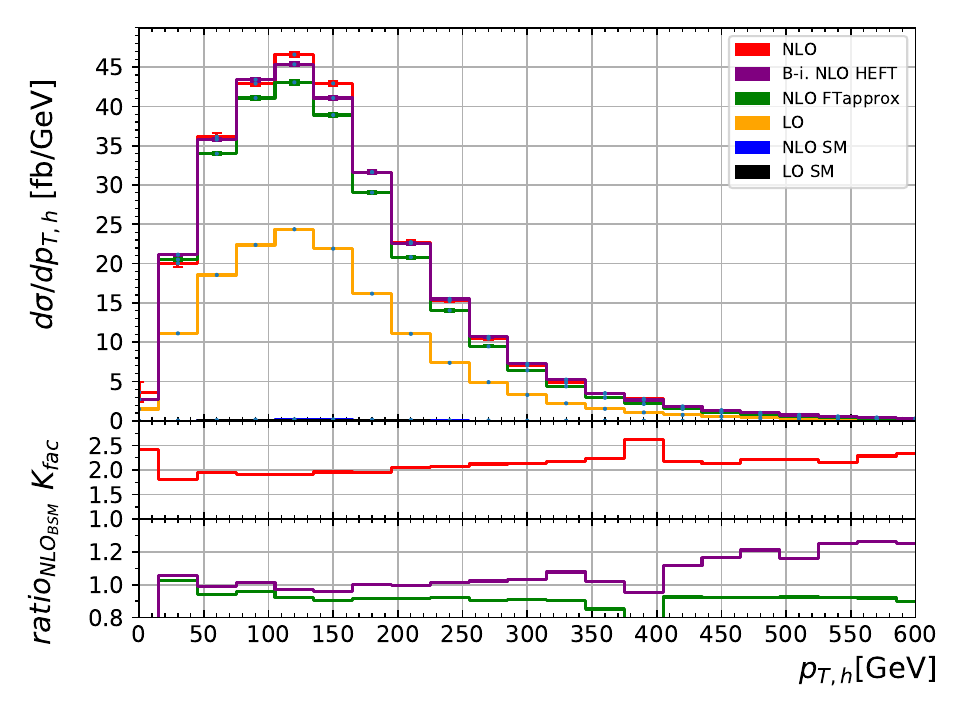}
    \vspace{\TwoFigBottom em}
    \caption{\label{fig:B4_pth}}
  \end{subfigure}
\caption{Same as Fig.~\ref{fig:benchmark1} but  for benchmark point 4, $\chhh=-3.5,\ct= 1.5, \ctt=-3, \cg=\cgg=0$.}
\label{fig:benchmark4}
\end{figure}

\begin{figure}[htb]
  \centering
  \begin{subfigure}{0.495\textwidth}
    \includegraphics[width=\textwidth]{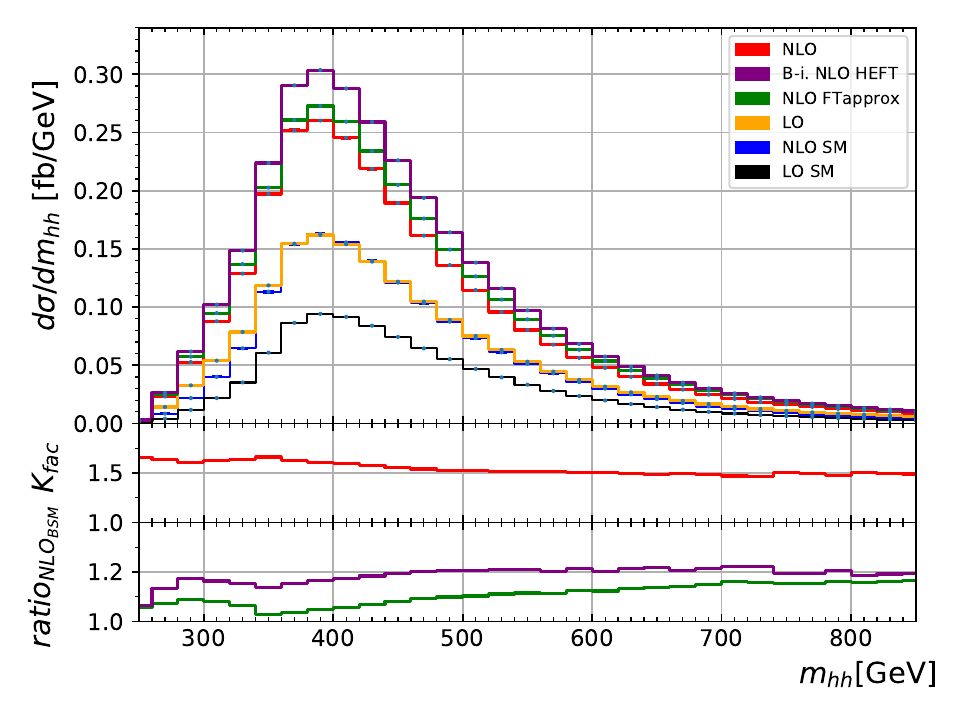}
    \vspace{\TwoFigBottom em}
    \caption{\label{fig:B4a_mhh}}
  \end{subfigure}
  \hfill
  \begin{subfigure}{0.495\textwidth}
    \includegraphics[width=\textwidth]{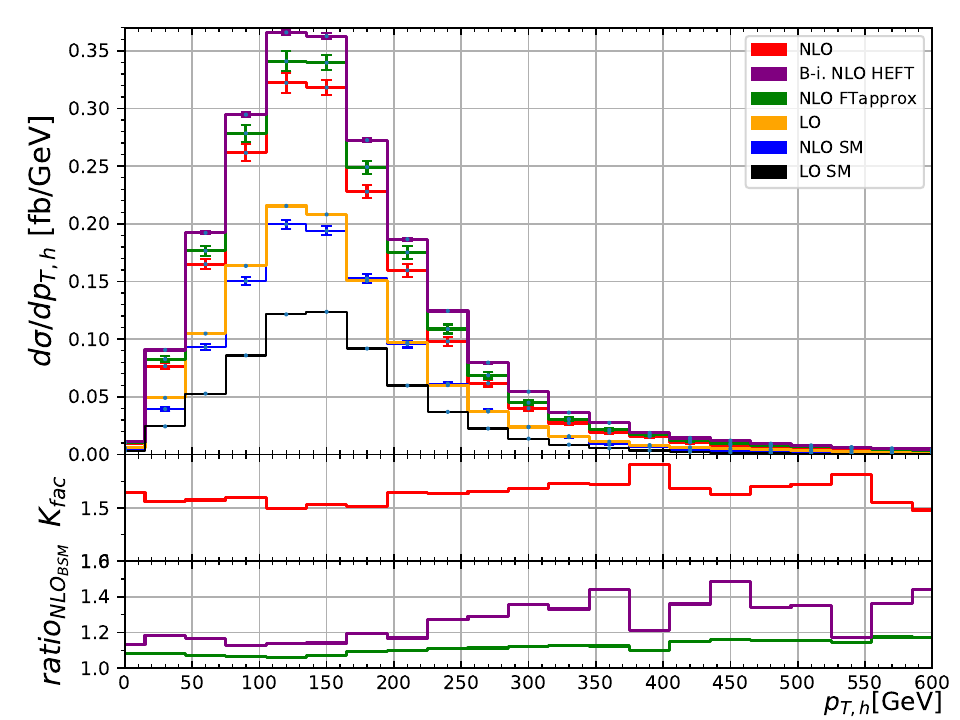}
    \vspace{\TwoFigBottom em}
    \caption{\label{fig:B4a_pth}}
  \end{subfigure}
\caption{A point from cluster 4,
  $\chhh=1,\ct=1,\ctt=0,\cg=4/15,\cgg=-0.2$, which leads to a similar
  shape as benchmark point 4, but to a much smaller cross section.}
\label{fig:benchmark4a}
\end{figure}

Benchmark point 4, shown in Fig.~\ref{fig:benchmark4}, has negative values for $\chhh$ and $\ctt$, a slightly encreased Yukawa coupling $\ct$, and no Higgs-gluon contact interactions. This combination removes the destructive interference between different types of diagrams present in the SM, and therefore leads to a very large cross section.
The differential K-factor is about 2, as for the other benchmarks, and rather constant over the whole $\mhh$ range 
(whereas for benchmark 2, the differential K-factor is far from being homogeneous).
Benchmark 4 is the one with the largest cross section of all the
considered benchmark points, with a total cross section of about 270
times the SM one. This point in parameter space is  excluded
already.
Therefore, in Fig.~\ref{fig:benchmark4a}, we also show results for another point from cluster 4, defined by 
  $\chhh=1,\ct=1,\ctt=0,\cg=4/15,\cgg=-0.2$, which leads to a similar
  shape as benchmark point 4, but to $\sigma/\sigma_{SM}=1.8$, and hence is not yet excluded. 
This parameter point also has the interesting feature that the distributions for NLO SM and LO BSM almost coincide. 
However, there is no degeneracy with the SM distribution once the NLO corrections are taken into account.

\begin{figure}[htb]
  \centering
  \begin{subfigure}{0.495\textwidth}
    \includegraphics[width=\textwidth]{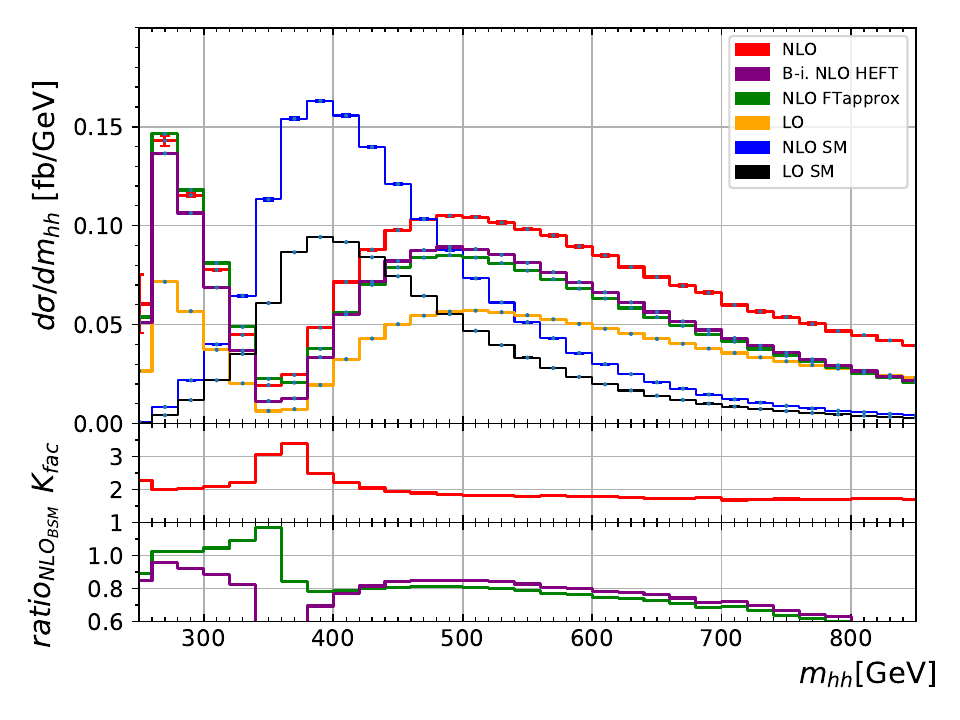}
    \vspace{\TwoFigBottom em}
    \caption{\label{fig:B5_mhh}}
  \end{subfigure}
  \hfill
  \begin{subfigure}{0.495\textwidth}
    \includegraphics[width=\textwidth]{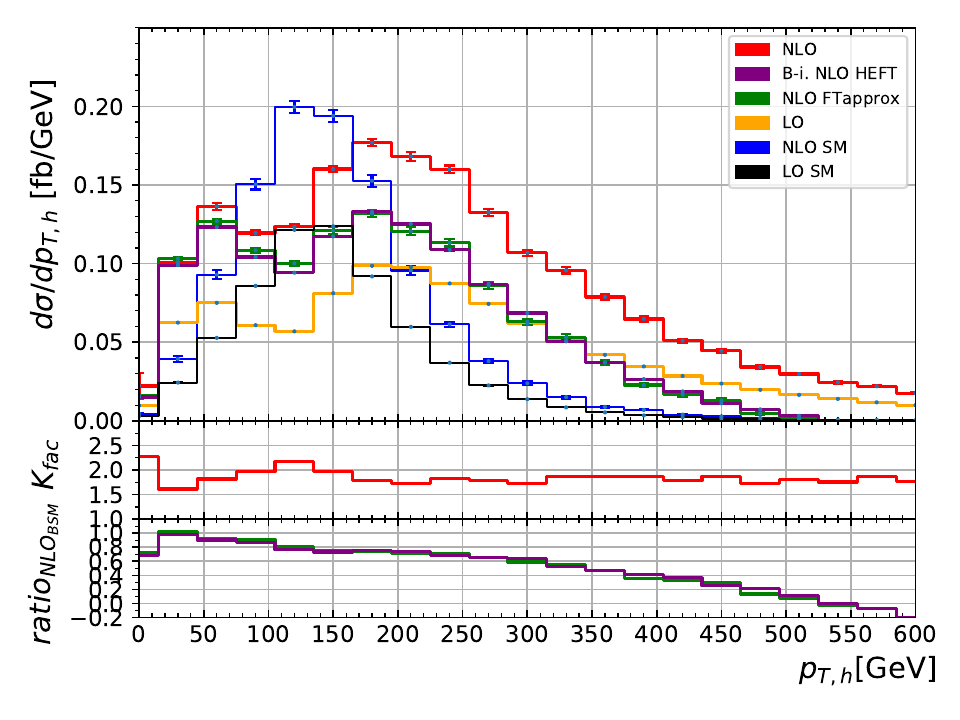}
    \vspace{\TwoFigBottom em}
    \caption{\label{fig:B5_pth}}
  \end{subfigure}
\caption{Same as Fig.~\ref{fig:benchmark1} but  for benchmark point 5, $\chhh=1,\ct= 1, \ctt=0, \cg=8/15,\cgg=1/3$.}
\label{fig:benchmark5}
\end{figure}
Fig.~\ref{fig:benchmark5} shows distributions for benchmark point 5, where $\ctt$ is zero and $\chhh$ and $\ct$ are as in the SM, while the Higgs-gluon interactions are nonzero.
Similar to benchmark point 2, we observe a dip near $\mhh=350$\,GeV, but the LO HEFT amplitude does not vanish there.
The total cross section for benchmark point 5 
is very similar to the SM one. This is an example where differential measurements are crucial to establish a clear BSM signal.
The $p_{T,h}$ distribution shows the rather unexpected behaviour that \ftapprox{} and Born-improved HEFT drop very rapidly at large values of $p_{T,h}$. The reason is that the rescaling factor $B_{FT}/B_{HEFT}$ becomes very large as the energy increases, because $B_{HEFT}$ does not grow with $\hat{s}$ for this combination of couplings,  but becomes very small. Therefore the negative virtual corrections are multiplied by a very large factor, leading to the fall-off of the green and purple curves in the tail of the $p_{T,h}$ distribution.

\begin{figure}[htb]
  \centering
  \begin{subfigure}{0.495\textwidth}
    \includegraphics[width=\textwidth]{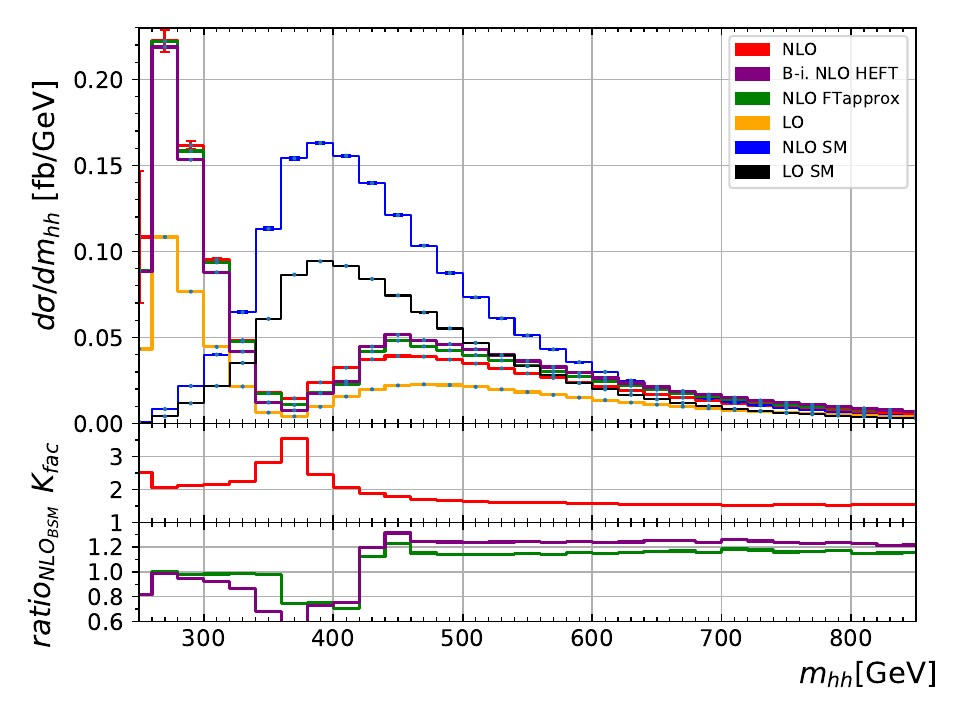}
    \vspace{\TwoFigBottom em}
    \caption{\label{fig:B6_mhh}}
  \end{subfigure}
  \hfill
  \begin{subfigure}{0.495\textwidth}
    \includegraphics[width=\textwidth]{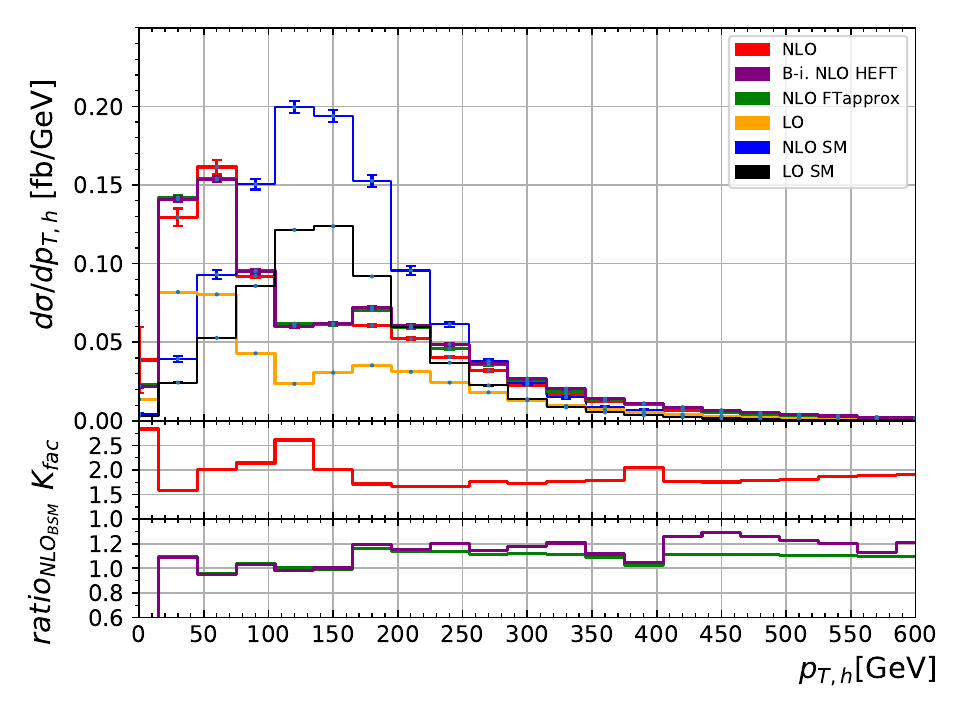}
    \vspace{\TwoFigBottom em}
    \caption{\label{fig:B6_pth}}
  \end{subfigure}
\caption{Same as Fig.~\ref{fig:benchmark1} but  for benchmark point 6, $\chhh=2.4,\ct= 1, \ctt=0, \cg=2/15,\cgg=1/15$.}
\label{fig:benchmark6}
\end{figure}

Benchmark point 6, shown in Fig.~\ref{fig:benchmark6}, also shows a dip, related to the fact that the LO HEFT amplitude exactly vanishes at $\mhh=429$\,GeV.
In addition it has a large enhancement of the low $\mhh$ region due to the value $\chhh=2.4$. Note  that this value for $\chhh$ is very close to the point where the 
total cross section as a function of $\chhh$ goes through a minimum if all other couplings are kept SM-like.

\begin{figure}[htb]
  \centering
  \begin{subfigure}{0.495\textwidth}
    \includegraphics[width=\textwidth]{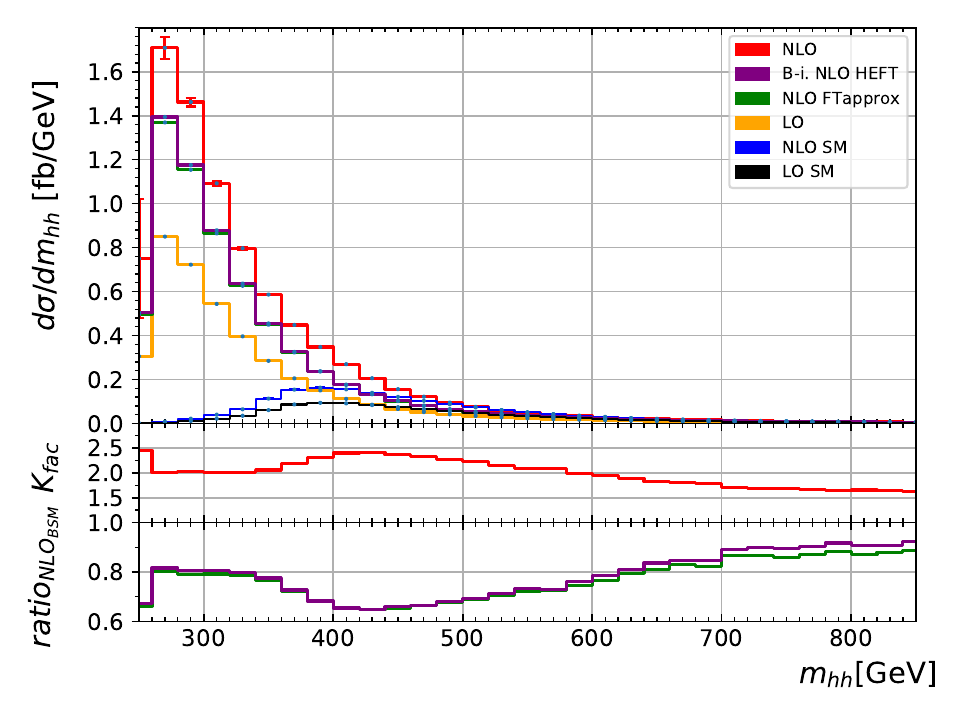}
    \vspace{\TwoFigBottom em}
    \caption{\label{fig:B7_mhh}}
  \end{subfigure}
  \hfill
  \begin{subfigure}{0.495\textwidth}
    \includegraphics[width=\textwidth]{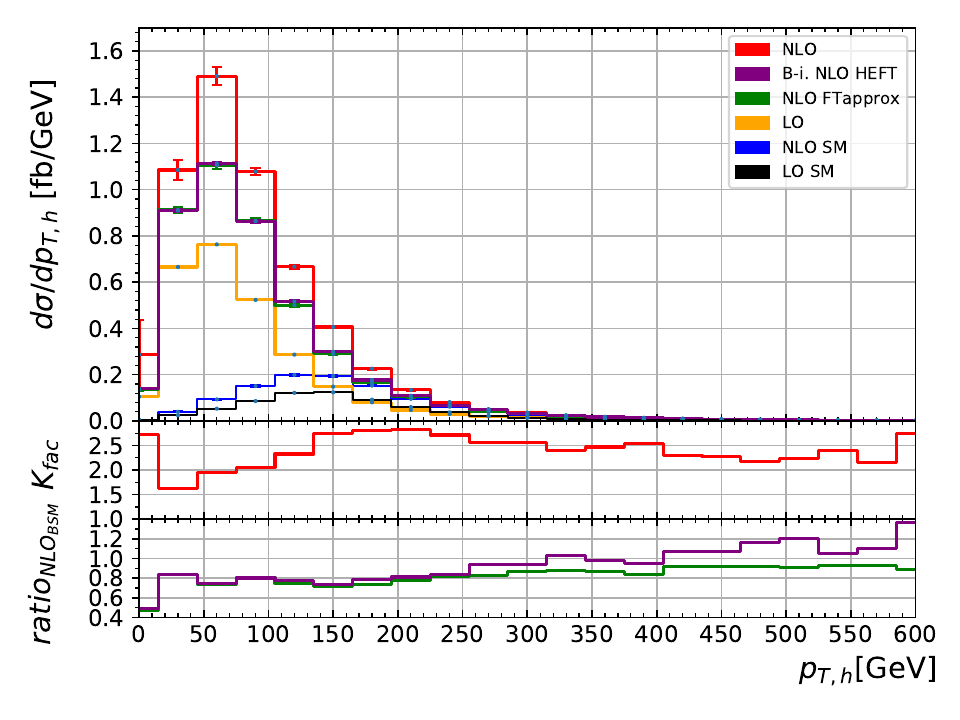}
    \vspace{\TwoFigBottom em}
    \caption{\label{fig:B7_pth}}
  \end{subfigure}
\caption{Same as Fig.~\ref{fig:benchmark1} but  for benchmark point 7, $\chhh=5,\ct= 1, \ctt=0, \cg=2/15,\cgg=1/15$.}
\label{fig:benchmark7}
\end{figure}
Benchmark point 7, shown in Fig.~\ref{fig:benchmark7}, has the same values for $\cg,\cgg,\ct$ and $\ctt$ as  benchmark point 6,
but a different value for $\chhh \,(\chhh=5)$. This makes the dip disappear completely, leading to a total cross section which is about 6.7 times larger than the one for benchmark 6, and a large enhancement of the low $\mhh$ and low $\pth$ regions.
The distributions also show that the full top quark mass dependence in the ``triangle-type'' diagrams containing $\chhh$, which dominate the low $\mhh$ region, seems to play a significant role, as the full NLO result is quite different from the approximate results.

\begin{figure}[htb]
  \centering
  \begin{subfigure}{0.495\textwidth}
    \includegraphics[width=\textwidth]{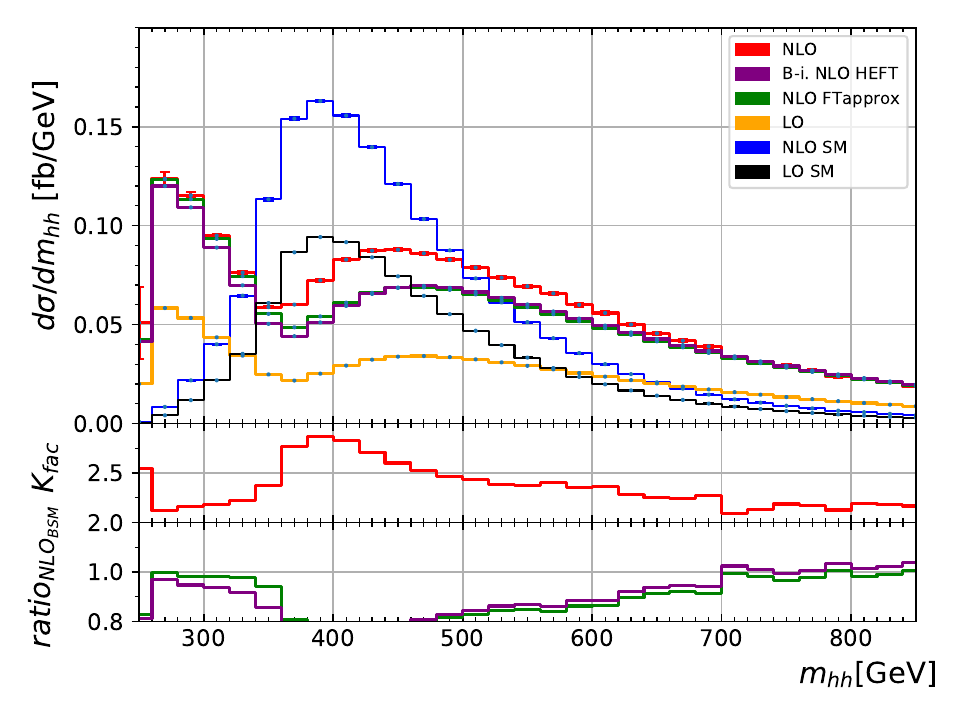}
    \vspace{\TwoFigBottom em}
    \caption{\label{fig:B8a_mhh}}
  \end{subfigure}
  \hfill
  \begin{subfigure}{0.495\textwidth}
    \includegraphics[width=\textwidth]{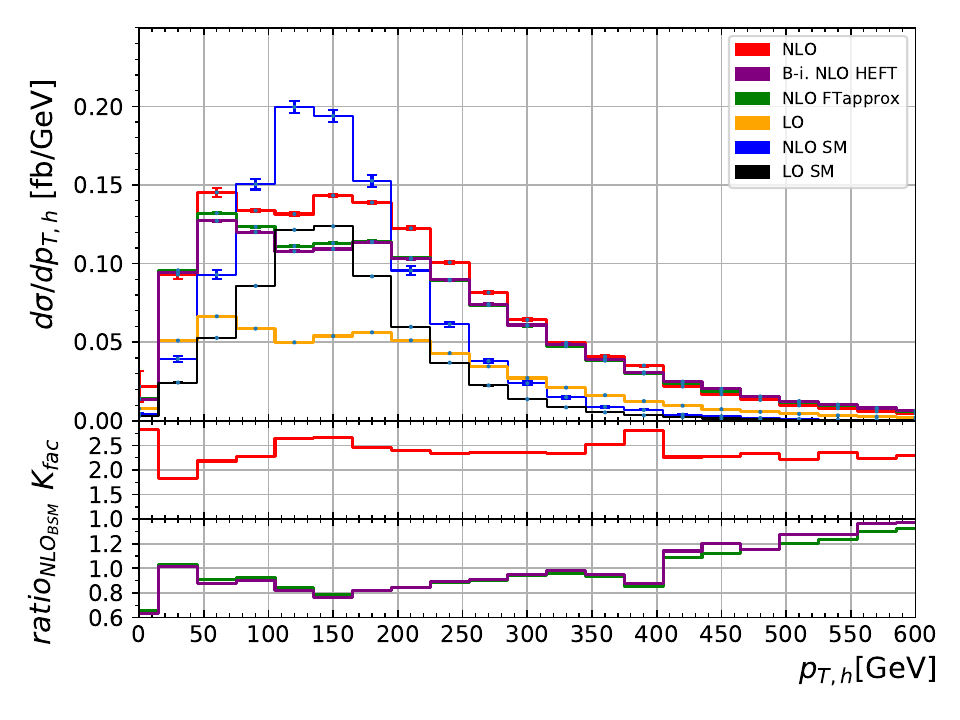}
    \vspace{\TwoFigBottom em}
    \caption{\label{fig:B8a_pth}}
  \end{subfigure}
\caption{Same as Fig.~\ref{fig:benchmark1} but  for benchmark point 8a, $\chhh=1,\ct= 1, \ctt=0.5, \cg=4/15,\cgg=0$.}
\label{fig:benchmark8a}
\end{figure}

Benchmark point 8a, displayed in Fig.~\ref{fig:benchmark8a}, again shows a characteristic dip just before the $2m_t$ threshold. It is also an example where the total cross section is very similar to the SM one, but the shape of both the $\mhh$ and the $\pth$ distributions clearly discriminates the SM from the BSM case.

\begin{figure}[htb]
  \centering
  \begin{subfigure}{0.495\textwidth}
    \includegraphics[width=\textwidth]{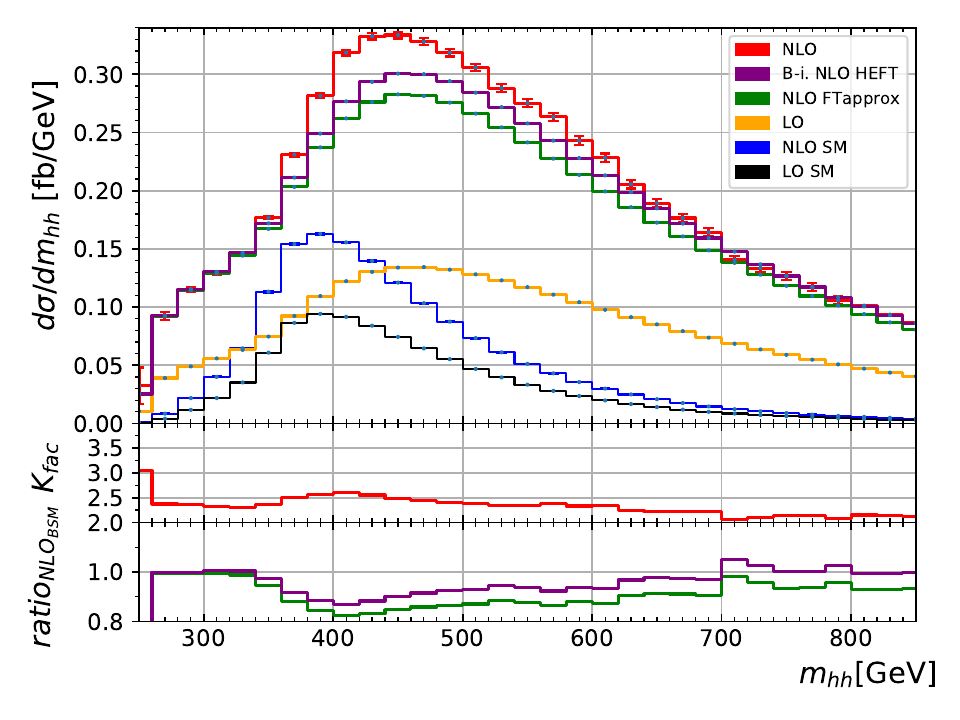}
    \vspace{\TwoFigBottom em}
    \caption{\label{fig:B9_mhh}}
  \end{subfigure}
  \hfill
  \begin{subfigure}{0.495\textwidth}
    \includegraphics[width=\textwidth]{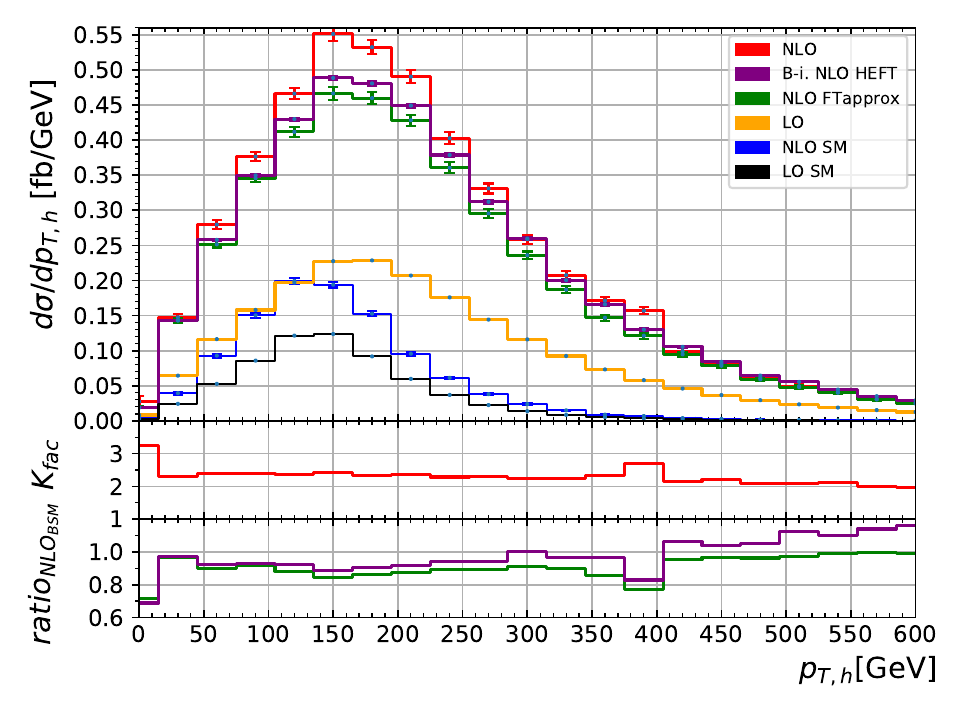}
    \vspace{\TwoFigBottom em}
    \caption{\label{fig:B9_pth}}
  \end{subfigure}
\caption{Same as Fig.~\ref{fig:benchmark1} but  for benchmark point 9, $\chhh=1,\ct= 1, \ctt=1, \cg=-0.4,\cgg=-0.2$.}
\label{fig:benchmark9}
\end{figure}
Benchmark point 9, displayed in Fig.~\ref{fig:benchmark9}, shows a large enhancement in the tails of the distributions, similar to benchmarks 2 and 3, which can be attributed mainly to the rather large value of $\cgg$, in combination with a non-zero value of $\ctt$.

\begin{figure}[htb]
  \centering
  \begin{subfigure}{0.495\textwidth}
    \includegraphics[width=\textwidth]{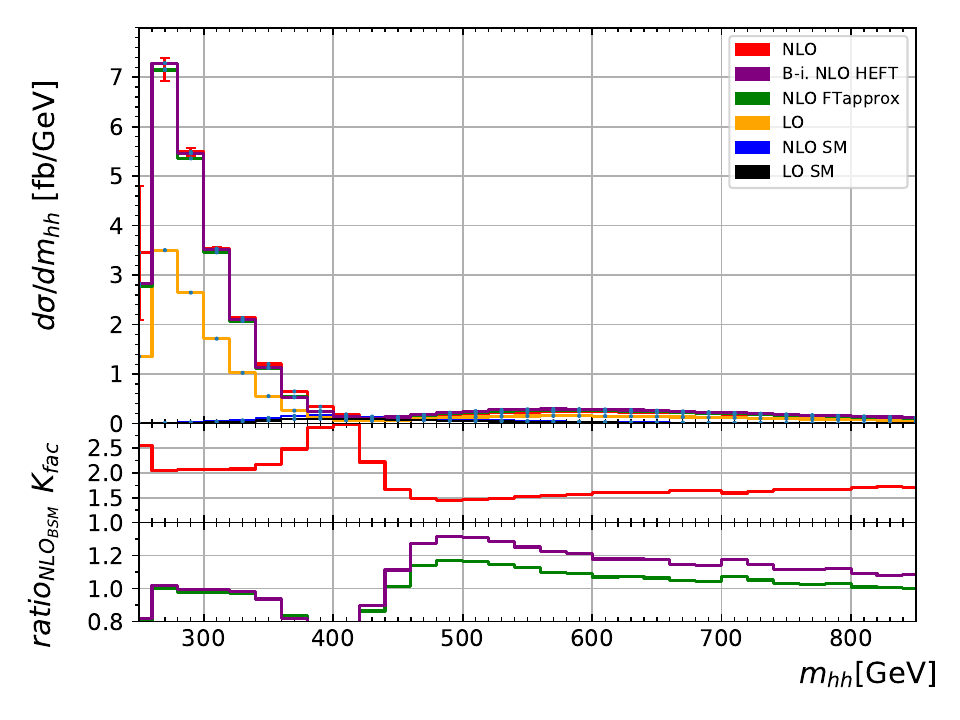}
    \vspace{\TwoFigBottom em}
    \caption{\label{fig:B10_mhh}}
  \end{subfigure}
  \hfill
  \begin{subfigure}{0.495\textwidth}
    \includegraphics[width=\textwidth]{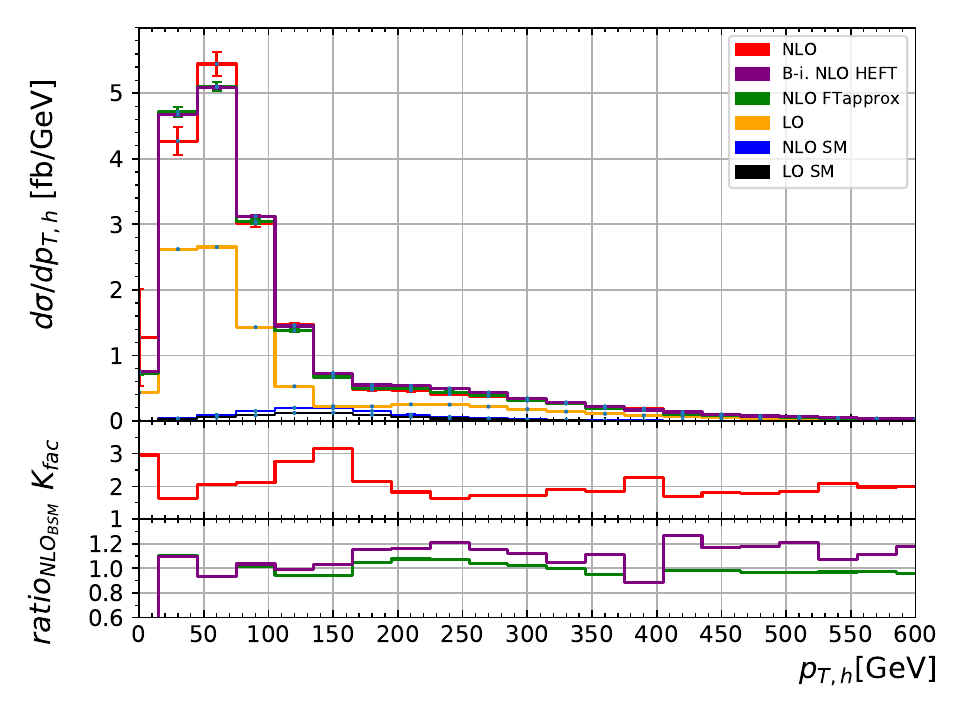}
    \vspace{\TwoFigBottom em}
    \caption{\label{fig:B10_pth}}
  \end{subfigure}
\caption{Same as Fig.~\ref{fig:benchmark1} but  for benchmark point 10, $\chhh=10,\ct= 1.5, \ctt=-1, \cg=\cgg=0$.}
\label{fig:benchmark10}
\end{figure}

For benchmark point 10, shown in Fig.~\ref{fig:benchmark10}, the large
value of $\chhh=10$ completely dominates the shape, leading to a large
enhancement in the low $\mhh$ and $\pth$ regions. With a value for the
total cross section which is about 17 times larger than the SM cross
section, benchmark point 10 is still allowed by the limits given by
CMS~\cite{Sirunyan:2018iwt}, where separate limits for the various benchmark points
are given.

\begin{figure}[htb]
  \centering
  \begin{subfigure}{0.495\textwidth}
    \includegraphics[width=\textwidth]{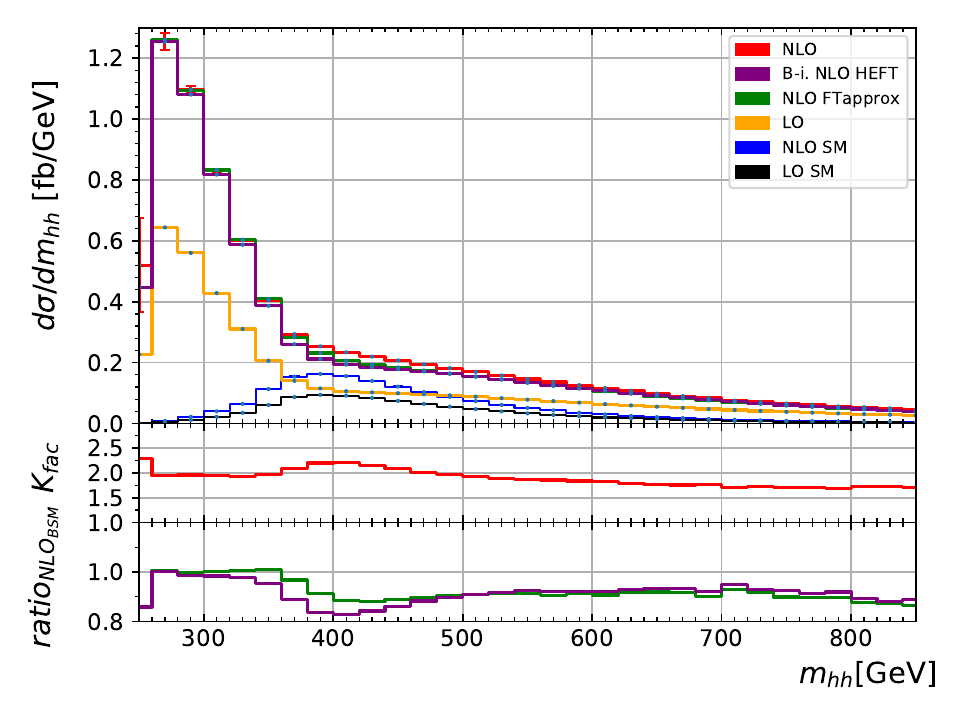}
    \vspace{\TwoFigBottom em}
    \caption{\label{fig:B11_mhh}}
  \end{subfigure}
  \hfill
  \begin{subfigure}{0.495\textwidth}
    \includegraphics[width=\textwidth]{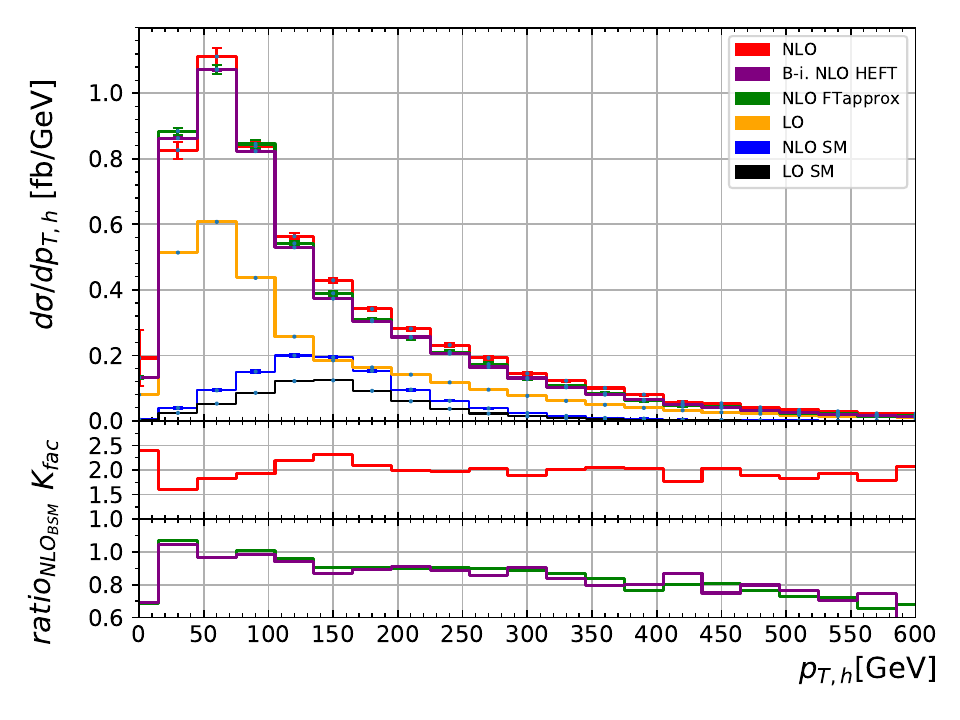}
    \vspace{\TwoFigBottom em}
    \caption{\label{fig:B11_pth}}
  \end{subfigure}
\caption{Same as Fig.~\ref{fig:benchmark1} but  for benchmark point 11, $\chhh=2.4,\ct= 1, \ctt=0, \cg=2/3,\cgg=1/3$.}
\label{fig:benchmark11}
\end{figure}
Benchmark point 11, displayed in Fig.~\ref{fig:benchmark11},  has the same value for $\chhh$ as benchmark 6, which is the one where the destructive interference would be maximal if all other couplings are kept SM-like.
However, the destructive interference is compensated by the large and non-zero values of $\cg$ and $\cgg$, such that the total cross section for benchmark 11 is about 5 times larger than the SM cross section. In view of the fact that this benchmark point is dominated by the Higgs-gluon contact interactions parametrised by $\cg$ and $\cgg$, 
it is not a surprise that the approximations \ftapprox\, and Born-improved HEFT agree quite well with the full calculation, as all three curves have these contributions in common, while the part which differs  is damped by the destructive interference. 

\begin{figure}[htb]
  \centering
  \begin{subfigure}{0.495\textwidth}
    \includegraphics[width=\textwidth]{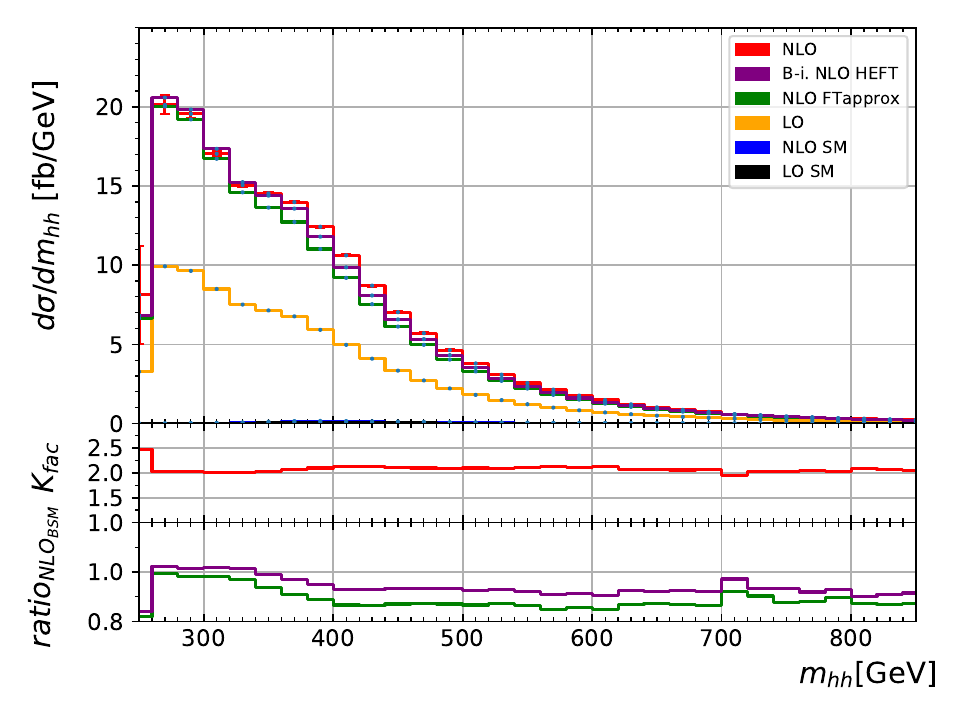}
    \vspace{\TwoFigBottom em}
    \caption{\label{fig:B12_mhh}}
  \end{subfigure}
  \hfill
  \begin{subfigure}{0.495\textwidth}
    \includegraphics[width=\textwidth]{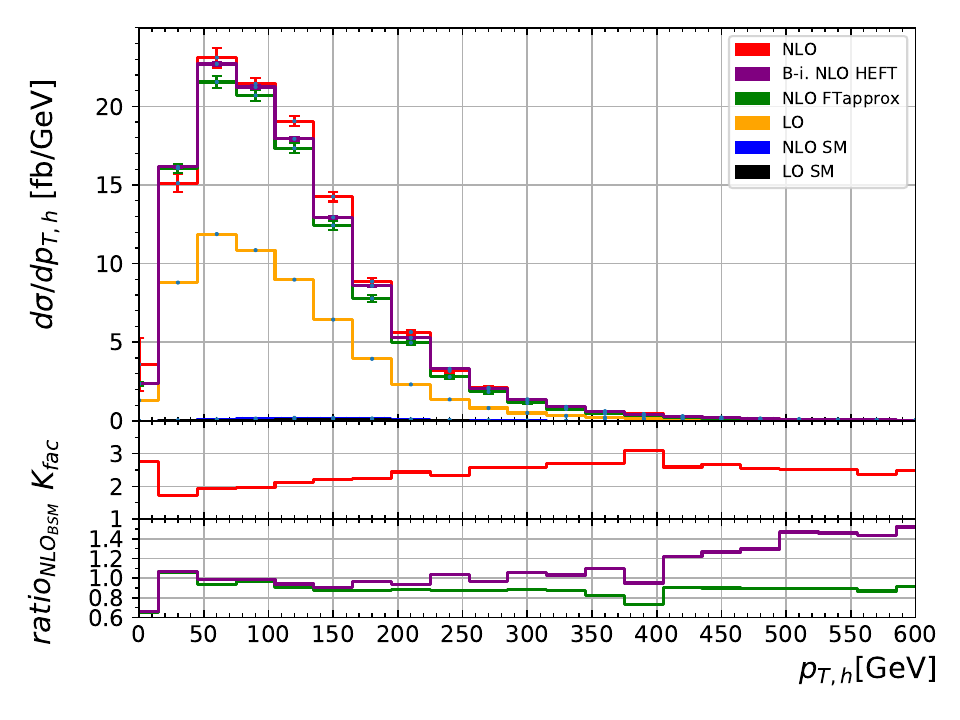}
    \vspace{\TwoFigBottom em}
    \caption{\label{fig:B12_pth}}
  \end{subfigure}
\caption{Same as Fig.~\ref{fig:benchmark1} but  for benchmark point 12, $\chhh=15,\ct= 1, \ctt=1, \cg=\cgg=0$.}
\label{fig:benchmark12}
\end{figure}

Benchmark point 12, shown in Fig.~\ref{fig:benchmark12}, has all
couplings SM-like except $\ctt=1$ and $\chhh$, where for the latter an extreme value of $\chhh=15$ is chosen, leading to a cross section about 100 times larger than the SM cross section.
This scenario is already ruled out by current LHC measurements.

All the distributions show that the NLO K-factors are large, being about a factor of two or larger. 
Therefore it is essential to take NLO corrections into account.
The approximations where the top quark mass dependence is only partly taken into account also differ substantially in the shape from the full result for some of the benchmark points, which emphasises the importance of including the full top quark mass dependence. 

\begin{figure}[htb]
  \centering
  \begin{subfigure}{0.495\textwidth}
    \includegraphics[width=\textwidth]{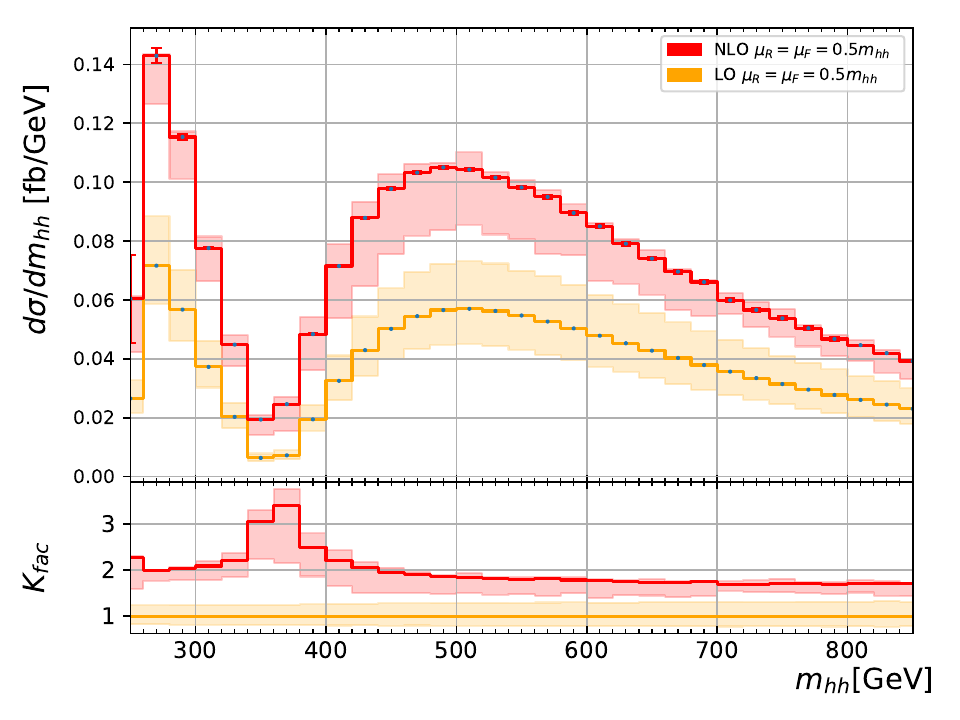}
    \vspace{\TwoFigBottom em}
    \caption{\label{fig:B5_mhh_scalevar}}
  \end{subfigure}
  \hfill
  \begin{subfigure}{0.495\textwidth}
    \includegraphics[width=\textwidth]{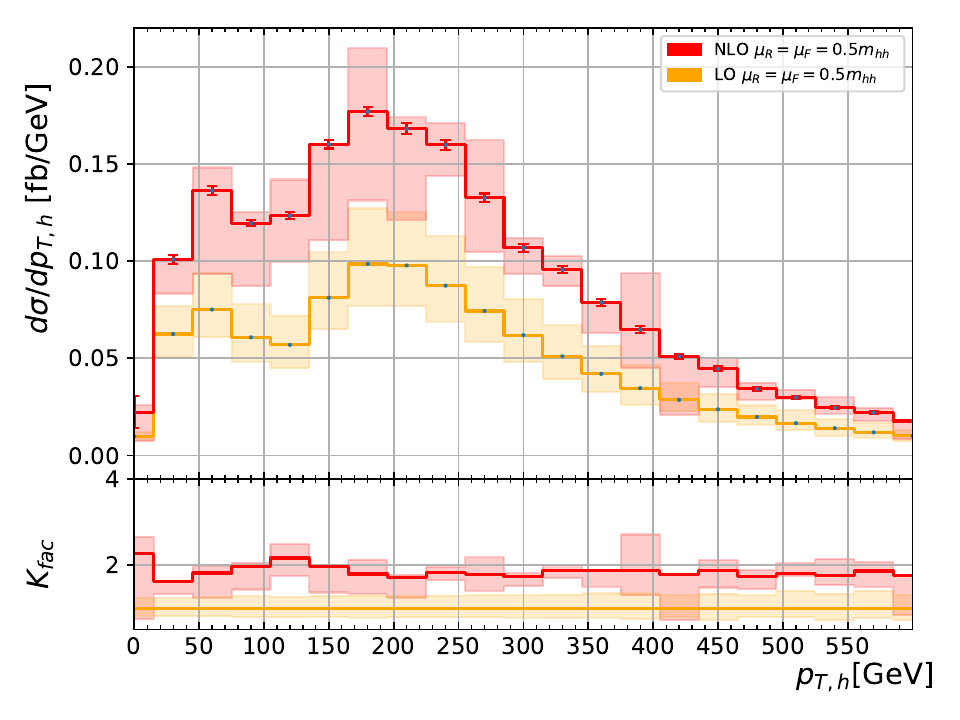}
    \vspace{\TwoFigBottom em}
    \caption{\label{fig:B5_pth_scalevar}}
  \end{subfigure}
\caption{Scale variations  for benchmark point 5.}
\label{fig:benchmark5_scalevar}
\end{figure}

In Fig.~\ref{fig:benchmark5_scalevar}, we show the LO and NLO scale
variation bands for benchmark point 5. This benchmark point is an
example where the scale variation band of the 7-point scale variation
mainly decreases the differential cross section over almost the whole
$\mhh$ range, where the upper limit of the scale variation band is
mostly given by the combination $\mu_F=\mu_0/2,\mu_R=\mu_0$, for some
of the bins also by  $\mu_F=\mu_0, \mu_R=2\mu_0$.
In the SM, the upper limit of the 7-point scale variation band is
given by $\mu_F=\mu_R=\mu_0$ for all bins of the $\mhh$ distribution.
We further notice that LO and NLO scale variation bands do not overlap
for the $\mhh$ distribution.
However, this feature is also present in the SM.

\subsubsection{Discussion of the benchmark points}

\begin{figure}[htb]
\begin{center}
\includegraphics[width=9cm]{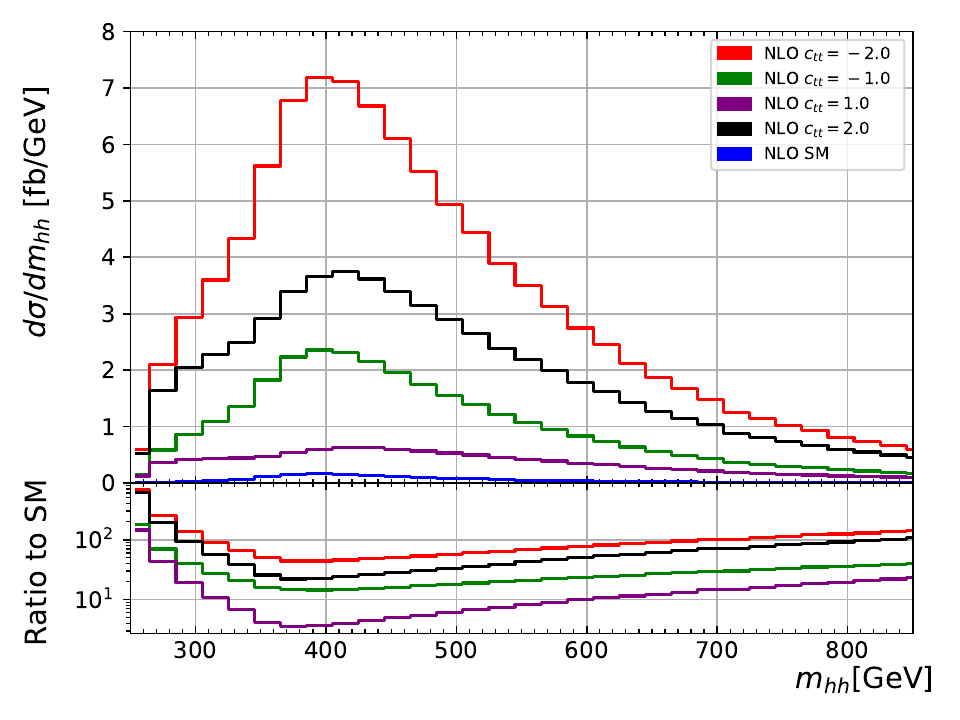}
\end{center}
\caption{Higgs boson pair invariant mass distributions for various values of $\ctt$.}
\label{fig:cttvariations}
\end{figure}

Attempting a more global view of the behaviour of the $\mhh$
distribution as a function of the five BSM parameters, we can identify
the following patterns: a dip in the $\mhh$ distributions is present
for benchmark points 1, 2, 5, 6 and 8a. 
The presence of a non-zero value for $\ctt$ or $\cg$ 
is a characteristic feature of many parameter space points that show a dip in the $\mhh$ distribution, 
but this is not a necessary condition for the presence of the dip.   
For instance, points with $\chhh \simeq  2.5 \ct$ and the other couplings vanishing also show such a dip.
For the subset (1, 2, 6) of the above
points there is a $\mhh$ value where the LO amplitude in the $m_t\to
\infty$ limit exactly vanishes, which is a feature that can cause the dip.
The low $\mhh$ region is enhanced for benchmark points 1, 6, 7, 10,
11, 12, which is mainly due to the large value of $\chhh$, as the
matrix element squared proportional to $\chhh^2$ for large $\hat{s}$ behaves like
$m_h^2/\hat{s}\,\log^2\left(m_t^2/\hat{s}\right)$~\cite{Azatov:2015oxa} and
therefore dominates at low values of $\hat{s}$. 
The term proportional to $\ctt^2$ for large $\hat{s}$ behaves like
$\log^2\left(m_t^2/\hat{s}\right)$ 
and seems to partially cancel the logarithmic terms from $\chhh$, such
that benchmark 4 has a SM-like shape even though the absolute value
for $\chhh$ is large.
The matrix element squared proportional to $\cgg$ grows like
$\hat{s}$, this is why for benchmark points which have large values of
$\cgg$, the tail of the $\mhh$ distribution is enhanced. 

In order to assess the effect of a variation of $\ctt$ while the other
couplings are fixed to their SM values, we show $\mhh$ distributions
for the $\ctt$ values $\ctt=-2,-1,0,1,2$ in Fig.~\ref{fig:cttvariations}.
The minimum of the cross section is at $\ctt\sim 0.25$.
We observe that the enhancement of the cross section as $|\ctt|$ increases is
growing more rapidly for negative values of $\ctt$, see also Fig.~\ref{fig:chhh_ctt}.
The shape changes compared to the SM are most pronounced in the low $\mhh$ region.

\section{Conclusions}

We have calculated the NLO QCD corrections with full $m_t$ dependence to Higgs boson
 pair production within the framework of the
 electroweak chiral Lagrangian, a non-linearly realised
 Effective Field Theory in the Higgs sector, which allows to focus on
 anomalous Higgs boson properties. 
This restricts the BSM parameter space to five possibly anomalous
 couplings, $\chhh,\ct,\ctt,\cg$ and $\cgg$.

We gave a parametrisation of the total NLO cross section and of the $\mhh$
 distribution in terms of 23 coefficients of all combinations of these
 couplings, and also showed iso-contours of LO and NLO cross section
 ratios $\sigma/\sigma_{SM}$ for two-dimensional projections of the
 parameter space. These studies showed that the cross sections are
 very sensitive to variations of $\ctt$, the effective $t\bar{t}hh$
 coupling, and that the K-factors can be large and non-uniform as the
 anomalous couplings are varied.

We have also shown differential cross sections for $\mhh$
and $\pth$ at several benchmark points which exhibit characteristic
shapes of the distributions. The differential K-factors for the
$\mhh$ distributions are of the order of two, but can reach up to
three and can be very non-uniform over the $\mhh$ range. 
This means that a rescaling of the LO distribution with a global
K-factor can be rather misleading. 

Some combinations of couplings lead to a huge enhancement of the cross
section, others lead to a total cross section which is nearly
degenerate to the SM one, but the corresponding $\mhh$ and $\pth$
distributions have a shape which is very different from the SM one, 
and therefore should have discriminating power even with low
statistics, which emphasises the importance of measuring distributions. 

Our analytical parametrisation of the total NLO cross section and of the
$\mhh$ distribution in terms of all possible combinations of anomalous
couplings should open the door to further studies of the considered
BSM parameter space and lead to refined limits on anomalous Higgs
boson couplings in the not too distant future.

\section*{Acknowledgements}
We would like to thank Luca Cadamuro, Oscar Cata, Nicolas Greiner, Ramona Gr{\"o}ber,
Stephen Jones, Matthias Kerner, Gionata
Luisoni, Joao Pires and Johannes Schlenk for many helpful discussions.
We also thank Alexandra Carvalho and Florian~Goertz for useful communication.
This research was supported in part by the COST Action CA16201 (`Particleface') of the European Union 
and by the DFG cluster of excellence EXC 153  `Origin and Structure of the Universe' and  DFG grant BU 1391/2-1.
We gratefully acknowledge resources provided by the Max Planck Computing and Data Facility (MPCDF).


\renewcommand \thesection{\Alph{section}}
\renewcommand{\theequation}{\Alph{section}.\arabic{equation}}
\setcounter{section}{0}
\setcounter{equation}{0}

\section{Appendix}

\subsection{Differential coefficients of all coupling combinations for $\mhh$}

In order to allow a flexible use of our results, we provide tables for
the differential coefficients of the various coupling combinations
contributing to the $\mhh$ distribution. 
They are given in {\tt .csv} format as ancillary files to the arXiv
submission and the JHEP publication.
The conventions are as follows. The given numbers are the coefficients
$\tilde{A}_i$ of the coupling combinations $c_i$ as given in
eq.~(\ref{eq:Acoeffs_all}). We call them $\tilde{A}_i$ rather than
$A_i$ to make clear that they are the {\em differential} coefficients.
In more detail, we provide the coefficients $\tilde{A}_i$ in
\begin{align}
\frac{d\sigma}{d\mhh}&=\sum_{i=1}^{23} \tilde{A}_i\,c_i
\end{align}
in units of fb/GeV, where the $c_i$ stand for the 23 possible
combinations of couplings, in the same order as in
Eq.~(\ref{eq:Acoeffs_all}).
We should point out that the $\tilde{A}_i$ are {\em not} normalised by the SM values.
The numbers in the 24 columns are 
$\mhh$ (bin center), $\tilde{A}_1, \ldots, \tilde{A}_{23}$.
Each line gives the numbers for one  bin in $\mhh$, with a bin size of
20\,GeV. We give results for 40 bins, corresponding to the range
$250\,{\rm GeV} \leq \mhh \leq 1030\,{\rm GeV}$ (bin center values).

The coefficients have been determined by 
evaluating 23 differential cross sections, obtained by changing the
values of the five coupling parameters. 
Then the system of 23 equations is solved to extract the value of each
coefficient $\tilde{A}_i$. 

\subsection{Relation between EWChL and SMEFT}

Accounting for deviations from the SM within a low-energy, bottom-up
effective field theory requires a power-counting prescription.
The power counting determines, in a systematic manner, which corrections
to include, and which ones to neglect, under certain assumptions that
need to be specified.
The power counting of the Higgs-electroweak chiral Lagrangian has been
reviewed in Sec. \ref{sec:ewchl}.
In this appendix we discuss how the analysis of $gg\to hh$
within the EWChL is related to a treatment of this process
using SMEFT. 

SMEFT is an EFT at the weak scale $v$, organised primarily by the canonical
dimension of operators. This corresponds to the assumption that New Physics 
enters at some generic scale $\Lambda$, presumably in the TeV range, 
and is weakly coupled to the SM fields. 
The renormalisable SM then represents the leading-order term.
The leading corrections are given by operators of canonical dimension 6
(apart from a single, lepton-number violating operator of dimension 5,
which is not relevant in the present context)
\cite{Buchmuller:1985jz,Grzadkowski:2010es}.
The dimension-6 terms relevant for $gg\to hh$ can be written as
\begin{align}\label{lsmeft}
\Delta{\cal L}_6 &=
\frac{\bar c_H}{2 v^2}\partial_\mu(\phi^\dagger\phi)\partial^\mu(\phi^\dagger\phi)
+\frac{\bar c_u}{v^2} y_t(\phi^\dagger\phi\, \bar q_L\tilde\phi t_R +{\rm h.c.})
-\frac{\bar c_6}{2 v^2}\frac{m^2_h}{v^2} (\phi^\dagger\phi)^3
\nonumber\\
&+\frac{\bar c_{ug}}{v^2} g_s
(\bar q_L\sigma^{\mu\nu}G_{\mu\nu}\tilde\phi t_R +{\rm h.c.})
+\frac{4\bar c_g}{v^2} g^2_s \phi^\dagger\phi\, G^a_{\mu\nu}G^{a\mu\nu}\;.
\end{align}
Here we follow the conventions used in \cite{Grober:2015cwa},
except for $\bar c_g$, which includes an extra factor of the electroweak 
coupling $g^2$ in the definition of \cite{Grober:2015cwa}.
We are assuming CP conservation to leading order in $\Delta{\cal L}_6$.
Then all coefficients in (\ref{lsmeft}) are real and the CP-odd operator
$\phi^\dagger\phi\, \tilde G^a_{\mu\nu}G^{a\mu\nu}$ can be omitted.

The SM amplitude for $gg\to hh$ (first and third diagram in the
top row of Fig.~\ref{fig:hprocess}) arises at one-loop order, counting
as $A_{SM}={\cal O}(1/16\pi^2)$. Considering the next order in SMEFT
based on (\ref{lsmeft}), we may distinguish two cases.
\begin{description}
\item
i) Pure dimensional counting: In this case we only assume that the dimension-6
operators in the SMEFT Lagrangian are suppressed by a factor of
$1/\Lambda^2$ from dimensional analysis. The coefficients in
(\ref{lsmeft}) are thus treated as $\bar c_i={\cal O}(v^2/\Lambda^2)$ 
by power counting.
It then follows that the dominant correction comes from the operator
$\phi^\dagger\phi\, G^a_{\mu\nu}G^{a\mu\nu}$ through the tree diagrams in the 
bottom row of Fig.~\ref{fig:hprocess}. This correction to the amplitude
counts as $\Delta A_g={\cal O}(v^2/\Lambda^2)$, a suppression that
is competing with the loop factor of $A_{SM}$. All other operators in
(\ref{lsmeft}) correct vertices within the SM loop diagrams and therefore
contribute $\Delta A_{other}={\cal O}((1/16\pi^2)(v^2/\Lambda^2))$.
This is a subleading effect, negligible in the scenario under consideration.
The dominant correction is thus described by a single parameter, $\bar c_g$.
In view of typical New-Physics models, such a scenario appears unrealistic.
To generalise the treatment, dimensional counting needs to be supplemented
by further assumptions, as discussed in the next item.
\item
ii) Dimensional counting including loop factors:
Supposing that the New Physics at scale $\Lambda$ is a weakly coupled gauge
theory, it can be shown that dimension-6 operators with field-strength 
factors are only generated through loop 
diagrams~\cite{Arzt:1994gp,Grzadkowski:2010es}.
Their coefficients then come with an extra factor of $1/16\pi^2$. In this case,
the coefficients $\bar c_{ug}$ and $\bar c_g$ in (\ref{lsmeft}) are counted
as order $(1/16\pi^2)(v^2/\Lambda^2)$, while $\bar c_H$, $\bar c_u$ and
$\bar c_6$ are still of order $v^2/\Lambda^2$.
As a consequence, the leading-order corrections to the SM amplitude for
$gg\to hh$ (see Fig. \ref{fig:hprocess}) come from the tree-diagrams with
$\bar c_g$, as well as from top-loop diagrams with vertices modified
by $\bar c_H$, $\bar c_u$ and $\bar c_6$. All these corrections count as order
$(1/16\pi^2)(v^2/\Lambda^2)$, a relative correction of order $v^2/\Lambda^2$
to $A_{SM}$. 
\end{description}

This discussion of applying SMEFT to $gg\to hh$ has several
implications:
\begin{description}
\item a)
Note that under both scenarios i) and ii) the magnetic-moment type
operator $\bar q_L\sigma^{\mu\nu}G_{\mu\nu}\tilde\phi t_R$ gives only a
subleading contribution (of order $1/16\pi^2$ times the leading correction)
and can be consistently neglected.
\item b)
We emphasise that the loop suppression of operators with field-strength
factors in scenario ii) follows the rules of chiral counting.
The Higgs-gluon operator, for example, is given by
$\kappa^2\phi^\dagger\phi\, g^2_s\, G^a_{\mu\nu}G^{a\mu\nu}$, when we include
the weak\footnote{In the context of power counting, ``weak coupling''
  means a coupling of order unity, as opposed to a strong coupling of
  order $4\pi$.}  couplings, $g_s$ for each gluon, and $\kappa$ for the coupling of
$\phi$ to the heavy sector. The chiral dimension of the Higgs-gluon operator
is then $d_\chi\equiv 2 L + 2=6$, corresponding to a loop order of $L=2$.
Taking into account the canonical dimension, the coefficient is estimated 
as \cite{Buchalla:2016sop}
\begin{equation}\label{cgchidim}
\frac{1}{v^2}\frac{1}{(16\pi^2)^2}=\frac{1}{\Lambda^2}\frac{1}{16\pi^2}\;,
\end{equation}
where we used the NDA relation $\Lambda=4\pi v$ \cite{Manohar:1983md}.
This implies a $\bar c_g$ of order $(1/16\pi^2)(v^2/\Lambda^2)$, in agreement
with \cite{Arzt:1994gp}. A similar argument holds for $\bar c_{ug}$.
\item c)
The coefficients $\bar c_i$ may be related to the couplings of the physical
Higgs field $h$ and compared with the corresponding parameters of the
chiral Lagrangian (\ref{eq:ewchl}).
After a field redefinition of $h$ to eliminate $\bar c_H$ from the
kinetic term one finds \cite{Azatov:2015oxa,Grober:2015cwa}
\begin{equation}\label{cthhh}
c_t=1-\frac{\bar c_H}{2}-\bar c_u\, ,\quad
c_{tt}=-\frac{\bar c_H + 3\bar c_u}{2}\, ,\quad
c_{hhh}=1-\frac{3}{2}\bar c_H +\bar c_6\;,
\end{equation}
\begin{equation}\label{cghh}
c_{ggh}=2 c_{gghh}=128\pi^2\bar c_g\;.
\end{equation}
\item d)
After taking into account factors from loop counting in scenario ii),
the parametrisation of New-Physics effects in $gg\to hh$ is similar in SMEFT 
and in the EWChL, as is apparent from (\ref{cthhh}) and (\ref{cghh}).
However, there are still notable differences.  
While the four SMEFT coefficients $\bar c_H$, $\bar c_u$, $\bar c_6$
and $16\pi^2\, \bar c_g$ are parametrically small, of order $v^2/\Lambda^2$,
the non-linear coefficients $c_t$, $c_{tt}$, $c_{hhh}$, $c_{ggh}$ and
$c_{gghh}$ may be treated as quantities of order one. No further expansion
in the latter coefficients is needed when computing cross sections,
whereas the SMEFT coefficients should only be kept to first order
when working at the level of dimension-6 operators.
It appears that the SMEFT treatment has only 4 parameters, instead of
5 for the EWChL, due to the relation in (\ref{cghh}).
It should, however, be kept in mind that the extraction of Higgs couplings
ultimately requires a global analysis, where other Higgs-related processes
are also taken into account. In this case, in particular for the important
single-Higgs observables, the EWChL has overall fewer parameters to
describe leading NP effects \cite{Buchalla:2015wfa}.
It has the advantage to focus on anomalous Higgs properties and to
naturally allow for large deviations from the SM in the Higgs sector.   
\end{description}

\section{Erratum}

After comparison with the authors of Ref.~\cite{Bagnaschi:2023rbx}, it turned out
that the two-loop amplitude used in
the original version of this work was
missing a term related to triangle-type diagrams, affecting the cases
where the ratio between trilinear Higgs coupling $c_{hhh}$ and Yukawa coupling modifier $c_t$ is different from 1 (i.e. the Standard Model (SM) value), 
or when the effective coupling of a $t\bar{t}$ pair to a Higgs pair, $c_{tt}$, is nonzero.
The SM results are unchanged.
Therefore, benchmark points with a value of $c_{hhh}/c_{t}$ or
$c_{tt}$ very different from the SM are the most affected.
We have recalculated the values for the cross sections at the 12
benchmark points shown in Table~\ref{tab:benchmarks} of the main text.
In Table \ref{tab:benchmarks_sigmatot}, we show a comparison of the corrected values for the cross sections to the previous values.

\begin{table}[htb]
  \centering
  \phantom{x}\medskip
  {\renewcommand{\arraystretch}{1.2} 
  \begin{tabular}{|c|c|c|c|c|c|}
    \bottomrule
    Benchmark & $\sigma_{\mathrm{NLO}}^{\rm{old}}$ [fb] & \multicolumn{3}{c|}{$\sigma_{\mathrm{NLO}}^{\rm{new}}$ [fb]} & $\sigma_{\mathrm{NLO}}^{\rm{new}}/ \sigma_{\mathrm{NLO},\mathrm{SM}}$\\
    & 14 TeV & 13 TeV & 13.6 TeV & 14 TeV & 14 TeV  \\
    \toprule
    \hline
$B_1$ & 194.89 & 150.80 & 168.35 & 180.53 & 5.48 \\ \hline
$B_2$ & 14.55 & 10.06 & 11.51 & 12.54 & 0.38 \\ \hline
$B_3$ & 1047.37 & 803.78 & 894.69 & 957.79 & 29.07 \\ \hline
$B_4$ & 8922.75 & 7050.62 & 7811.76 & 8338.07 & 253.05 \\ \hline
$B_5$ & 59.325 & 48.66 & 54.93 & 59.33 & 1.80 \\ \hline
$B_6$ & 24.69 & 20.73 & 22.97 & 24.53 & 0.74 \\ \hline
$B_7$ & 169.41 & 140.97 & 154.92 & 164.52 & 4.99 \\ \hline
$B_{8a}$ & 41.70 & 30.36 & 33.87 & 36.32 & 1.10 \\ \hline
$B_9$ & 146.00 & 101.63 & 114.01 & 122.66 & 3.72 \\ \hline
$B_{10}$ & 575.86 & 481.17 & 529.65 & 563.00 & 17.09 \\ \hline
$B_{11}$ & 174.70 & 145.84 & 161.91 & 173.06 & 5.25 \\ \hline
$B_{12}$ & 3618.53 & 2925.69 & 3223.98 & 3429.40 & 104.08 \\ \hline

  \end{tabular}
  }
\caption{Comparison of the total cross section values at NLO before and after the
correction at a centre-of-mass energy of $\sqrt{s}=14$ TeV and ratio of the new values to the SM cross section, $\sigma_\mathrm{NLO,SM}(14 \mathrm{\,TeV})=32.95 \mathrm{\,fb}$. In addition, we provide corrected cross-section values at
$\sqrt{s}=13$ TeV and $13.6$ TeV.
\label{tab:benchmarks_sigmatot}}
\end{table}

\begin{figure}[htb]
\includegraphics[width=0.49\textwidth,page=1]{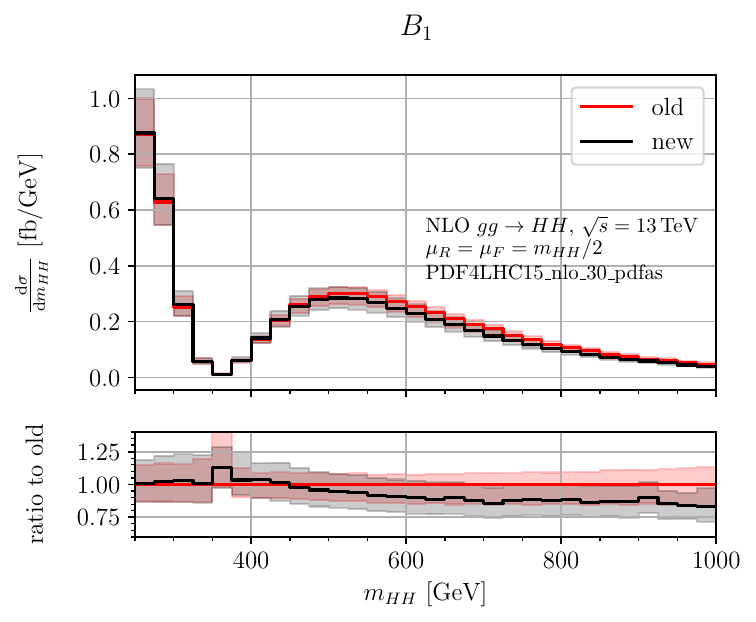}\hspace{2pc}%
\includegraphics[width=0.49\textwidth,page=2]{plot_1806.05162.pdf}\hspace{2pc}%
  \caption{\label{fig:compare_old_new} 
     Comparison of old and new results for the cross sections
     differential in $\mhh$ for benchmark points 1 and 10 of Table~3
     in the main text, at $\sqrt{s}=13$\,TeV.}
\end{figure}

In Fig.~\ref{fig:compare_old_new} we show the effects of the correction on the $\mhh$ distribution for benchmark points 1 and 10, which are affected most due to their large value of $\chhh$.
The differences are found to be below $\sim 20\%$ and therefore within the scale and top mass scheme uncertainties. In general, we have observed that the relative size of the scale uncertainty bands is not significantly affected by the correction.

We also provide a new fit of the $A_i$ coefficients at NLO,
\setcounter{section}{1}
\begin{align}
    \frac{\sigma_{\mathrm{NLO}}}{\sigma_{\mathrm{NLO,SM}}} &= A_1 c_t^4 + A_2 c_{tt}^2 + A_3 c_t^2 c_{hhh}^2 + A_4 c_{ggh}^2 c_{hhh}^2 + A_5 c_{gghh}^2 + A_6 c_{tt} c_t^2 + A_7 c_t^3 c_{hhh} \nonumber\\
    &+ A_8 c_{tt} c_t c_{hhh} + A_9 c_{tt} c_{ggh} c_{hhh} + A_{10} c_{tt} c_{gghh} + A_{11} c_t^2 c_{ggh} c_{hhh} + A_{12} c_t^2 c_{gghh} \nonumber \\
    &+ A_{13} c_t c_{hhh}^2 c_{ggh} + A_{14} c_t c_{hhh} c_{gghh} + A_{15} c_{ggh} c_{hhh} c_{gghh} \nonumber\\
    &+ A_{16} c_t^3 c_{ggh} + A_{17} c_t c_{tt} c_{ggh} + A_{18} c_t c_{ggh}^2 c_{hhh} + A_{19} c_t c_{ggh} c_{gghh} \nonumber \\
    &+ A_{20} c_t^2 c_{ggh}^2 + A_{21} c_{tt} c_{ggh}^2 + A_{22} c_{ggh}^3 c_{hhh} + A_{23} c_{ggh}^2 c_{gghh}\,,
\label{eq:Ai}
\end{align}

as given in Table~\ref{tab:Ai}.
For the corrected values, our treatment of the uncertainties has also improved, now including statistical uncertainties from the sample of BSM points as well as correlations among the coefficients. We provide them in Table~\ref{tab:Ai_sigmatot} for $\sqrt{s}=13$ and $13.6$\,TeV, with $\sigma_{\mathrm{NLO,SM}}(13\,\mathrm{TeV})=27.80\mathrm{\,fb}$ and $\sigma_{\mathrm{NLO,SM}}(13.6\,\mathrm{TeV})=30.82\mathrm{\,fb}$.
\begin{table}[htb]
  \centering
  \phantom{x}\medskip
  {\renewcommand{\arraystretch}{1.2} 
  \begin{tabular}{|c|c|c|}
    \bottomrule
    Coefficient & 13 TeV & 13.6 TeV \\
    \toprule
    \hline
$A_{1}$ & $2.20913 \pm 0.00034$  & $2.20259 \pm 0.00014$   \\
$A_{2}$ & $11.2754 \pm 0.0041$  & $11.31544 \pm 0.00062$   \\
$A_{3}$ & $0.334152 \pm 0.000073$  & $0.331430 \pm 0.000029$   \\
$A_{4}$ & $0.3520 \pm 0.0011$  & $0.34943 \pm 0.00030$   \\
$A_{5}$ & $12.631 \pm 0.036$  & $12.83225 \pm 0.00066$   \\
$A_{6}$ & $-9.1965 \pm 0.0046$  & $-9.18628 \pm 0.00060$   \\
$A_{7}$ & $-1.54327 \pm 0.00035$  & $-1.53405 \pm 0.00014$   \\
$A_{8}$ & $3.26347 \pm 0.00076$  & $3.25036 \pm 0.00023$   \\
$A_{9}$ & $2.811 \pm 0.011$  & $2.7974 \pm 0.0014$   \\
$A_{10}$ & $16.139 \pm 0.025$  & $16.12925 \pm 0.00096$   \\
$A_{11}$ & $-1.2628 \pm 0.0077$  & $-1.2534 \pm 0.0011$   \\
$A_{12}$ & $-5.818 \pm 0.016$  & $-5.7712 \pm 0.0012$   \\
$A_{13}$ & $0.6485 \pm 0.0015$  & $0.64328 \pm 0.00021$   \\
$A_{14}$ & $2.8127 \pm 0.0025$  & $2.79661 \pm 0.00042$   \\
$A_{15}$ & $3.1813 \pm 0.0098$  & $3.16880 \pm 0.00089$   \\
$A_{16}$ & $-0.0075 \pm 0.0052$  & $-0.00877 \pm 0.00084$   \\
$A_{17}$ & $0.023 \pm 0.012$  & $0.0219 \pm 0.0017$   \\
$A_{18}$ & $0.0171 \pm 0.0034$  & $0.01792 \pm 0.00037$   \\
$A_{19}$ & $0.023 \pm 0.030$  & $0.0271 \pm 0.0014$   \\
$A_{20}$ & $-0.0279 \pm 0.0011$  & $-0.02741 \pm 0.00017$   \\
$A_{21}$ & $0.079 \pm 0.027$  & $0.07335 \pm 0.00064$   \\
$A_{22}$ & $0.0150 \pm 0.0033$  & $0.01547 \pm 0.00043$   \\
$A_{23}$ & $0.117 \pm 0.036$  & $0.11712 \pm 0.00082$   \\

  \toprule
  \end{tabular}
  }
\caption{Updated values of the $A_i$ coefficients at NLO, as per Eq.~(\ref{eq:Ai}). The uncertainties quoted here are statistical and include correlations between coefficients.
\label{tab:Ai_sigmatot}}
\end{table}

We also updated the ancillary files available on arXiv for the $A_i$ coefficients, both for the inclusive cross sections and the cross sections differential in $m_{HH}$, at $\sqrt{s}=13$\,TeV and $\sqrt{s}=13.6$\,TeV.

\vspace*{5mm}

We would like to thank the authors of Ref.~\cite{Bagnaschi:2023rbx} for pointing us to the discrepancy with their result.


\bibliographystyle{JHEP}
 
\bibliography{refs_EWChL}

\providecommand{\href}[2]{#2}\begingroup\raggedright\begin{thebibliography}{100}

\bibitem{Khachatryan:2016vau}
{\bf ATLAS, CMS} Collaboration, G.~Aad et~al., {\it {Measurements of the Higgs
  boson production and decay rates and constraints on its couplings from a
  combined ATLAS and CMS analysis of the LHC pp collision data at $ \sqrt{s}=7
  $ and 8 TeV}},  {\em JHEP} {\bf 08} (2016) 045,
  [\href{http://arxiv.org/abs/1606.02266}{{\tt arXiv:1606.02266}}].

\bibitem{Brooijmans:2018xbu}
G.~Brooijmans et~al., {\it {Les Houches 2017: Physics at TeV Colliders New
  Physics Working Group Report}},  in {\em {10th Les Houches Workshop on
  Physics at TeV Colliders (PhysTeV 2017) Les Houches, France, June 5-23,
  2017}}, 2018.
\newblock \href{http://arxiv.org/abs/1803.10379}{{\tt arXiv:1803.10379}}.

\bibitem{Buchmuller:1985jz}
W.~Buchm{\"u}ller and D.~Wyler, {\it {Effective Lagrangian Analysis of New
  Interactions and Flavor Conservation}},  {\em Nucl. Phys.} {\bf B268} (1986)
  621--653.

\bibitem{Grzadkowski:2010es}
B.~Grzadkowski, M.~Iskrzynski, M.~Misiak, and J.~Rosiek, {\it {Dimension-Six
  Terms in the Standard Model Lagrangian}},  {\em JHEP} {\bf 10} (2010) 085,
  [\href{http://arxiv.org/abs/1008.4884}{{\tt arXiv:1008.4884}}].

\bibitem{Berthier:2015oma}
L.~Berthier and M.~Trott, {\it {Towards consistent Electroweak Precision Data
  constraints in the SMEFT}},  {\em JHEP} {\bf 05} (2015) 024,
  [\href{http://arxiv.org/abs/1502.02570}{{\tt arXiv:1502.02570}}].

\bibitem{Ghezzi:2015vva}
M.~Ghezzi, R.~Gomez-Ambrosio, G.~Passarino, and S.~Uccirati, {\it {NLO Higgs
  effective field theory and $\kappa$-framework}},  {\em JHEP} {\bf 07} (2015)
  175, [\href{http://arxiv.org/abs/1505.03706}{{\tt arXiv:1505.03706}}].

\bibitem{deBlas:2017wmn}
J.~de~Blas, M.~Ciuchini, E.~Franco, S.~Mishima, M.~Pierini, L.~Reina, and
  L.~Silvestrini, {\it {The Global Electroweak and Higgs Fits in the LHC era}},
   {\em PoS} {\bf EPS-HEP2017} (2017) 467,
  [\href{http://arxiv.org/abs/1710.05402}{{\tt arXiv:1710.05402}}].

\bibitem{Brivio:2017bnu}
I.~Brivio and M.~Trott, {\it {Scheming in the SMEFT... and a reparameterization
  invariance!}},  {\em JHEP} {\bf 07} (2017) 148,
  [\href{http://arxiv.org/abs/1701.06424}{{\tt arXiv:1701.06424}}].

\bibitem{Ellis:2018gqa}
J.~Ellis, C.~W. Murphy, V.~Sanz, and T.~You, {\it {Updated Global SMEFT Fit to
  Higgs, Diboson and Electroweak Data}},
  \href{http://arxiv.org/abs/1803.03252}{{\tt arXiv:1803.03252}}.

\bibitem{Feruglio:1992wf}
F.~Feruglio, {\it {The Chiral approach to the electroweak interactions}},  {\em
  Int. J. Mod. Phys.} {\bf A8} (1993) 4937--4972,
  [\href{http://arxiv.org/abs/hep-ph/9301281}{{\tt hep-ph/9301281}}].

\bibitem{Bagger:1993zf}
J.~Bagger, V.~D. Barger, K.-m. Cheung, J.~F. Gunion, T.~Han, G.~A. Ladinsky,
  R.~Rosenfeld, and C.~P. Yuan, {\it {The Strongly interacting W W system: Gold
  plated modes}},  {\em Phys. Rev.} {\bf D49} (1994) 1246--1264,
  [\href{http://arxiv.org/abs/hep-ph/9306256}{{\tt hep-ph/9306256}}].

\bibitem{Koulovassilopoulos:1993pw}
V.~Koulovassilopoulos and R.~S. Chivukula, {\it {The Phenomenology of a
  nonstandard Higgs boson in W(L) W(L) scattering}},  {\em Phys. Rev.} {\bf
  D50} (1994) 3218--3234, [\href{http://arxiv.org/abs/hep-ph/9312317}{{\tt
  hep-ph/9312317}}].

\bibitem{Burgess:1999ha}
C.~P. Burgess, J.~Matias, and M.~Pospelov, {\it {A Higgs or not a Higgs? What
  to do if you discover a new scalar particle}},  {\em Int. J. Mod. Phys.} {\bf
  A17} (2002) 1841--1918, [\href{http://arxiv.org/abs/hep-ph/9912459}{{\tt
  hep-ph/9912459}}].

\bibitem{Wang:2006im}
L.-M. Wang and Q.~Wang, {\it {Electroweak chiral Lagrangian for neutral Higgs
  boson}},  {\em Chin. Phys. Lett.} {\bf 25} (2008) 1984,
  [\href{http://arxiv.org/abs/hep-ph/0605104}{{\tt hep-ph/0605104}}].

\bibitem{Grinstein:2007iv}
B.~Grinstein and M.~Trott, {\it {A Higgs-Higgs bound state due to new physics
  at a TeV}},  {\em Phys. Rev.} {\bf D76} (2007) 073002,
  [\href{http://arxiv.org/abs/0704.1505}{{\tt arXiv:0704.1505}}].

\bibitem{Contino:2010mh}
R.~Contino, C.~Grojean, M.~Moretti, F.~Piccinini, and R.~Rattazzi, {\it {Strong
  Double Higgs Production at the LHC}},  {\em JHEP} {\bf 05} (2010) 089,
  [\href{http://arxiv.org/abs/1002.1011}{{\tt arXiv:1002.1011}}].

\bibitem{Contino:2010rs}
R.~Contino, {\it {The Higgs as a Composite Nambu-Goldstone Boson}},  in {\em
  {Physics of the large and the small, TASI 09, proceedings of the Theoretical
  Advanced Study Institute in Elementary Particle Physics, Boulder, Colorado,
  USA, 1-26 June 2009}}, pp.~235--306, 2011.
\newblock \href{http://arxiv.org/abs/1005.4269}{{\tt arXiv:1005.4269}}.

\bibitem{Alonso:2012px}
R.~Alonso, M.~B. Gavela, L.~Merlo, S.~Rigolin, and J.~Yepes, {\it {The
  Effective Chiral Lagrangian for a Light Dynamical "Higgs Particle"}},  {\em
  Phys. Lett.} {\bf B722} (2013) 330--335,
  [\href{http://arxiv.org/abs/1212.3305}{{\tt arXiv:1212.3305}}]. [Erratum:
  Phys. Lett.B726,926(2013)].

\bibitem{Buchalla:2013rka}
G.~Buchalla, O.~Cata, and C.~Krause, {\it {Complete Electroweak Chiral
  Lagrangian with a Light Higgs at NLO}},  {\em Nucl. Phys.} {\bf B880} (2014)
  552--573, [\href{http://arxiv.org/abs/1307.5017}{{\tt arXiv:1307.5017}}].
  [Erratum: Nucl. Phys.B913,475(2016)].

\bibitem{Delgado:2013hxa}
R.~L. Delgado, A.~Dobado, and F.~J. Llanes-Estrada, {\it {One-loop $W_LW_L$ and
  $Z_LZ_L$ scattering from the electroweak Chiral Lagrangian with a light
  Higgs-like scalar}},  {\em JHEP} {\bf 02} (2014) 121,
  [\href{http://arxiv.org/abs/1311.5993}{{\tt arXiv:1311.5993}}].

\bibitem{Buchalla:2013eza}
G.~Buchalla, O.~Cata, and C.~Krause, {\it {On the Power Counting in Effective
  Field Theories}},  {\em Phys. Lett.} {\bf B731} (2014) 80--86,
  [\href{http://arxiv.org/abs/1312.5624}{{\tt arXiv:1312.5624}}].

\bibitem{Buchalla:2015qju}
G.~Buchalla, O.~Cata, A.~Celis, and C.~Krause, {\it {Fitting Higgs Data with
  Nonlinear Effective Theory}},  {\em Eur. Phys. J.} {\bf C76} (2016), no.~5
  233, [\href{http://arxiv.org/abs/1511.00988}{{\tt arXiv:1511.00988}}].

\bibitem{deBlas:2018tjm}
J.~de~Blas, O.~Eberhardt, and C.~Krause, {\it {Current and future constraints
  on Higgs couplings in the nonlinear Effective Theory}},
  \href{http://arxiv.org/abs/1803.00939}{{\tt arXiv:1803.00939}}.

\bibitem{Pierce:2006dh}
A.~Pierce, J.~Thaler, and L.-T. Wang, {\it {Disentangling Dimension Six
  Operators through Di-Higgs Boson Production}},  {\em JHEP} {\bf 05} (2007)
  070, [\href{http://arxiv.org/abs/hep-ph/0609049}{{\tt hep-ph/0609049}}].

\bibitem{Contino:2012xk}
R.~Contino, M.~Ghezzi, M.~Moretti, G.~Panico, F.~Piccinini, and A.~Wulzer, {\it
  {Anomalous Couplings in Double Higgs Production}},  {\em JHEP} {\bf 08}
  (2012) 154, [\href{http://arxiv.org/abs/1205.5444}{{\tt arXiv:1205.5444}}].

\bibitem{Baglio:2012np}
J.~Baglio, A.~Djouadi, R.~Gr{\"o}ber, M.~M. M{\"u}hlleitner, J.~Quevillon, and
  M.~Spira, {\it {The measurement of the Higgs self-coupling at the LHC:
  theoretical status}},  {\em JHEP} {\bf 04} (2013) 151,
  [\href{http://arxiv.org/abs/1212.5581}{{\tt arXiv:1212.5581}}].

\bibitem{Dawson:2012mk}
S.~Dawson, E.~Furlan, and I.~Lewis, {\it {Unravelling an extended quark sector
  through multiple Higgs production?}},  {\em Phys. Rev.} {\bf D87} (2013),
  no.~1 014007, [\href{http://arxiv.org/abs/1210.6663}{{\tt arXiv:1210.6663}}].

\bibitem{Dolan:2012rv}
M.~J. Dolan, C.~Englert, and M.~Spannowsky, {\it {Higgs self-coupling
  measurements at the LHC}},  {\em JHEP} {\bf 10} (2012) 112,
  [\href{http://arxiv.org/abs/1206.5001}{{\tt arXiv:1206.5001}}].

\bibitem{Goertz:2014qta}
F.~Goertz, A.~Papaefstathiou, L.~L. Yang, and J.~Zurita, {\it {Higgs boson pair
  production in the D=6 extension of the SM}},  {\em JHEP} {\bf 04} (2015) 167,
  [\href{http://arxiv.org/abs/1410.3471}{{\tt arXiv:1410.3471}}].

\bibitem{Barr:2014sga}
A.~J. Barr, M.~J. Dolan, C.~Englert, D.~E. Ferreira~de Lima, and M.~Spannowsky,
  {\it {Higgs Self-Coupling Measurements at a 100 TeV Hadron Collider}},  {\em
  JHEP} {\bf 02} (2015) 016, [\href{http://arxiv.org/abs/1412.7154}{{\tt
  arXiv:1412.7154}}].

\bibitem{Azatov:2015oxa}
A.~Azatov, R.~Contino, G.~Panico, and M.~Son, {\it {Effective field theory
  analysis of double Higgs boson production via gluon fusion}},  {\em Phys.
  Rev.} {\bf D92} (2015), no.~3 035001,
  [\href{http://arxiv.org/abs/1502.00539}{{\tt arXiv:1502.00539}}].

\bibitem{Dolan:2015zja}
M.~J. Dolan, C.~Englert, N.~Greiner, K.~Nordstrom, and M.~Spannowsky, {\it
  {$hhjj$ production at the LHC}},  {\em Eur. Phys. J.} {\bf C75} (2015), no.~8
  387, [\href{http://arxiv.org/abs/1506.08008}{{\tt arXiv:1506.08008}}].

\bibitem{Behr:2015oqq}
J.~K. Behr, D.~Bortoletto, J.~A. Frost, N.~P. Hartland, C.~Issever, and
  J.~Rojo, {\it {Boosting Higgs pair production in the $b\bar{b}b\bar{b}$ final
  state with multivariate techniques}},  {\em Eur. Phys. J.} {\bf C76} (2016),
  no.~7 386, [\href{http://arxiv.org/abs/1512.08928}{{\tt arXiv:1512.08928}}].

\bibitem{Maltoni:2016yxb}
F.~Maltoni, E.~Vryonidou, and C.~Zhang, {\it {Higgs production in association
  with a top-antitop pair in the Standard Model Effective Field Theory at NLO
  in QCD}},  {\em JHEP} {\bf 10} (2016) 123,
  [\href{http://arxiv.org/abs/1607.05330}{{\tt arXiv:1607.05330}}].

\bibitem{Kling:2016lay}
F.~Kling, T.~Plehn, and P.~Schichtel, {\it {Maximizing the significance in
  Higgs boson pair analyses}},  {\em Phys. Rev.} {\bf D95} (2017), no.~3
  035026, [\href{http://arxiv.org/abs/1607.07441}{{\tt arXiv:1607.07441}}].

\bibitem{Cao:2016zob}
Q.-H. Cao, G.~Li, B.~Yan, D.-M. Zhang, and H.~Zhang, {\it {Double Higgs
  production at the 14 TeV LHC and a 100 TeV $pp$ collider}},  {\em Phys. Rev.}
  {\bf D96} (2017), no.~9 095031, [\href{http://arxiv.org/abs/1611.09336}{{\tt
  arXiv:1611.09336}}].

\bibitem{DiVita:2017eyz}
S.~Di~Vita, C.~Grojean, G.~Panico, M.~Riembau, and T.~Vantalon, {\it {A global
  view on the Higgs self-coupling}},  {\em JHEP} {\bf 09} (2017) 069,
  [\href{http://arxiv.org/abs/1704.01953}{{\tt arXiv:1704.01953}}].

\bibitem{DiLuzio:2017tfn}
L.~Di~Luzio, R.~Gr{\"ob}er, and M.~Spannowsky, {\it {Maxi-sizing the trilinear
  Higgs self-coupling: how large could it be?}},  {\em Eur. Phys. J.} {\bf C77}
  (2017), no.~11 788, [\href{http://arxiv.org/abs/1704.02311}{{\tt
  arXiv:1704.02311}}].

\bibitem{Corbett:2017ieo}
T.~Corbett, A.~Joglekar, H.-L. Li, and J.-H. Yu, {\it {Exploring Extended
  Scalar Sectors with Di-Higgs Signals: A Higgs EFT Perspective}},  {\em JHEP}
  {\bf 05} (2018) 061, [\href{http://arxiv.org/abs/1705.02551}{{\tt
  arXiv:1705.02551}}].

\bibitem{Dawson:2017vgm}
S.~Dawson and C.~W. Murphy, {\it {Standard Model EFT and Extended Scalar
  Sectors}},  {\em Phys. Rev.} {\bf D96} (2017), no.~1 015041,
  [\href{http://arxiv.org/abs/1704.07851}{{\tt arXiv:1704.07851}}].

\bibitem{Alves:2017ued}
A.~Alves, T.~Ghosh, and K.~Sinha, {\it {Can We Discover Double Higgs Production
  at the LHC?}},  {\em Phys. Rev.} {\bf D96} (2017), no.~3 035022,
  [\href{http://arxiv.org/abs/1704.07395}{{\tt arXiv:1704.07395}}].

\bibitem{Adhikary:2017jtu}
A.~Adhikary, S.~Banerjee, R.~K. Barman, B.~Bhattacherjee, and S.~Niyogi, {\it
  {Revisiting the non-resonant Higgs pair production at the HL-LHC}},  {\em
  Physics} {\bf 2018} (2018) 116, [\href{http://arxiv.org/abs/1712.05346}{{\tt
  arXiv:1712.05346}}].

\bibitem{Kim:2018uty}
J.~H. Kim, Y.~Sakaki, and M.~Son, {\it {Combined analysis of double Higgs
  production via gluon fusion at the HL-LHC in the effective field theory
  approach}},  \href{http://arxiv.org/abs/1801.06093}{{\tt arXiv:1801.06093}}.

\bibitem{Goncalves:2018qas}
D.~Gonçalves, T.~Han, F.~Kling, T.~Plehn, and M.~Takeuchi, {\it {Higgs Pair
  Production at Future Hadron Colliders: From Kinematics to Dynamics}},
  \href{http://arxiv.org/abs/1802.04319}{{\tt arXiv:1802.04319}}.

\bibitem{Eboli:1987dy}
O.~J.~P. Eboli, G.~C. Marques, S.~F. Novaes, and A.~A. Natale, {\it {Twin Higgs
  Boson Production}},  {\em Phys. Lett.} {\bf B197} (1987) 269.

\bibitem{Glover:1987nx}
E.~W.~N. Glover and J.~J. van~der Bij, {\it {Higgs Boson Pair Production via
  Gluon Fusion}},  {\em Nucl. Phys.} {\bf B309} (1988) 282.

\bibitem{Plehn:1996wb}
T.~Plehn, M.~Spira, and P.~M. Zerwas, {\it {Pair production of neutral Higgs
  particles in gluon-gluon collisions}},  {\em Nucl. Phys.} {\bf B479} (1996)
  46--64, [\href{http://arxiv.org/abs/hep-ph/9603205}{{\tt hep-ph/9603205}}].
  [Erratum: Nucl. Phys.B531,655(1998)].

\bibitem{Dawson:1998py}
S.~Dawson, S.~Dittmaier, and M.~Spira, {\it {Neutral Higgs boson pair
  production at hadron colliders: QCD corrections}},  {\em Phys. Rev.} {\bf
  D58} (1998) 115012, [\href{http://arxiv.org/abs/hep-ph/9805244}{{\tt
  hep-ph/9805244}}].

\bibitem{Frederix:2014hta}
R.~Frederix, S.~Frixione, V.~Hirschi, F.~Maltoni, O.~Mattelaer, P.~Torrielli,
  E.~Vryonidou, and M.~Zaro, {\it {Higgs pair production at the LHC with NLO
  and parton-shower effects}},  {\em Phys. Lett.} {\bf B732} (2014) 142--149,
  [\href{http://arxiv.org/abs/1401.7340}{{\tt arXiv:1401.7340}}].

\bibitem{Maltoni:2014eza}
F.~Maltoni, E.~Vryonidou, and M.~Zaro, {\it {Top-quark mass effects in double
  and triple Higgs production in gluon-gluon fusion at NLO}},  {\em JHEP} {\bf
  11} (2014) 079, [\href{http://arxiv.org/abs/1408.6542}{{\tt
  arXiv:1408.6542}}].

\bibitem{Grigo:2013rya}
J.~Grigo, J.~Hoff, K.~Melnikov, and M.~Steinhauser, {\it {On the Higgs boson
  pair production at the LHC}},  {\em Nucl. Phys.} {\bf B875} (2013) 1--17,
  [\href{http://arxiv.org/abs/1305.7340}{{\tt arXiv:1305.7340}}].

\bibitem{Grigo:2014jma}
J.~Grigo, K.~Melnikov, and M.~Steinhauser, {\it {Virtual corrections to Higgs
  boson pair production in the large top quark mass limit}},  {\em Nucl. Phys.}
  {\bf B888} (2014) 17--29, [\href{http://arxiv.org/abs/1408.2422}{{\tt
  arXiv:1408.2422}}].

\bibitem{Grigo:2015dia}
J.~Grigo, J.~Hoff, and M.~Steinhauser, {\it {Higgs boson pair production: top
  quark mass effects at NLO and NNLO}},  {\em Nucl. Phys.} {\bf B900} (2015)
  412, [\href{http://arxiv.org/abs/1508.00909}{{\tt arXiv:1508.00909}}].

\bibitem{Degrassi:2016vss}
G.~Degrassi, P.~P. Giardino, and R.~Gr{\"o}ber, {\it {On the two-loop virtual
  QCD corrections to Higgs boson pair production in the Standard Model}},  {\em
  Eur. Phys. J.} {\bf C76} (2016), no.~7 411,
  [\href{http://arxiv.org/abs/1603.00385}{{\tt arXiv:1603.00385}}].

\bibitem{deFlorian:2013uza}
D.~de~Florian and J.~Mazzitelli, {\it {Two-loop virtual corrections to Higgs
  pair production}},  {\em Phys. Lett.} {\bf B724} (2013) 306--309,
  [\href{http://arxiv.org/abs/1305.5206}{{\tt arXiv:1305.5206}}].

\bibitem{deFlorian:2013jea}
D.~de~Florian and J.~Mazzitelli, {\it {Higgs Boson Pair Production at
  Next-to-Next-to-Leading Order in QCD}},  {\em Phys. Rev. Lett.} {\bf 111}
  (2013) 201801, [\href{http://arxiv.org/abs/1309.6594}{{\tt
  arXiv:1309.6594}}].

\bibitem{deFlorian:2016uhr}
D.~de~Florian, M.~Grazzini, C.~Hanga, S.~Kallweit, J.~M. Lindert,
  P.~Maierhöfer, J.~Mazzitelli, and D.~Rathlev, {\it {Differential Higgs Boson
  Pair Production at Next-to-Next-to-Leading Order in QCD}},  {\em JHEP} {\bf
  09} (2016) 151, [\href{http://arxiv.org/abs/1606.09519}{{\tt
  arXiv:1606.09519}}].

\bibitem{Shao:2013bz}
D.~Y. Shao, C.~S. Li, H.~T. Li, and J.~Wang, {\it {Threshold resummation
  effects in Higgs boson pair production at the LHC}},  {\em JHEP} {\bf 07}
  (2013) 169, [\href{http://arxiv.org/abs/1301.1245}{{\tt arXiv:1301.1245}}].

\bibitem{deFlorian:2015moa}
D.~de~Florian and J.~Mazzitelli, {\it {Higgs pair production at
  next-to-next-to-leading logarithmic accuracy at the LHC}},  {\em JHEP} {\bf
  09} (2015) 053, [\href{http://arxiv.org/abs/1505.07122}{{\tt
  arXiv:1505.07122}}].

\bibitem{Borowka:2016ehy}
S.~Borowka, N.~Greiner, G.~Heinrich, S.~Jones, M.~Kerner, J.~Schlenk,
  U.~Schubert, and T.~Zirke, {\it {Higgs Boson Pair Production in Gluon Fusion
  at Next-to-Leading Order with Full Top-Quark Mass Dependence}},  {\em Phys.
  Rev. Lett.} {\bf 117} (2016), no.~1 012001, erratum ibid 079901,
  [\href{http://arxiv.org/abs/1604.06447}{{\tt arXiv:1604.06447}}].

\bibitem{Borowka:2016ypz}
S.~Borowka, N.~Greiner, G.~Heinrich, S.~P. Jones, M.~Kerner, J.~Schlenk, and
  T.~Zirke, {\it {Full top quark mass dependence in Higgs boson pair production
  at NLO}},  {\em JHEP} {\bf 10} (2016) 107,
  [\href{http://arxiv.org/abs/1608.04798}{{\tt arXiv:1608.04798}}].

\bibitem{Ferrera:2016prr}
G.~Ferrera and J.~Pires, {\it {Transverse-momentum resummation for Higgs boson
  pair production at the LHC with top-quark mass effects}},  {\em JHEP} {\bf
  02} (2017) 139, [\href{http://arxiv.org/abs/1609.01691}{{\tt
  arXiv:1609.01691}}].

\bibitem{Heinrich:2017kxx}
G.~Heinrich, S.~P. Jones, M.~Kerner, G.~Luisoni, and E.~Vryonidou, {\it {NLO
  predictions for Higgs boson pair production with full top quark mass
  dependence matched to parton showers}},  {\em JHEP} {\bf 08} (2017) 088,
  [\href{http://arxiv.org/abs/1703.09252}{{\tt arXiv:1703.09252}}].

\bibitem{Jones:2017giv}
S.~Jones and S.~Kuttimalai, {\it {Parton Shower and NLO-Matching uncertainties
  in Higgs Boson Pair Production}},  {\em JHEP} {\bf 02} (2018) 176,
  [\href{http://arxiv.org/abs/1711.03319}{{\tt arXiv:1711.03319}}].

\bibitem{Grober:2017uho}
R.~Gr{\"o}ber, A.~Maier, and T.~Rauh, {\it {Reconstruction of top-quark mass
  effects in Higgs pair production and other gluon-fusion processes}},  {\em
  JHEP} {\bf 03} (2018) 020, [\href{http://arxiv.org/abs/1709.07799}{{\tt
  arXiv:1709.07799}}].

\bibitem{Davies:2018ood}
J.~Davies, G.~Mishima, M.~Steinhauser, and D.~Wellmann, {\it {Double-Higgs
  boson production in the high-energy limit: planar master integrals}},  {\em
  JHEP} {\bf 03} (2018) 048, [\href{http://arxiv.org/abs/1801.09696}{{\tt
  arXiv:1801.09696}}].

\bibitem{Grazzini:2018bsd}
M.~Grazzini, G.~Heinrich, S.~Jones, S.~Kallweit, M.~Kerner, J.~M. Lindert, and
  J.~Mazzitelli, {\it {Higgs boson pair production at NNLO with top quark mass
  effects}},  {\em JHEP} {\bf 05} (2018) 059,
  [\href{http://arxiv.org/abs/1803.02463}{{\tt arXiv:1803.02463}}].

\bibitem{Grober:2015cwa}
R.~Gr{\"o}ber, M.~M{\"u}hlleitner, M.~Spira, and J.~Streicher, {\it {NLO QCD
  Corrections to Higgs Pair Production including Dimension-6 Operators}},  {\em
  JHEP} {\bf 09} (2015) 092, [\href{http://arxiv.org/abs/1504.06577}{{\tt
  arXiv:1504.06577}}].

\bibitem{Grober:2017gut}
R.~Gr{\"o}ber, M.~M{\"u}hlleitner, and M.~Spira, {\it {Higgs Pair Production at
  NLO QCD for CP-violating Higgs Sectors}},  {\em Nucl. Phys.} {\bf B925}
  (2017) 1--27, [\href{http://arxiv.org/abs/1705.05314}{{\tt
  arXiv:1705.05314}}].

\bibitem{deFlorian:2017qfk}
D.~de~Florian, I.~Fabre, and J.~Mazzitelli, {\it {Higgs boson pair production
  at NNLO in QCD including dimension 6 operators}},  {\em JHEP} {\bf 10} (2017)
  215, [\href{http://arxiv.org/abs/1704.05700}{{\tt arXiv:1704.05700}}].

\bibitem{Carvalho:2015ttv}
A.~Carvalho, M.~Dall'Osso, T.~Dorigo, F.~Goertz, C.~A. Gottardo, and M.~Tosi,
  {\it {Higgs Pair Production: Choosing Benchmarks With Cluster Analysis}},
  {\em JHEP} {\bf 04} (2016) 126, [\href{http://arxiv.org/abs/1507.02245}{{\tt
  arXiv:1507.02245}}].

\bibitem{Carvalho:2016rys}
A.~Carvalho, M.~Dall'Osso, P.~De~Castro~Manzano, T.~Dorigo, F.~Goertz,
  M.~Gouzevich, and M.~Tosi, {\it {Analytical parametrization and shape
  classification of anomalous HH production in the EFT approach}},
  \href{http://arxiv.org/abs/1608.06578}{{\tt arXiv:1608.06578}}.

\bibitem{Carvalho:2017vnu}
A.~Carvalho, F.~Goertz, K.~Mimasu, M.~Gouzevitch, and A.~Aggarwal, {\it {On the
  reinterpretation of non-resonant searches for Higgs boson pairs}},
  \href{http://arxiv.org/abs/1710.08261}{{\tt arXiv:1710.08261}}.

\bibitem{Buchalla:2017jlu}
G.~Buchalla, O.~Cata, A.~Celis, M.~Knecht, and C.~Krause, {\it {Complete
  One-Loop Renormalization of the Higgs-Electroweak Chiral Lagrangian}},  {\em
  Nucl. Phys.} {\bf B928} (2018) 93--106,
  [\href{http://arxiv.org/abs/1710.06412}{{\tt arXiv:1710.06412}}].

\bibitem{ALEPH:2005ab}
{\bf SLD Electroweak Group, DELPHI, ALEPH, SLD, SLD Heavy Flavour Group, OPAL,
  LEP Electroweak Working Group, L3} Collaboration, S.~Schael et~al., {\it
  {Precision electroweak measurements on the $Z$ resonance}},  {\em Phys.
  Rept.} {\bf 427} (2006) 257--454,
  [\href{http://arxiv.org/abs/hep-ex/0509008}{{\tt hep-ex/0509008}}].

\bibitem{Buchalla:2015wfa}
G.~Buchalla, O.~Cata, A.~Celis, and C.~Krause, {\it {Note on Anomalous
  Higgs-Boson Couplings in Effective Field Theory}},  {\em Phys. Lett.} {\bf
  B750} (2015) 298--301, [\href{http://arxiv.org/abs/1504.01707}{{\tt
  arXiv:1504.01707}}].

\bibitem{deFlorian:2016spz}
{\bf LHC Higgs Cross Section Working Group} Collaboration, D.~de~Florian
  et~al., {\it {Handbook of LHC Higgs Cross Sections: 4. Deciphering the Nature
  of the Higgs Sector}},  \href{http://arxiv.org/abs/1610.07922}{{\tt
  arXiv:1610.07922}}.

\bibitem{Deutschmann:2017qum}
N.~Deutschmann, C.~Duhr, F.~Maltoni, and E.~Vryonidou, {\it {Gluon-fusion Higgs
  production in the Standard Model Effective Field Theory}},  {\em JHEP} {\bf
  12} (2017) 063, [\href{http://arxiv.org/abs/1708.00460}{{\tt
  arXiv:1708.00460}}]. [Erratum: JHEP02,159(2018)].

\bibitem{Cullen:2011ac}
G.~Cullen, N.~Greiner, G.~Heinrich, G.~Luisoni, P.~Mastrolia, G.~Ossola,
  T.~Reiter, and F.~Tramontano, {\it {Automated One-Loop Calculations with
  GoSam}},  {\em Eur. Phys. J.} {\bf C72} (2012) 1889,
  [\href{http://arxiv.org/abs/1111.2034}{{\tt arXiv:1111.2034}}].

\bibitem{Cullen:2014yla}
G.~Cullen et~al., {\it {GoSam-2.0: a tool for automated one-loop calculations
  within the Standard Model and beyond}},  {\em Eur. Phys. J.} {\bf C74}
  (2014), no.~8 3001, [\href{http://arxiv.org/abs/1404.7096}{{\tt
  arXiv:1404.7096}}].

\bibitem{Degrande:2011ua}
C.~Degrande, C.~Duhr, B.~Fuks, D.~Grellscheid, O.~Mattelaer, and T.~Reiter,
  {\it {UFO - The Universal FeynRules Output}},  {\em Comput. Phys. Commun.}
  {\bf 183} (2012) 1201--1214, [\href{http://arxiv.org/abs/1108.2040}{{\tt
  arXiv:1108.2040}}].

\bibitem{Alloul:2013bka}
A.~Alloul, N.~D. Christensen, C.~Degrande, C.~Duhr, and B.~Fuks, {\it
  {FeynRules 2.0 - A complete toolbox for tree-level phenomenology}},  {\em
  Comput. Phys. Commun.} {\bf 185} (2014) 2250--2300,
  [\href{http://arxiv.org/abs/1310.1921}{{\tt arXiv:1310.1921}}].

\bibitem{Catani:1996vz}
S.~Catani and M.~H. Seymour, {\it {A General algorithm for calculating jet
  cross-sections in NLO QCD}},  {\em Nucl. Phys.} {\bf B485} (1997) 291--419,
  [\href{http://arxiv.org/abs/hep-ph/9605323}{{\tt hep-ph/9605323}}]. [Erratum:
  Nucl. Phys.B510,503(1998)].

\bibitem{Nagy:2003tz}
Z.~Nagy, {\it {Next-to-leading order calculation of three jet observables in
  hadron hadron collision}},  {\em Phys. Rev.} {\bf D68} (2003) 094002,
  [\href{http://arxiv.org/abs/hep-ph/0307268}{{\tt hep-ph/0307268}}].

\bibitem{Lepage:1980dq}
G.~P. Lepage, {\it {VEGAS: An Adaptive Multidimensional Integration Program}},
  {\em CLNS-80/447} (1980).

\bibitem{Hahn:2004fe}
T.~Hahn, {\it {CUBA: A Library for multidimensional numerical integration}},
  {\em Comput. Phys. Commun.} {\bf 168} (2005) 78--95,
  [\href{http://arxiv.org/abs/hep-ph/0404043}{{\tt hep-ph/0404043}}].

\bibitem{Martin:2009iq}
A.~D. Martin, W.~J. Stirling, R.~S. Thorne, and G.~Watt, {\it {Parton
  distributions for the LHC}},  {\em Eur. Phys. J.} {\bf C63} (2009) 189--285,
  [\href{http://arxiv.org/abs/0901.0002}{{\tt arXiv:0901.0002}}].

\bibitem{Butterworth:2015oua}
J.~Butterworth et~al., {\it {PDF4LHC recommendations for LHC Run II}},  {\em J.
  Phys.} {\bf G43} (2016) 023001, [\href{http://arxiv.org/abs/1510.03865}{{\tt
  arXiv:1510.03865}}].

\bibitem{CT14}
S.~Dulat, T.-J. Hou, J.~Gao, M.~Guzzi, J.~Huston, P.~Nadolsky, J.~Pumplin,
  C.~Schmidt, D.~Stump, and C.~P. Yuan, {\it {New parton distribution functions
  from a global analysis of quantum chromodynamics}},  {\em Phys. Rev.} {\bf
  D93} (2016), no.~3 033006, [\href{http://arxiv.org/abs/1506.07443}{{\tt
  arXiv:1506.07443}}].

\bibitem{MMHT14}
L.~A. Harland-Lang, A.~D. Martin, P.~Motylinski, and R.~S. Thorne, {\it {Parton
  distributions in the LHC era: MMHT 2014 PDFs}},  {\em Eur. Phys. J.} {\bf
  C75} (2015), no.~5 204, [\href{http://arxiv.org/abs/1412.3989}{{\tt
  arXiv:1412.3989}}].

\bibitem{NNPDF}
{\bf NNPDF} Collaboration, R.~D. Ball et~al., {\it {Parton distributions for
  the LHC Run II}},  {\em JHEP} {\bf 04} (2015) 040,
  [\href{http://arxiv.org/abs/1410.8849}{{\tt arXiv:1410.8849}}].

\bibitem{Buckley:2014ana}
A.~Buckley, J.~Ferrando, S.~Lloyd, K.~Nordstr{\"o}m, B.~Page, M.~R{\"u}fenacht,
  M.~Sch{\"o}nherr, and G.~Watt, {\it {LHAPDF6: parton density access in the
  LHC precision era}},  {\em Eur. Phys. J.} {\bf C75} (2015) 132,
  [\href{http://arxiv.org/abs/1412.7420}{{\tt arXiv:1412.7420}}].

\bibitem{Aad:2015gba}
{\bf ATLAS} Collaboration, G.~Aad et~al., {\it {Measurements of the Higgs boson
  production and decay rates and coupling strengths using pp collision data at
  $\sqrt{s}=7$ and 8 TeV in the ATLAS experiment}},  {\em Eur. Phys. J.} {\bf
  C76} (2016), no.~1 6, [\href{http://arxiv.org/abs/1507.04548}{{\tt
  arXiv:1507.04548}}].

\bibitem{CMS-PAS-HIG-17-031}
{\bf CMS} Collaboration, {\it {Combined measurements of the Higgs boson's
  couplings at $\sqrt{s}=13$ TeV}},  Tech. Rep. CMS-PAS-HIG-17-031, CERN,
  Geneva, 2018.

\bibitem{CMS:2018obr}
{\bf CMS} Collaboration, C.~Collaboration, {\it {Combination of searches for
  Higgs boson pair production in proton-proton collisions at $\sqrt{s} =
  13~\mathrm{TeV}$}}, .

\bibitem{Aaboud:2018knk}
{\bf ATLAS} Collaboration, M.~Aaboud et~al., {\it {Search for pair production
  of Higgs bosons in the $b\bar{b}b\bar{b}$ final state using proton-proton
  collisions at $\sqrt{s} = 13$ TeV with the ATLAS detector}},
  \href{http://arxiv.org/abs/1804.06174}{{\tt arXiv:1804.06174}}.

\bibitem{Sirunyan:2018iwt}
{\bf CMS} Collaboration, A.~M. Sirunyan et~al., {\it {Search for Higgs boson
  pair production in the $\gamma\gamma\mathrm{b\overline{b}}$ final state in pp
  collisions at $\sqrt{s}=$ 13 TeV}},
  \href{http://arxiv.org/abs/1806.00408}{{\tt arXiv:1806.00408}}.

\bibitem{Aaboud:2018ftw}
{\bf ATLAS} Collaboration, M.~Aaboud et~al., {\it {Search for Higgs boson pair
  production in the $\gamma\gamma b\bar{b}$ final state with 13 TeV $pp$
  collision data collected by the ATLAS experiment}},
  \href{http://arxiv.org/abs/1807.04873}{{\tt arXiv:1807.04873}}.

\bibitem{Sirunyan:2018hoz}
{\bf CMS} Collaboration, A.~M. Sirunyan et~al., {\it {Observation of
  $\mathrm{t\overline{t}}$H production}},  {\em Phys. Rev. Lett.} {\bf 120}
  (2018), no.~23 231801, [\href{http://arxiv.org/abs/1804.02610}{{\tt
  arXiv:1804.02610}}].

\bibitem{Aaboud:2018urx}
{\bf ATLAS} Collaboration, M.~Aaboud et~al., {\it {Observation of Higgs boson
  production in association with a top quark pair at the LHC with the ATLAS
  detector}},  \href{http://arxiv.org/abs/1806.00425}{{\tt arXiv:1806.00425}}.

\bibitem{Arzt:1994gp}
C.~Arzt, M.~B. Einhorn, and J.~Wudka, {\it {Patterns of deviation from the
  standard model}},  {\em Nucl. Phys.} {\bf B433} (1995) 41--66,
  [\href{http://arxiv.org/abs/hep-ph/9405214}{{\tt hep-ph/9405214}}].

\bibitem{Buchalla:2016sop}
G.~Buchalla, O.~Cata, A.~Celis, and C.~Krause, {\it {Comment on "Analysis of
  General Power Counting Rules in Effective Field Theory"}},
  \href{http://arxiv.org/abs/1603.03062}{{\tt arXiv:1603.03062}}.

\bibitem{Manohar:1983md}
A.~Manohar and H.~Georgi, {\it {Chiral Quarks and the Nonrelativistic Quark
  Model}},  {\em Nucl. Phys.} {\bf B234} (1984) 189--212.

\bibitem{Bagnaschi:2023rbx}
E.~Bagnaschi, G.~Degrassi and R.~Gr\"ober, \emph{{Higgs boson pair production
  at NLO in the POWHEG approach and the top quark mass uncertainties}},
  \href{http://dx.doi.org/10.1140/epjc/s10052-023-12238-8}{\emph{Eur. Phys. J.
  C} {\bfseries 83} (2023) 1054},
  [\href{https://arxiv.org/abs/2309.10525}{{\ttfamily 2309.10525}}].

\bibitem{Buchalla:2018yce}
G.~Buchalla, M.~Capozi, A.~Celis, G.~Heinrich and L.~Scyboz, \emph{{Higgs boson
  pair production in non-linear Effective Field Theory with full
  $m_t$-dependence at NLO QCD}},
  \href{http://dx.doi.org/10.1007/JHEP09(2018)057}{\emph{JHEP} {\bfseries 09}
  (2018) 057}, [\href{https://arxiv.org/abs/1806.05162}{{\ttfamily
  1806.05162}}].

\end{thebibliography}\endgroup

\end{document}